\pdfoutput=1
\documentclass{ws-rv9x6}

\ifpdf
\else
  \usepackage{ws-crop9x6}
\fi
\usepackage{ws-rv-thm}   
\usepackage{ws-rv-van}   

\makeindex

\usepackage[T1]{fontenc}
\usepackage[utf8]{inputenc}
\usepackage{lmodern}
\usepackage{float}

\usepackage{amsmath,amssymb,amsfonts,mathtools,bm}
\numberwithin{equation}{section}

\usepackage[english]{babel}
\usepackage{csquotes}

\usepackage[most]{tcolorbox} 
\tcbuselibrary{breakable}
\newtcolorbox{takeawaybox}{
  breakable,
  colback=gray!4,
  colframe=gray!55!black,
  title=Key Takeaways: Risk--Neutrality under Common Cause,
  fonttitle=\bfseries,
  sharp corners,
  left=8pt,
  right=8pt,
  top=6pt,
  bottom=6pt
}

\usepackage{graphicx}
\usepackage{subcaption} 

\makeatletter
\@ifpackageloaded{algorithm}{%

}{}
\makeatother

\usepackage{array,tabularx,booktabs}
\newcolumntype{C}{>{\centering\arraybackslash}X}
\usepackage[strict]{changepage}
\newlength{\extralength}
\usepackage{pdflscape}
\usepackage{adjustbox}
\usepackage[dvipsnames]{xcolor}

\usepackage{tikz}
\usetikzlibrary{arrows.meta,positioning,calc,fit,shapes.geometric,decorations.markings}
\usepackage{pgfplots}
\pgfplotsset{compat=1.18}

\newtheorem{Theorem}{Theorem}[section]
\newtheorem{Lemma}[Theorem]{Lemma}
\newtheorem{Proposition}[Theorem]{Proposition}
\newtheorem{Corollary}[Theorem]{Corollary}
\newtheorem{Definition}[Theorem]{Definition}
\newtheorem{Assumption}[Theorem]{Assumption}
\newtheorem{Remark}[Theorem]{Remark}

\usepackage{bbm}

\newcommand{\E}{\mathbb{E}}

\newcommand{\diag}{\operatorname{diag}}

\usepackage{titlesec}

\makeatletter
\AtBeginDocument{%
  \g@addto@macro\appendix{%
    \setcounter{section}{0}%
    \renewcommand\thesection{\Alph{section}}%
    \renewcommand\thesubsection{\thesection.\arabic{subsection}}%
    \renewcommand\theequation{\thesection.\arabic{equation}}%
    \numberwithin{equation}{section}%
  }%
}
\makeatother

\allowdisplaybreaks

\begin{document}

\chapter[Causal PDE-Control Models for Dynamic Portfolio Optimization with Latent Drivers]{Causal PDE-Control Models for Dynamic Portfolio Optimization with Latent Drivers\label{ra_ch1}}

\author[Alejandro Rodriguez Dominguez]{Alejandro Rodriguez Dominguez}

\address{Miraltabank, Head of Quantitative Analysis and Artificial Intelligence,\\
Plaza Manuel Gomez Moreno 2, Madrid 28043, \\
arodriguez@miraltabank.com}

\begin{abstract}

Classical portfolio models degrade under structural breaks, whereas flexible machine–learning allocation methods often lack arbitrage consistency and interpretability. We propose \emph{Causal PDE–Control Models} (CPCMs), a framework that integrates structural causal drivers, nonlinear filtering, and forward–backward PDE control to produce robust and transparent allocation rules under partial information. We construct driver–conditional risk-neutral measures on the observable filtration via filtering together with the corresponding martingale representation, linking pricing, hedging, and portfolio choice under a common information set. We further establish a projection–divergence duality showing that restricting portfolios to the causal driver span selects the feasible allocation closest to the unconstrained optimum under a convex divergence, thereby quantifying the stability cost of deviations from the causal manifold, and derive a causal completeness condition identifying when a finite driver span captures systematic premia. Markowitz, CAPM/APT, and Black–Litterman arise as limiting cases, while reinforcement learning and deep hedging appear as unconstrained approximations within the same pricing–control geometry. Empirically, on a U.S.\ equity panel with more than 300 candidate drivers, CPCM solvers achieve higher Sharpe ratios, lower turnover, and more persistent premia than econometric and machine–learning benchmarks.

\end{abstract}

\markboth{Alejandro Rodriguez Dominguez}{Causal PDE-Control Models for Dynamic Portfolio Optimization}

\body


\section{Introduction}

Classical portfolio theory, from mean–variance optimization \citep{10.2307/2975974} through the Capital Asset Pricing Model (CAPM) \citep{sharpe1964,lintner1965} and Arbitrage Pricing Theory (APT) \citep{ross1976}, rests on single-period or stationary settings with stable covariances and fixed factors. These assumptions are frequently violated by regime shifts and latent macro drivers, making allocations brittle. Extensions such as the Black–Litterman model (BL) \citep{black1992global}, Bayesian stress allocation 
\citep{RebonatoDenev}, and entropy pooling (EP) \citep{Meucci2008} improve estimation stability, 
while intertemporal formulations \citep{merton1973intertemporal} account for hedging demand. 
However, these approaches largely operate through scenario or prior specification rather 
than structural dynamic portfolio control.

A second strand of methodology uses stochastic control, filtering, and learning. Reinforcement learning (RL) \citep{moody1998performance,bellemare2023distributional}, deep hedging (DH) \citep{buehler2019deep}, and goal-based wealth management (GBWM) \citep{Dixon2020,ZHANG2024121578} capture rich nonlinearities but typically lack an explicit arbitrage-free pricing link or causal invariance to regime changes. Filtering models \citep{duffie2009frailty} and dynamic dependence constructions \citep{pareek2023copula,ITO2025103724} target hidden structure but are rarely integrated with valuation. Mathematical control approaches \citep{OksendalSulem2011,KARATZAS200989,GuZheng2020} provide rigor via Hamilton–Jacobi–Bellman equations (HJB) and Backward Stochastic Differential Equations (BSDEs) yet face dimensionality and calibration burdens.

Despite these advances, the literature still lacks a unified framework that simultaneously combines causal driver identification, arbitrage-consistent pricing, partial-information filtering, and dynamic portfolio control within a single tractable system.

In contrast, we take a structural route grounded in causality. The Commonality Principle (CP) \citep{RODRIGUEZDOMINGUEZ2023100447} posits that cross-sectional dependence is mediated by a small set of common, causal, and persistent drivers. Causal Partial Differential Equation–Control Models (CPCMs) embed these drivers into a pricing–control calculus as follows. First, conditional on a driver state, we construct a risk-neutral measure and extend the martingale representation on the observable filtration via nonlinear filtering. This ties pricing and hedging to the same information set used for allocation. Second, dynamic portfolio choice is posed as a forward–backward problem: a Fokker–Planck equation (FP) for state propagation coupled with a Hamilton–Jacobi–Bellman equation (HJB) for control. Third, we characterize the geometry of feasible allocations through the driver span. A projection–divergence duality shows that restricting portfolios to the causal driver span is equivalent to selecting, among feasible portfolio laws, the allocation closest to the unconstrained optimum under a convex divergence. The projection therefore admits an information–geometric interpretation: the feasible portfolio is the least informationally distorted element relative to the ideal unconstrained allocation, providing a quantitative measure of the stability cost of departing from the causal manifold. Finally, conformal transport and smooth evolution of the driver subspace ensure that co-movement patterns are preserved and exposures vary continuously as drivers are re-estimated.

Taken together, these components yield three practical outcomes. Pricing and hedging are arbitrage-consistent under evolving information; diversification and risk sharing are controlled by a finite, interpretable driver span (causal completeness); and allocations inherit geometric stability from conformal and smooth transport on the driver manifold. Classical methods (mean–variance optimization, CAPM/APT, and BL) emerge as constrained or limiting cases, while learning-based methods (RL and DH) can be viewed as unconstrained approximations once expressed in the same geometry.

The contributions of the paper are threefold. First, we develop a causal pricing–control framework linking filtered driver dynamics with arbitrage-consistent valuation under partial information. Second, we establish a projection–divergence duality that characterizes feasible portfolio allocations within the causal driver span and provides an information–geometric interpretation of portfolio constraints. Third, we implement a modular empirical architecture combining driver identification, nonlinear filtering via the Extended Kalman Filter (EKF) and Particle Filter (PF), and forward–backward partial differential equation (PDE) control for large-scale asset universes.

Empirically, we build a modular solver stack—driver identification under the Commonality Principle, EKF/PF filtering, coupled Fokker–Planck–Hamilton–Jacobi–Bellman control, and projection onto the driver span or its tangent space—and evaluate it on a U.S.\ equity panel with more than 300 candidate drivers. Across crisis, expansion, and transition regimes, CPCMs deliver higher Sharpe ratios with lower turnover and more persistent premia relative to econometric and machine-learning baselines, while preserving interpretability.

The remainder of the paper develops the framework in stages. Section~\ref{ResearchBackground} situates the approach within the portfolio optimization, stochastic control, and causal inference literatures. Section~\ref{sec:three} introduces the CPCM framework and establishes the theoretical results, including existence, replicability, and the projection–divergence characterization of feasible portfolios. Section~\ref{sec:design} describes the empirical architecture for driver identification, filtering, and numerical control. Section~\ref{sec:empirical} evaluates the approach on a large U.S.\ equity panel across multiple market regimes, and Section~\ref{sec:conclusion} summarizes implications for deployment.

\section{Literature Review}
\label{ResearchBackground}
Classical portfolio theory is rooted in static, single-period models. Mean--variance optimization \citep{10.2307/2975974}, the Capital Asset Pricing Model (CAPM) \citep{sharpe1964,lintner1965}, and the Arbitrage Pricing Theory (APT) \citep{ross1976} rely on stationary distributions, constant covariances, and linear factor structures. Although analytically elegant, these assumptions break down under structural shifts, leaving optimized portfolios fragile and highly sensitive to estimation error \citep{Fama1992,fama1993common,DeMiguel2009}. Extensions such as the Black–Litterman model \citep{black1992global,KolmRitter2021}, 
entropy pooling \citep{Meucci2008}, Bayesian nets for stress allocation 
\citep{RebonatoDenev}, and Bayesian model averaging \citep{HansenSargent2008} mitigate 
instability. Sequential entropy pooling approaches \citep{Vorobets2021SEP} further relax 
distributional restrictions by updating scenario weights over time. However, these 
methods remain focused on probability reweighting rather than structural dynamic 
portfolio control.

Dynamic approaches, beginning with Merton’s intertemporal formulation \citep{merton1973intertemporal}, recast portfolio choice as a stochastic control problem. The resulting Hamilton--Jacobi--Bellman (HJB) equations \citep{bellman1957,FlemingSoner2006,YongZhou1999} provide theoretical rigor but suffer from the curse of dimensionality. Classical treatments of stochastic control \citep{CvitKaratzas1992,KARATZAS200989,OksendalSulem2011} establish convex duality, stochastic portfolio theory, and Backward Stochastic Differential Equation (BSDE)-based formulations, later extended to constrained settings \citep{GuZheng2020}. Model predictive control and kinetic approaches \citep{Trimborn2019} further bridge portfolio optimization with control theory. Numerical approximations \citep{kushner2001numerical} and robust control \citep{HansenSargent2008} extend tractability, yet often yield overly conservative allocations in practice. Distributionally robust optimization (DRO) offers a related strand, where ambiguity sets defined by $f$-divergences or Wasserstein distances hedge against worst-case distributions \citep{delage2010distributionally,esfahani2018data}. While DRO provides strong guarantees, it typically treats all perturbations as adversarial and can become excessively pessimistic. CPCMs differ from the above by relying on structured state-space dynamics and causal drivers, while filtering latent states rather than extremizing over arbitrary distributions, thereby achieving robustness through adaptivity rather than uniform worst-case protection.

Machine learning has introduced flexible tools such as reinforcement learning \citep{moody1998performance,bellemare2023distributional,Dixon2020,ZHANG2024121578,Nakayama2023CausalIO}, deep hedging \citep{buehler2019deep,ruf2020hedging,Fu2022SolvingBO}, and related methods. These approaches learn adaptive strategies and capture nonlinearities, but generally lack interpretability, causal semantics, and robustness under regime changes. Sector- and causality-based approaches \citep{BishtKumar2023,RODRIGUEZDOMINGUEZ2023100447} show the promise of causal selection rules and structural driver identification, but remain limited to static or semi-dynamic designs. Parallel advances in applied mathematics show that deep neural architectures can approximate high-dimensional partial differential equations (PDEs): neural network solvers \citep{han2018,sirignano2018,raissi2019physics} and physics-informed neural networks (PINN) \citep{NoguerAlonso2023PINN} mitigate dimensionality barriers, linking stochastic control with scalable computation.

Research in causal inference and filtering provides complementary perspectives. Structural causal models formalize interventions and counterfactuals in economics and finance \citep{pearl2009causality,peters2017elements,chernozhukov2022causal}, while nonlinear filtering addresses partial observability through Kalman-type and particle methods \citep{bensoussan1992stochastic,bain2009fundamentals}. Recent causal machine learning approaches extend these tools through invariant prediction and generative causal models that learn counterfactual distributions or environment-invariant features \citep{peters2017elements,Arjovsky2019InvariantRM}. These methods are effective at identifying stable relationships across domains but typically do not produce sequential control policies. CPCMs differ by embedding causal structure directly into a forward--backward PDE system, producing dynamically optimal allocations under filtered latent states.

Dynamic portfolio choice under partial information also has been developed via HJB/viscosity and (forward--)backward SDE methodologies, including mean-field and separated (filtered) formulations \citet{Pham2009,CarmonaDelarue2018}. In this context, we use the term Causal PDE--Control to denote a structural framework with explicit causal drivers, conditional risk-neutral measures, and filtered martingale/replicability linked to FP--HJB/BSDE systems, distinct from physics-informed RL/data-driven PDE control \citep{han2018,sirignano2018,raissi2019physics}, which provide solver technologies but do not, by themselves, impose causal semantics or no-arbitrage structure.

PDE analysis contributes existence and uniqueness guarantees \citep{evans2010,JacodShiryaev2003,protter2005stochastic}, anchoring computational advances in rigorous mathematics. In economics and finance, causal ML combines flexible learners with Neyman-orthogonal moments to mitigate omitted-variable bias \citep{chernozhukov2022causal}, while SCMs employ the do-operator and d-separation to formalize interventions \citep{pearl2009causality,peters2017elements}. In parallel, latent-state inference has evolved from classical Kalman and particle filters \citep{bensoussan1992stochastic,bain2009fundamentals} to generative posteriors that scale beyond linear--Gaussian settings. These strands primarily target identification or state estimation and are typically decoupled from no-arbitrage valuation and dynamic control. CPCMs connect these developments into a unified framework by linking interventions on market drivers with arbitrage-consistent pricing and by integrating forward--backward PDEs with nonlinear filtering.

\subsection*{Notation}
\label{sec:notation}

\noindent
The main symbols used throughout the paper are summarised below.

\medskip

\begin{table}[H]
\centering
\scriptsize
\setlength{\tabcolsep}{4pt}
\begin{tabular}{@{}cl@{}}
\toprule
\textbf{Symbol} & \textbf{Description} \\
\midrule
$n,\, m,\, k,\, d_Y$ & assets, drivers ($m\!\ll\!n$), asset BM dim., observation dim. \\
$T,\, r,\, \gamma$ & horizon, short rate, risk aversion \\
$\mathbf{F}_t,\;\mathbf{f}$ & latent driver process $\in\mathbb{R}^m$; realization \\
$\mathbf{S}_t,\;\mathbf{r}_t$ & asset prices, instantaneous returns $\in\mathbb{R}^n$ \\
$\mathbf{Y}_t$ & observation process $\in\mathbb{R}^{d_Y}$ \\
$\mathbf{W}_t,\;\mathbf{W}^F_t,\;\mathbf{V}_t$ & Brownian motions (assets, drivers, observations) \\
$\mathbf{z}_t$ & driver-aligned coordinates, $\Psi_t\mathbf{r}_t\in\mathbb{R}^m$ \\
$p_t,\;\tilde{p}_t$ & portfolio return $\boldsymbol{\theta}_t^\top\mathbf{r}_t$; discounted value \\
$\boldsymbol{\theta}_t$ & portfolio weights $\in\mathbb{R}^n$ \\
$\boldsymbol{\phi}_\tau,\;\boldsymbol{\eta}_t$ & driver-space tilt $\in\mathbb{R}^m$; driver exposure $B_t^\top\boldsymbol{\theta}_t$ \\
$\boldsymbol{\mu}(\mathbf{F}_t,t)$ & asset drift vector $\in\mathbb{R}^n$; scalar component $\mu_i$ \\
$\boldsymbol{\lambda}_t$ & market price of risk $\in\mathbb{R}^k$ \\
$\boldsymbol{\beta}_i(t)$ & driver sensitivity $\in\mathbb{R}^m$, $\nabla_{\boldsymbol{D}^\star}\mathbb{E}[r_i\mid\boldsymbol{D}^\star]$ \\
$\sigma(\mathbf{F}_t,t)$ & diffusion loading $\in\mathbb{R}^{k\times n}$ \\
$\Sigma(\mathbf{F}_t,t)$ & conditional covariance $\sigma^\top\sigma\in\mathbb{R}^{n\times n}$ \\
$\Sigma^{\pi_t}(t)$ & posterior-integrated covariance $\int\Sigma(\mathbf{f},t)\,\pi_t(\mathrm{d}\mathbf{f})$ \\
$\boldsymbol{\Sigma}_F,\;\boldsymbol{\Sigma}_D,\;\boldsymbol{\Sigma}_\varepsilon,\;\boldsymbol{\Sigma}_Y$ & named covariances (drivers, candidates, idiosyncratic, obs.) \\
$B_t,\;U_t,\;\Lambda_t,\;\Psi_t$ & Jacobian, orth.\ basis, eigenvalues, whitening map $\Lambda_t^{1/2}U_t^\top$ \\
$K_t$ & Kalman gain \\
$\mathcal{U}\!=\!\{D_1,\ldots,D_M\}$ & candidate driver universe; $D_k$ scalar driver \\
$\boldsymbol{D}\in\mathbb{R}^m$ & driver random vector (theoretical sections) \\
$\boldsymbol{D}^\star$ & optimal driver vector; $\boldsymbol{D}^\star\!\subseteq\!\sigma(\mathbf{F}_t)$ \\
$D,\;D^\star$ & driver subset / selected set (empirical sections) \\
$\mathcal{A},\;\mathcal{W},\;\mathcal{K}$ & admissible set, feasible set, budget-constrained set \\
$\mathcal{V}_t,\;\mathcal{M}_\tau$ & driver span $\mathrm{Range}(\Sigma)$; estimated subspace $\mathrm{span}(U_\tau)$ \\
$\mathbb{P},\;\mathbb{Q}^{\mathbf{f}},\;\mathbb{Q}^{\pi_t}$ & physical, driver-conditional, posterior-integrated measures \\
$\pi_t,\;\mathcal{F}^Y_t,\;M_t^{\mathcal{F}^Y}$ & filtering posterior, observation filtration, innovation martingale \\
$\mathcal{G}(\boldsymbol{D}),\;\Delta_{ij}$ & screening objective $\sum_{i<j}\Delta_{ij}$; dependence functional \\
$\mathcal{J}(\boldsymbol{D}),\;\Psi(D)$ & coverage-overlap surrogate; residual commonality statistic \\
$\rho(\tilde{p},t\mid\mathbf{f}),\;u(\tilde{p},t\mid\mathbf{f})$ & conditional density and value function \\
$\Phi,\;s_\tau,\;\omega$ & terminal payoff, HJB amplitude, soft-PDE weight \\
$c_1(t),\;c_2(t),\;K_t$ & conformal scale factors; quasi-conformal distortion \\
$\varphi_{ij},\;\alpha_{ij}(t),\;\alpha'_{ij}(t)$ & unconditional, conditional, sensitivity-space angles \\
\bottomrule
\end{tabular}
\end{table}

\section{CPCM Framework and Theoretical Foundations}
\label{sec:three}

\subsection{Setup and Assumptions}
\label{sec:preliminaries}

This section introduces the building blocks of the framework:
(i) stochastic drivers and induced asset dynamics;
(ii) filtering and observation structures;
(iii) forward--backward PDE formulations; and
(iv) structural causal models.
Together, these elements provide the foundation for the Causal
PDE--Control Meta--Class (CPCM) in Section~\ref{sec:CPCM}.

\subsubsection{Stochastic Drivers and Asset Dynamics}
\label{sec:drivers-assets}

Let $(\Omega,\mathcal F,\{\mathcal F_t\}_{t\ge0},\mathbb P)$ be a filtered
probability space satisfying the usual conditions. We consider $n$ tradable assets with price vector
\[
\mathbf S_t=(S_t^{(1)},\ldots,S_t^{(n)})^\top,
\]
and $m$ common drivers $\mathbf F_t\in\mathbb R^m$. The drivers evolve as an It\^o diffusion

\begin{equation}
\mathrm d\mathbf F_t
=
\boldsymbol{\mu}_F(\mathbf F_t,t)\,\mathrm dt
+
\boldsymbol{\Sigma}_F(\mathbf F_t,t)\,\mathrm d\mathbf W_t^F,
\qquad
\mathbf F_0\in L^2(\Omega),
\label{eq:driver-sde}
\end{equation}
where $\mathbf W^F$ is a $d_F$--dimensional Brownian motion on
$(\Omega,\mathcal F,\mathbb P)$, and $\boldsymbol{\Sigma}_F$ are measurable maps
with local Lipschitz and linear--growth bounds ensuring a unique strong
solution\footnote{The diffusion
specification is adopted for analytical tractability and to align with
the continuous-time asset-pricing literature. The framework can be
extended to more general semimartingale drivers, including jump--diffusion
processes. In that case, the Fokker--Planck equation is replaced by the
corresponding integro--PDE, and the change of measure proceeds via
Girsanov transformations for semimartingales. The causal-span projection
and portfolio control structure remain unchanged.}. Conditional on $\mathbf F_t$, asset prices follow

\begin{equation}
\mathrm dS_t^{(i)}
=
S_t^{(i)}
\Big(
\mu_i(\mathbf F_t,t)\,\mathrm dt
+
\boldsymbol{\sigma}_i^\top(\mathbf F_t,t)\,\mathrm d\mathbf W_t
\Big),
\qquad i=1,\ldots,n,
\label{eq:asset-sde1}
\end{equation}
where $\mathbf W_t$ is a $k$--dimensional Brownian motion. Define the vector of instantaneous asset returns
\[
\mathbf r_t
=
\left(
\frac{\mathrm dS_t^{(1)}}{S_t^{(1)}},
\ldots,
\frac{\mathrm dS_t^{(n)}}{S_t^{(n)}}
\right)^\top.
\]

We allow instantaneous correlation between driver and asset shocks via
\[
\mathrm d\langle \mathbf W^F,\mathbf W\rangle_t
=
\Gamma(\mathbf F_t,t)\,\mathrm dt,
\qquad
\Gamma\in\mathbb R^{d_F\times k},
\quad
\|\Gamma\|\le 1,
\]
so the special case of independence is covered by $\Gamma\equiv 0$. Define the (column) loading matrix
\[
\boldsymbol \sigma(\mathbf F_t,t)
=
[\boldsymbol{\sigma}_1(\mathbf F_t,t),\ldots,\boldsymbol{\sigma}_n(\mathbf F_t,t)]
\in\mathbb R^{k\times n},
\]
and the conditional covariance
\[
\boldsymbol \sigma(\mathbf F_t,t)
=
\boldsymbol \sigma(\mathbf F_t,t)^\top\boldsymbol \sigma(\mathbf F_t,t)
\in\mathbb R^{n\times n}.
\]

Let $\mathcal A$ be the set of admissible self--financing strategies
$\boldsymbol{\theta}_t\in\mathbb R^n$ that are progressively measurable
with respect to $\{\mathcal F_t\}$ and satisfy
\[
\mathbb E\!\int_0^T
\boldsymbol{\theta}_t^\top
\boldsymbol \sigma(\mathbf F_t,t)
\boldsymbol{\theta}_t\,\mathrm dt
<\infty.
\]

The instantaneous portfolio return is
\[
p_t=\boldsymbol{\theta}_t^\top\mathbf r_t,
\]
with conditional variance
\[
\sigma_p^2(\mathbf F_t,t)
=
\boldsymbol{\theta}_t^\top\boldsymbol \sigma(\mathbf F_t,t)\boldsymbol{\theta}_t.
\]

Cross--sectional dependence is mediated by $\mathbf F_t$ in the sense
that $\{\mathrm dS_t^{(i)}/S_t^{(i)}\}_{i=1}^n$ are conditionally
independent given $\mathbf F_t$ up to the shared exposure captured by
$\boldsymbol \sigma(\mathbf F_t,t)$. This embeds CAPM/APT in continuous time with
explicit driver dynamics and permits latent or observed drivers.

\subsubsection{Filtering and Observation Structures}
\label{sec:filtering}

In practice, the driver process $\mathbf F_t$ is only partially observable. Market participants infer latent states from noisy measurements, such as asset prices and market drivers (i.e., factors, derivative pricing data, indexes, macroeconomic indicators). Let the observation process be
\begin{equation}
\mathrm d\mathbf Y_t = h(\mathbf F_t,t)\,\mathrm dt + \Sigma_Y^{1/2}\,\mathrm d\mathbf V_t,
\label{eq:observation-sde}
\end{equation}
where $\mathbf Y_t\in\mathbb R^{d_Y}$ collects $d_Y$ observable quantities, $h:\mathbb R^m\times[0,T]\to\mathbb R^{d_Y}$ is the observation function, $\Sigma_Y$ is a positive definite covariance matrix, and $\mathbf V_t$ is a $d_Y$–dimensional Brownian motion independent of the driver noise $\mathbf W^F$. The goal of filtering is to form the posterior distribution of $\mathbf F_t$ conditional on the observation history $\mathcal F^Y_t=\sigma(\mathbf Y_s:0\le s\le t)$. Denote this posterior by
\begin{equation}
\pi_t(\mathrm df) := \mathbb P\big(\mathbf F_t\in\mathrm df \,\big|\, \mathcal F^Y_t\big).
\label{eq:filtering-posterior}
\end{equation}
Under standard Assumptions (boundedness and Lipschitz continuity of the drift and diffusion coefficients of $\mathbf F_t$, and linear growth of $h$), the posterior is well defined and evolves according to nonlinear stochastic PDEs of Zakai or Kushner–Stratonovich type \citep{Kallianpur1980,bain2009fundamentals}. For a test function $\varphi:\mathbb R^m\to\mathbb R$, the Zakai equation reads
\[
\mathrm d\pi_t(\varphi) = \pi_t(\mathcal L^F \varphi)\,\mathrm dt 
+ \pi_t(\varphi h^\top)\,\Sigma_Y^{-1}\big(\mathrm d\mathbf Y_t - \pi_t(h)\,\mathrm dt\big),
\]
where $\mathcal L^F$ is the generator of $\mathbf F_t$. The second term captures the innovation process $\mathrm d\mathbf Y_t - \pi_t(h)\,\mathrm dt$, which is a $\mathcal F^Y_t$–Brownian motion under mild regularity. This decomposition makes explicit how new information flows into posterior beliefs.

From an economic perspective, $\pi_t$ summarizes investors’ evolving beliefs about the latent causal drivers of asset returns. Because portfolios are functions of $\pi_t$, uncertainty about $\mathbf F_t$ propagates directly into valuations and trading rules. This link between latent drivers, filtering, and portfolio choice unifies stochastic control under partial information \citep{LiptserShiryaev1977,Xiong2008} with the structural interpretation of causal modeling \citep{RODRIGUEZDOMINGUEZ2023100447}. Belief uncertainty becomes an explicit state variable in CPCMs, shaping both forward dynamics and backward optimization.

\subsubsection{PDE Formulation}
\label{sec:pde-formulation}
Forward--backward PDEs form the analytic backbone of CPCMs, linking probabilistic dynamics of returns to optimal portfolio choice. To avoid overloading notation, we write $\rho(\tilde p,t\mid \mathbf f)$ for a discounted portfolio value density conditional on driver state $\mathbf f$, reserving $\mathbf f$ exclusively for driver realizations rather than densities. Given $\mathbf F_t=\mathbf f$, the discounted portfolio value density evolves under the Fokker--Planck equation
\begin{equation}
\partial_t\rho(\tilde p,t\mid \mathbf f) = -\partial_{\tilde p}\!\left(\mu_{\tilde p}(\mathbf f)\,\rho(\tilde p,t\mid \mathbf f)\right)
         + \frac{1}{2}\partial_{\tilde p\tilde p}\!\left(\sigma^2_{\tilde p}(\mathbf f)\,\rho(\tilde p,t\mid \mathbf f)\right),
\label{eq:fokker-planck}
\end{equation}
where $\mu_{\tilde p}(\mathbf f)$ is the conditional drift of the discounted portfolio value and $\sigma^2_{\tilde p}(\mathbf f)$ the conditional variance. This forward PDE describes how the distribution of portfolio outcomes shifts and spreads over time, given the current driver state. Intuitively, $\mu_{\tilde p}(\mathbf f)$ governs the directional pull of the density, while $\sigma_{\tilde p}^2(\mathbf f)$ encodes risk dispersion. The associated value function $u(\tilde p,t\mid \mathbf f)$ satisfies the Hamilton--Jacobi--Bellman equation
\begin{equation}
\partial_t u + \sup_{\boldsymbol{\theta}\in\mathcal{W}}\left\{
         \mu_{\tilde p}(\mathbf f,\boldsymbol{\theta})\,\partial_{\tilde p} u + \frac{1}{2}\sigma^2_{\tilde p}(\mathbf f,\boldsymbol{\theta})\,\partial_{\tilde p\tilde p}u - r\,u\right\} = 0,
         \quad u(\tilde p,T\mid \mathbf f) = \Phi(\tilde p),
\label{eq:hjb}
\end{equation}
where $\mathcal W$ denotes the set of admissible self-financing controls and $\Phi$ the terminal payoff. The backward PDE encodes the optimization problem: the investor chooses weights $\boldsymbol{\theta}$ that maximize expected utility subject to risk and discounting. The Feynman--Kac formula provides the probabilistic representation of this solution, showing that $u$ equals the expected discounted utility under the forward law. When drivers are latent, the forward and backward PDEs are averaged against the filtering posterior $\pi_t$, yielding
\[
\bar \rho(\tilde p,t) 
= 
\int \rho(\tilde p,t\mid \mathbf f)\,\pi_t(\mathrm d\mathbf f),
\qquad
\bar u(\tilde p,t) 
= 
\int u(\tilde p,t\mid \mathbf f)\,\pi_t(\mathrm d\mathbf f).
\]
Portfolio dynamics and policies therefore reflect both intrinsic uncertainty and belief uncertainty, ensuring that pricing and control remain coherent when drivers are only partially observed. The forward equation describes ``what can happen'' to portfolio outcomes given driver dynamics, while the backward equation prescribes ``what should be done'' in response. Filtering integrates these layers by replacing unknown states with belief distributions, so optimal allocations are based on the best available information rather than on unobservable variables. This forward--backward--filtering triad constitutes the defining analytic structure of CPCMs.

\subsection{Optimal Portfolio Common--Cause Drivers Identification}
\label{Subsection41}

This section develops a complete identification program for optimal portfolio common-cause drivers. 
The program proceeds in several steps:
(i) SCM semantics and a star–DAG, which formalize the causal structure linking drivers and asset returns;
(ii) Reichenbach screening and $\varepsilon$-SCCS, which eliminate spurious cross-asset dependence via conditional independence;
(iii) a screening-off objective with consistent estimators and finite-sample guarantees, which provides a statistical criterion for selecting candidate driver sets;
(iv) the linear–Gaussian equivalence of screening metrics, connecting information-theoretic and covariance-based criteria;
(v) a solver suite for efficiently computing optimal driver subsets in high-dimensional libraries;
(vi) a principled choice of driver cardinality via a condition-number guard balancing explanatory power and numerical stability; and
(vii) conformal relations among unconditional, conditional, and sensitivity embeddings that preserve geometric consistency across portfolio representations.
Complete proofs, estimation details, and algorithmic specifications are collected in Appendix~\ref{app:commonality-opt}.

We consider \(n\) assets with returns \(r_1,\ldots,r_n\) and a large market of candidate drivers \(\mathcal U=\{D_1,\ldots,D_M\}\), with \(M \gg m\). The Commonality Principle posits that cross-sectional dependence in returns is mediated by a reduced set of common, causal, and persistent drivers \(\boldsymbol{D}^\star\subseteq\sigma(\mathbf F_t)\) \citep{RODRIGUEZDOMINGUEZ2023100447}.

\subsubsection{Preliminaries: SCM semantics and star--DAG}

Let $(Z_t,\mathbf F_t,\mathbf r_t)$ denote latent regimes, observable drivers,
and asset returns
\[
\mathbf r_t=(r_1(t),\ldots,r_n(t)).
\]
A structural system is
\[
Z_t=f_Z(\varepsilon^Z_t),\qquad 
\mathbf F_t=f_F(Z_t,\varepsilon^F_t),\qquad 
\mathbf r_t=f_r(\mathbf F_t,\varepsilon^r_t),
\]
with independent exogenous shocks $\varepsilon^Z_t,\varepsilon^F_t,\varepsilon^r_t$. Portfolio returns are
\[
p_t=\boldsymbol{\theta}_t^\top \mathbf r_t ,
\]
where $\boldsymbol{\theta}_t\in\mathbb{R}^n$ denotes the vector of portfolio weights. The candidate driver vector $\boldsymbol D_t$ is taken to be a measurable
function of the structural driver process $\mathbf F_t$, i.e.
\[
\boldsymbol D_t \subseteq \sigma(\mathbf F_t),
\]
so that the screening procedure selects observable proxies or
subsets of the underlying driver state.

At a fixed time, the operative star SCM asserts that the driver vector
\(\boldsymbol D_t\) (components \(D_1,\ldots,D_m\))
points to all asset returns \(r_i(t)\), and idiosyncratic residuals are
mutually independent conditional on \(\boldsymbol D_t\). The star--DAG encodes the idea that a small vector of causal drivers
simultaneously influences all assets, while any remaining dependence
disappears after conditioning on these drivers.
This is exactly the graphical expression of the Commonality Principle
and provides the logical backbone for the Reichenbach Common Cause
Principle (RCCP)-based screening \citep{reichenbach-time}.

\begin{figure}[htbp]
\centering
\begin{tikzpicture}[
  node distance=1.6cm and 1.6cm,
  drv/.style={circle,draw,fill=gray!10,minimum size=9mm,inner sep=1pt},
  ast/.style={rectangle,rounded corners,draw,fill=blue!6,minimum width=14mm,minimum height=7mm,inner sep=2pt},
  arr/.style={-{Latex[length=2mm]},thick}
]
\node[drv] (d1) {$D_1$};
\node[drv, right=of d1] (d2) {$D_2$};
\node[drv, right=of d2] (d3) {$D_3$};
\node[drv, right=of d3] (dm) {$D_m$};

\node[ast, below=1.8cm of d1] (a1) {$r_1$};
\node[ast, below=1.8cm of d2] (a2) {$r_2$};
\node[ast, below=1.8cm of d3] (a3) {$r_3$};
\node[ast, below=1.8cm of dm] (an) {$r_n$};

\foreach \x in {d1,d2,d3,dm}{
  \foreach \y in {a1,a2,a3,an}{
    \draw[arr] (\x) -- (\y);
  }
}
\end{tikzpicture}

\caption{Star--DAG representation of the Commonality Principle.
Drivers \(D_1,\ldots,D_m\) causally influence all asset returns
\(r_1,\ldots,r_n\); conditional on the driver vector \(\boldsymbol D_t\),
idiosyncratic residuals are mutually independent.}

\label{fig:star-dag-commonality}
\end{figure}

Interventions \(\mathrm{do}(\boldsymbol D_t=\mathbf d)\) sever incoming edges
into \(\boldsymbol D_t\) and yield interventional laws for returns
\(\mathbf r_t\) and portfolio returns \(p_t\). With partial observability, a filtering posterior \(\pi_t\) over
\(\boldsymbol D_t\) produces mixtures
\[
\bar\rho(p,t)
=
\int
\rho(p,t\mid \mathrm{do}(\boldsymbol D_t=\mathbf d))
\,\pi_t(\mathrm d\mathbf d).
\]

\begin{Definition}[Commonality Principle
\citep{RODRIGUEZDOMINGUEZ2023100447}]
\label{def:commonality}
A driver vector \(\boldsymbol{D}^\star\subseteq\sigma(\mathbf F_t)\) is optimal if it is common across
assets, causal for returns, and persistent. Conditioning on \(\boldsymbol{D}^\star\) eliminates spurious correlations while
retaining idiosyncratic diversification.
\end{Definition}

"Common" guarantees cross-sectional scope (each asset is affected);
"causal" keeps us on the correct side of the Reichenbach Common Cause
Principle (screening-off after conditioning); "persistent" rules out
transient co-movements that fail to deliver stable exposures and premia over the estimation horizon, with the driver vector
re-estimated as regimes evolve.

\subsubsection{Reichenbach Common Cause Principle (RCCP) based screening}
\label{subsec:rccp-sccs}

For events \(A,B\), Reichenbach's principle states that if
\(
P(A\cap B)\neq P(A)P(B),
\)
then there exists a variable \(C\) such that
\[
\begin{aligned}
P(A\cap B\mid C) &= P(A\mid C)P(B\mid C),\\
P(A\mid C) &\ne P(A\mid \neg C),\qquad
P(B\mid C) \ne P(B\mid \neg C).
\end{aligned}
\]

For real-valued asset returns we translate this principle into a
conditional screening requirement: conditioning on the correct driver
vector \(\boldsymbol D\) should eliminate cross--asset dependence.
We employ three equivalent screening lenses:
\[
\begin{aligned}
\text{(Info)}\quad 
& I(r_i;r_j\mid \boldsymbol D)=0,\\
\text{(Cov)}\quad 
& \mathrm{Cov}(r_i,r_j\mid \boldsymbol D)=0,\\
\text{(Event)}\quad 
& \bigl|P(E_i,E_j\mid D_\ell)-P(E_i\mid D_\ell)P(E_j\mid D_\ell)\bigr|\le\varepsilon,
\end{aligned}
\]
where \(E_i=\{r_i>\tau_i\}\) and \(\{D_\ell\}\) denotes an empirical
partition of \(\boldsymbol D\). Each lens asserts that conditioning on \(\boldsymbol D\) removes
cross--asset dependence. The three lenses above are particular instances
of dependence functionals that vanish under conditional independence.
To treat these cases within a unified framework we introduce a
general notion of an admissible screening functional capturing
the screening-off property independently of the specific metric used.

\begin{Definition}[Admissible screening metric]
\label{def:admissiblescreeningmetric}
Let
\[
\Delta_{ij}(\boldsymbol D)
:=
\Delta(r_i,r_j \mid \boldsymbol D)
\]
denote a pairwise conditional dependence functional between asset
returns \(i\) and \(j\) given the candidate driver vector \(\boldsymbol D\).
The functional \(\Delta_{ij}\) is \emph{admissible} for driver identification if:
\begin{enumerate}
\item \(\Delta_{ij}(\boldsymbol D)\ge0\);
\item \(\Delta_{ij}(\boldsymbol D)=0\) whenever \(r_i\perp r_j\mid \boldsymbol D\);
\item \(\Delta_{ij}\) admits a consistent estimator with the concentration
guarantees required for the empirical screening objective defined below.
\end{enumerate}
\end{Definition}

With this definition, the identification program does not depend on a
specific dependence metric but only on the existence of an admissible
screening functional \(\Delta_{ij}(\boldsymbol D)\).
Conditional mutual information provides a canonical general-purpose
choice because it vanishes if and only if conditional independence
holds under standard regularity conditions.
Covariance-based criteria become equivalent under stronger structural
assumptions. The admissible screening functional therefore induces the global
screening objective
\[
\mathcal G(\boldsymbol D)=\sum_{i<j}\Delta_{ij}(\boldsymbol D),
\]
which aggregates conditional dependence across asset pairs and
provides the theoretical criterion for driver identification.
Empirical procedures used later should therefore be interpreted as
operational proxies for this objective, implemented through practical
dependence or evidence measures that aim to identify driver vectors
rendering asset returns conditionally independent.

\subsubsection{Screening--off objective under the Commonality Principle, estimators, and guarantees}
\label{subsec:objective}

Let \(\Delta_{ij}(\boldsymbol{D})\) denote the admissible pairwise screening
functional of Definition~\ref{def:admissiblescreeningmetric}. For \(\boldsymbol{D}\subseteq\mathcal{U}\) with \(|\boldsymbol{D}|=m\),

\begin{equation}
\label{eq:G-functional-rekey}
\begin{aligned}
\mathcal G(\boldsymbol{D})
&:=\sum_{1\le i<j\le n}\Delta_{ij}(\boldsymbol{D}),\\[4pt]
\Delta_{ij}(\boldsymbol{D})
&\in
\Big\{
I(r_i;r_j\mid\boldsymbol{D}),\;
|\mathrm{Cov}(r_i,r_j\mid\boldsymbol{D})|,\;
\text{event deviations}
\Big\}.
\end{aligned}
\end{equation}

\begin{Proposition}[Identification via screening--off]
\label{prop:identification-screening}

Let \(\boldsymbol{D}^\star\subseteq \mathcal{U}\) denote the true
common driver vector satisfying the screening--off property.
If \(\Delta_{ij}(\boldsymbol{D})\) is admissible in the sense of
Definition~\ref{def:admissiblescreeningmetric}, then
\[
\mathcal G(\boldsymbol{D}^\star)=0,
\]
and any candidate driver vector \(\boldsymbol{D}\) satisfying
\(\mathcal G(\boldsymbol{D})=0\) renders all asset pairs conditionally
independent. Consequently, minimizing \(\mathcal G(\boldsymbol{D})\) identifies the
smallest driver vector that renders asset returns conditionally independent.

\end{Proposition}

\begin{proof}

By admissibility,
\(\Delta_{ij}(\boldsymbol{D}^\star)=0\) whenever
\(r_i\perp r_j\mid\boldsymbol{D}^\star\).
Summing over all \(i<j\) yields

\[
\mathcal G(\boldsymbol{D}^\star)
=
\sum_{i<j}\Delta_{ij}(\boldsymbol{D}^\star)
=
0.
\]

Conversely, if \(\mathcal G(\boldsymbol{D})=0\),
nonnegativity of each \(\Delta_{ij}\) implies
\(\Delta_{ij}(\boldsymbol{D})=0\) for all \(i<j\),
which by admissibility implies conditional independence of asset
returns given \(\boldsymbol{D}\). Minimality follows by restricting attention to candidate driver vectors of
smallest cardinality achieving \(\mathcal G(\boldsymbol{D})=0\).

\end{proof}

The optimization objective becomes

\begin{equation}
\label{eq:selection-rekey}
\min_{\boldsymbol{D}\subseteq \mathcal{U},\;|\boldsymbol{D}|=m}
\mathcal G(\boldsymbol{D})
\quad\text{s.t.}\quad
\mathcal G(\boldsymbol{D})\le m\varepsilon .
\end{equation}

Given data \((r_t,X_t)\), where
\(X_t=(D_{1,t},\ldots,D_{m,t})\):

\begin{itemize}

\item \textbf{Information form:} estimate conditional mutual information
\(I(r_i;r_j\mid\boldsymbol{D})\) using kNN/KSG, copula, or neural
mutual-information estimators with bias correction.

\item \textbf{Covariance form:} regress \(r_i\) on \(\boldsymbol{D}\),
obtain residuals \(\tilde r_i\), and compute
\(|\widehat{\mathrm{Cov}}(\tilde r_i,\tilde r_j)|\).

\item \textbf{Event form:} use empirical partitions \(\{D_\ell\}\) and
deviations of conditional event probabilities.

\end{itemize}

Define \(\widehat{\mathcal G}_T(\boldsymbol{D})\) by plugging these
estimators into \eqref{eq:G-functional-rekey}.
If the sequence \((r_t,X_t)\) is strictly stationary and exponentially
\(\alpha\)-mixing, then there exist constants \(C_1,C_2>0\) such that

\[
\Pr\!\left(
|\widehat{\mathcal G}_T(\boldsymbol{D})-\mathcal G(\boldsymbol{D})|>\eta
\right)
\le C_1\exp(-C_2T\eta^2),
\]

providing finite--sample guarantees.
Further details on estimating \(\mathcal G(\boldsymbol{D})\),
concentration bounds, and implementation appear in
Appendix~\ref{app:estimating-G-rekey}.
To convert the statistical screen into an implementable driver
selection, we introduce a coverage--overlap surrogate that favors broad
cross--asset coverage while penalizing redundant variables:

\begin{align}
\label{eq:breadth}
\mathrm{Breadth}(D_k)
  &:= \sum_i
     \mathbf{1}\!\left\{\,|\mathrm{Corr}(D_k,r_i)|\ge \tau\,\right\},\\
\label{eq:overlap}
\mathrm{Overlap}(D_k,D_\ell)
  &:= \sum_i
     \mathbf{1}\!\left\{\,(D_k,r_i)\land(D_\ell,r_i)\,\right\},
\end{align}

and optimize

\begin{equation}
\label{eq:J-objective-rekey}
\mathcal J(\boldsymbol{D})
=
\alpha\,\mathcal G(\boldsymbol{D})
-\beta\!\sum_{D_k\in\boldsymbol{D}}\mathrm{Breadth}(D_k)
+\gamma\!\sum_{k<\ell}\mathrm{Overlap}(D_k,D_\ell),
\end{equation}

subject to \(\mathcal G(\boldsymbol{D})\le M\varepsilon\).

\subsubsection{Linear--Gaussian specialization and equivalence}
\label{subsec:lin-gauss-rekey}

Assume the linear--Gaussian model
\[
\begin{aligned}
r_i &= \boldsymbol{\beta}_i^\top \boldsymbol{D} + \epsilon_i, \qquad
\epsilon \perp \boldsymbol{D},\\
\epsilon &\sim \mathcal N(0,\Psi), \qquad
\Psi = \diag(\sigma_1^2,\ldots,\sigma_n^2),\\
\boldsymbol{D} &\sim \mathcal N(0,\boldsymbol{\Sigma}_D).
\end{aligned}
\]

\begin{Proposition}[Equivalence of screening metrics]
\label{prop:eqv}

Under the above model, for all \(i\neq j\):

\[
I(r_i;r_j\mid\boldsymbol{D})=0
\iff
\mathrm{Cov}(r_i,r_j\mid\boldsymbol{D})=0
\iff
\boldsymbol{\beta}_i^\top\boldsymbol{\Sigma}_D\boldsymbol{\beta}_j=0.
\]

Moreover,

\[
\arg\min_{\boldsymbol{D}}\sum_{i<j}I(r_i;r_j\mid\boldsymbol{D})
=
\arg\min_{\boldsymbol{D}}\sum_{i<j}|\boldsymbol{\beta}_i^\top\boldsymbol{\Sigma}_D\boldsymbol{\beta}_j|.
\]

\end{Proposition}

\noindent
The proof follows standard Gaussian--information equivalence arguments and is provided in Appendix~\ref{app:proof-prop-eqv}. In this widely used benchmark, all three lenses collapse to the requirement that fitted components be orthogonal across assets, measured by \(\boldsymbol{\beta}_i^\top\boldsymbol{\Sigma}_D\boldsymbol{\beta}_j\).

\subsubsection{Condition--number guard and driver cardinality}
\label{subsec:condnum}

Let \(\boldsymbol{X}\in\mathbb R^{T\times m}\) denote the standardized driver matrix and
\(\boldsymbol{R}\in\mathbb R^{T\times n}\) the matrix of asset returns
\((r_1,\ldots,r_n)\).
For the driver vector \(\boldsymbol{D}_m\) minimizing \(\widehat{\mathcal J}_T\),
compute sensitivities \(\widehat B(M)\) from regression of \(\boldsymbol{R}\) on
\(\boldsymbol{X}_{\boldsymbol{D}_m}\) and define

\[
\kappa(M)=\mathrm{cond}\!\left(\widehat B(M)^\top\widehat B(M)\right),\qquad
\Delta\widehat{\mathcal G}_T(M)
=
\widehat{\mathcal G}_T(\boldsymbol{D}_{M-1})
-
\widehat{\mathcal G}_T(\boldsymbol{D}_m).
\]

The stopping rule is to choose the smallest \(\widehat M\) such that

\[
\Delta\widehat{\mathcal G}_T(M)<\eta
\quad\text{or}\quad
\kappa(M)>\kappa_{\max}.
\]

A full derivation of the stability bound and an expanded design--matrix
penalized program are presented in
Appendix~\ref{subsec:condnum-opt}.
Aggressively increasing \(M\) can inflate \(\kappa(M)\), making
\(\widehat B(M)\) unstable even if \(\mathcal G\) continues to improve.
The guard therefore trades off screening strength against numerical
robustness, preventing spurious sensitivity blow-ups.

The Commonality Principle provides necessary and sufficient conditions
for selecting drivers that yield optimal diversification\citep{RODRIGUEZDOMINGUEZ2023100447}.
When moving from the unconditional Markowitz geometry (without
exogenous variables) to a driver representation, idiosyncratic
diversification must be preserved while systematic diversification is
added.
These favorable conditions hold precisely when the drivers are common
and causal.
In that case, the transformation from the unconditional space to the
driver span is a conformal map, and the subsequent transformation from
the driver span to the local sensitivity (tangent) space is also a
conformal map.
Because conformality preserves angle proportions while allowing
time-varying rescaling, optimization carried out either directly on the
driver span or on its tangent space is basis-invariant and maintains
maximal idiosyncratic diversification alongside added systematic
exposure \citep{RODRIGUEZDOMINGUEZ2023100447}.

\subsubsection{Conformal maps: unconditional {$\rightarrow$} conditional {$\rightarrow$} sensitivity}
\label{sectionGeometry-rekey}

Under common and causal drivers, the embedding from the unconditional geometry to the conditional geometry (driver span) and the subsequent embedding from the driver span to the local sensitivity (tangent) space are both conformal. Conformality preserves angle proportions (and thus pairwise co-movement patterns) while permitting time-varying rescaling; this is the precise condition under which optimization on the driver span or on its tangent space remains invariant to the choice of basis. We now set notation:
\[
\mu_i=\E[r_i],\qquad
\mu_i^{\boldsymbol{D}^\star}(t)=\E[r_i(t)\mid \boldsymbol{D}^\star_t],\qquad
\boldsymbol{\beta}_i(t)=\nabla_{\boldsymbol{D}^\star}\E[r_i(t)\mid \boldsymbol{D}^\star_t]\in\mathbb R^m.
\]

For a window $\mathcal T$, set centered embeddings
\[
\mathbf m_i := (\mu_i(s)-\bar\mu_i)_{s\in\mathcal T},\ 
\mathbf m^{\boldsymbol{D}^\star}_i(t)=(\mu_i^{\boldsymbol{D}^\star}(s)-\overline{\mu_i^{\boldsymbol{D}^\star}})_{s\in\mathcal T}.
\] 

Define the cosines

\begin{align}
\cos\varphi_{ij}&=\frac{\langle\mathbf m_i,\mathbf m_j\rangle}{\|\mathbf m_i\|\,\|\mathbf m_j\|},\nonumber\\
\cos\alpha_{ij}(t)&=\frac{\langle\mathbf m_i^{\boldsymbol{D}^\star}(t),\mathbf m_j^{\boldsymbol{D}^\star}(t)\rangle}{\|\mathbf m_i^{\boldsymbol{D}^\star}(t)\|\,\|\mathbf m_j^{\boldsymbol{D}^\star}(t)\|},\nonumber\\
\cos\alpha'_{ij}(t)&=\frac{\langle\boldsymbol{\beta}_i(t),\boldsymbol{\beta}_j(t)\rangle}{\|\boldsymbol{\beta}_i(t)\|\,\|\boldsymbol{\beta}_j(t)\|}.
\label{eq:cosine-definitions}
\end{align}

\begin{Assumption}[Common, causal, persistent drivers]
\label{ass:ccd-rekey}
There exists \(\boldsymbol{D}^\star\subseteq\sigma(\mathbf F_t)\) such that, for all \(i\neq j\):  
(i) \(r_i\perp r_j\mid \boldsymbol{D}^\star\) (common cause);  
(ii) $\mu_i^{\boldsymbol{D}^\star}(t)$ are path--differentiable in the drivers;  
(iii) each driver is weakly persistent (e.g., OU/AR(1) with positive spectral density at zero).  
Moreover, the scale ratio between embeddings is allowed to be time--varying.
\end{Assumption}

\begin{Theorem}[Conformality: Unconditional $\to$ Conditional]
\label{thm:conf1-rekey}
Under Assumption~\ref{ass:ccd-rekey}, there exists a time--dependent scale $c_1(t)>0$ such that
\[
\cos\alpha_{ij}(t)=\cos\varphi_{ij},
\qquad
\|\mathbf m^{\boldsymbol{D}^\star}_i(t)\|=c_1(t)\,\|\mathbf m_i\|
\quad \text{for all } i,j.
\]
Hence, the map from the unconditional to the conditional embedding is conformal: it preserves angles (co-movement patterns) while allowing time--varying rescaling.
\end{Theorem}

\begin{Theorem}[Conformality: Conditional $\to$ Sensitivity]
\label{thm:conf2-rekey}
Under Assumption~\ref{ass:ccd-rekey} with differentiable $\mu_i^{\boldsymbol{D}^\star}$, there exists a time--dependent scale $c_2(t)>0$ such that
\[
\cos\alpha'_{ij}(t)=\cos\alpha_{ij}(t),
\qquad
\|\boldsymbol{\beta}_i(t)\|=c_2(t)\,\|\mathbf m^{\boldsymbol{D}^\star}_i(t)\|
\quad \text{for all } i,j.
\]
Thus, the map from the conditional embedding to the sensitivity (beta) space is conformal.
\end{Theorem}

\begin{Corollary}[Two--step conformality and time--varying ratio]
\label{cor:two-step-rekey}
Combining Theorems~\ref{thm:conf1-rekey}--\ref{thm:conf2-rekey}, the unconditional and sensitivity embeddings are conformally equivalent with \(c(t)=c_1(t)c_2(t)\):
\[
\cos\alpha'_{ij}(t)=\cos\varphi_{ij},
\qquad
\|\boldsymbol{\beta}_i(t)\|=c(t)\,\|\mathbf m_i\|.
\]
\end{Corollary}

\noindent
Proofs of Theorems~\ref{thm:conf1-rekey}--\ref{thm:conf2-rekey} are given in Appendix~\ref{app:commonality-opt}, with persistence details in Appendix~\ref{app:persistence-detail-rekey} and empirical diagnostics in Appendix~\ref{app:conf-estimation-rekey}.

The following proposition formalizes the link between the conformal embeddings established above and the invariance of portfolio optimization performed on the driver span or on its tangent space.

\begin{Proposition}[Conformality and invariance of span/tangent representations]
\label{prop:necessity-span}
Let the mappings from the unconditional return space to the driver span and from the driver span to the local sensitivity space be the embeddings established in Theorems~\ref{thm:conf1-rekey}--\ref{thm:conf2-rekey}. Preservation of pairwise angle proportions under these mappings is necessary and sufficient for objectives defined on the driver span or its tangent space to remain invariant under orthonormal reparametrizations of the driver basis. Equivalently, span or tangent-plane optimization is basis-invariant if and only if these embeddings are conformal transformations whose scale factors depend only on time.
Consequently, to maintain maximum idiosyncratic diversification while adding systematic exposure in driver space, both embeddings must be conformal.
\end{Proposition}

\begin{proof}
By Theorems~\ref{thm:conf1-rekey}--\ref{thm:conf2-rekey}, the maps from the unconditional geometry to the driver span and from the driver span to the local sensitivity space are conformal, preserving pairwise angles up to time-varying scalar factors. Portfolio objectives defined on the driver span or its tangent space depend on the relative geometry of exposures, encoded through inner products between embedding vectors. Under orthonormal reparametrizations of the driver basis, conformal maps preserve these geometric relationships and therefore leave such objectives invariant. Conversely, if angle proportions were not preserved, orthonormal changes of basis would alter the relative geometry of exposures and therefore the optimization objective. Hence conformality of both embeddings is necessary and sufficient for basis-invariant span/tangent-plane optimization.
\end{proof}

\begin{figure}[htbp]
  \centering
  \includegraphics[width=\linewidth]{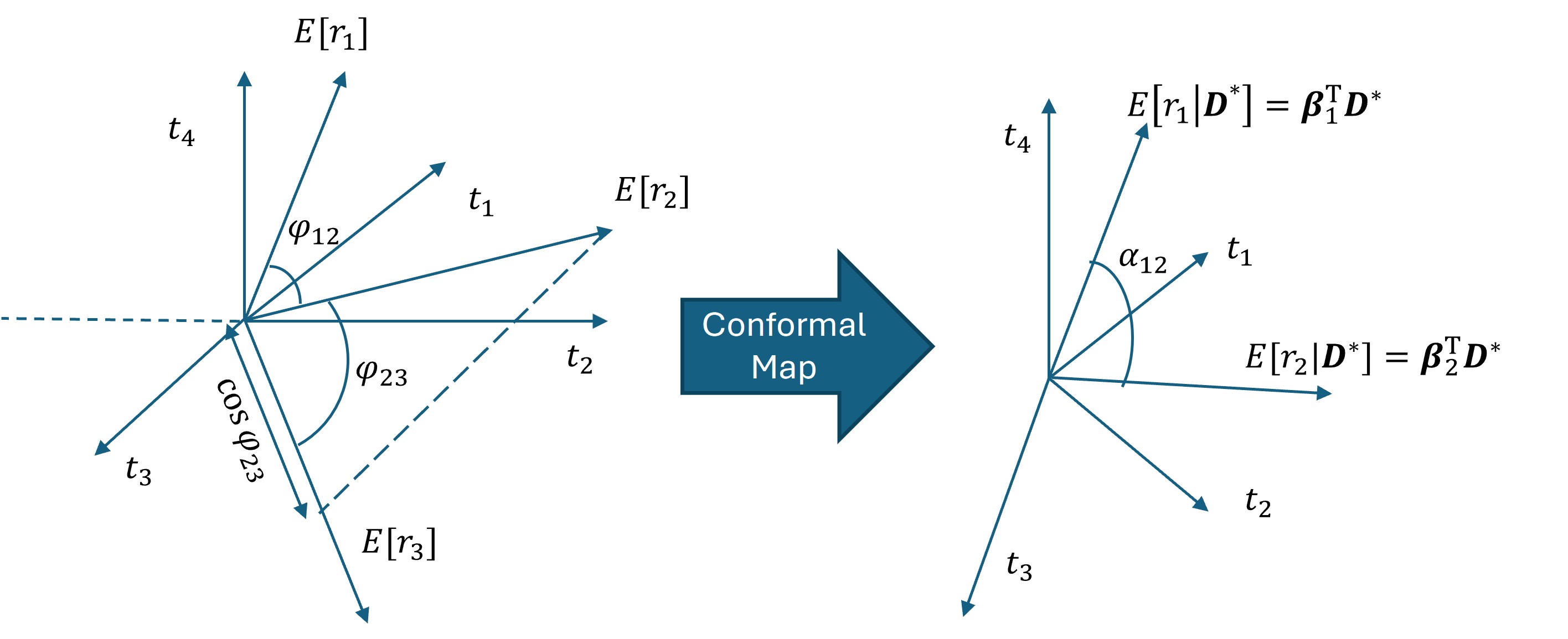}
  \caption{First conformal map: unconditional $\to$ conditional.
  The figure illustrates the conformal embedding from the unconditional
  expected-return time embedding to the conditional time embedding
  induced by the driver subspace spanned by
  $\boldsymbol{D}^\star$. Expected-return vectors are mapped into their
  conditional counterparts
  $\mathbb{E}[r_i \mid \boldsymbol{D}^\star]
    = \boldsymbol{\beta}_i^\top \boldsymbol{D}^\star$,
  while pairwise angle proportions are preserved. In particular, if
  $\varphi_{ij}$ denotes the unconditional angle and $\alpha_{ij}(t)$
  the conditional angle, then
  $\cos\alpha_{ij}(t) = \cos\varphi_{ij}$, with embedding norms
  rescaled by the time-varying conformal factor~$c_1(t)$, i.e.\
  $\lVert \boldsymbol{m}^{\boldsymbol{D}^\star}_i(t)\rVert
    = c_1(t)\,\lVert \boldsymbol{m}_i\rVert$.}
  \label{fig:enter-1stconformalmap}
\end{figure}

\begin{figure}[htbp]
  \centering
  \includegraphics[width=\linewidth]{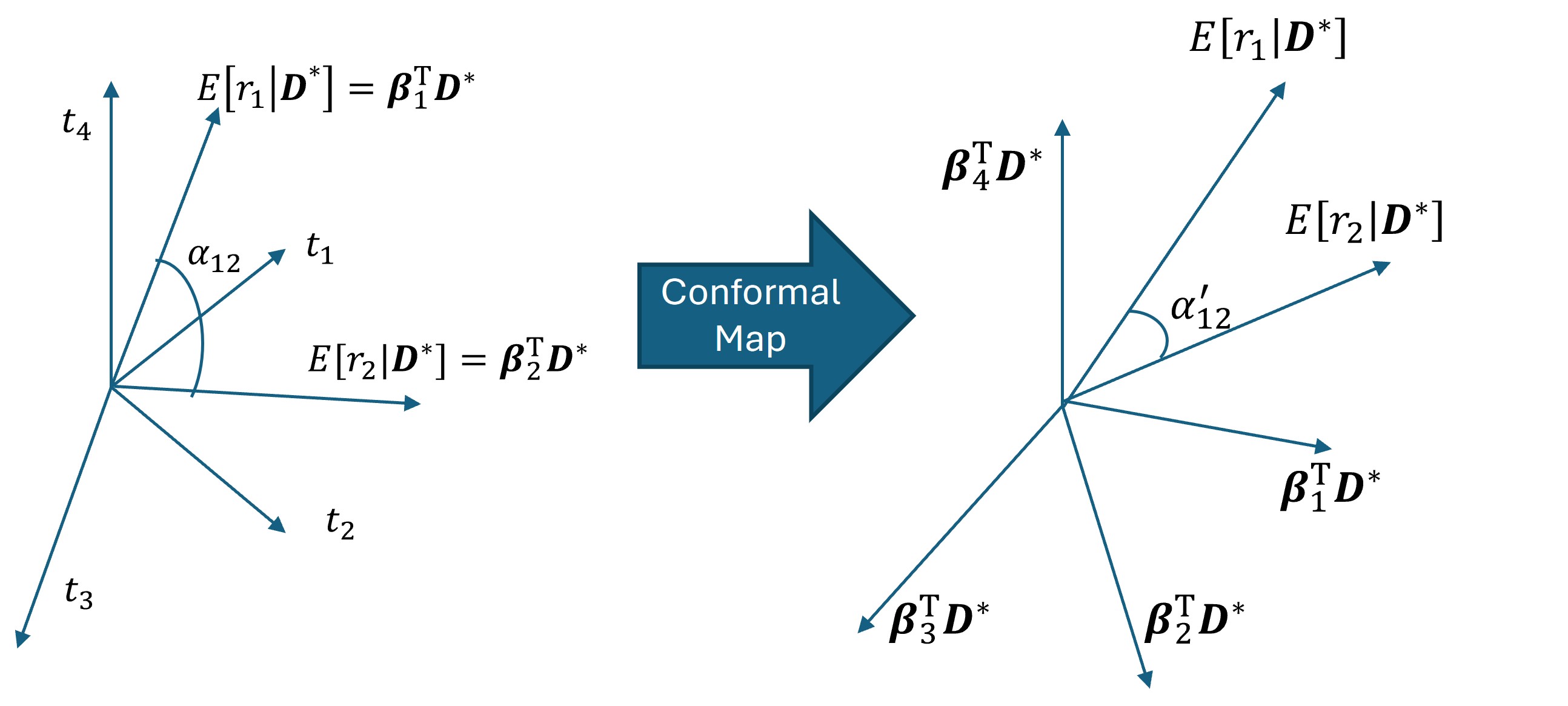}
  \caption{Second conformal map: conditional $\to$ sensitivity (beta)
  space. The figure illustrates the conformal embedding from the
  conditional time embedding induced by $\boldsymbol{D}^\star$ to the
  local sensitivity subspace spanned by the gradients
  $\boldsymbol{\beta}_i(t)
    = \nabla_{\boldsymbol{D}^\star}
      \mathbb{E}[r_i(t) \mid \boldsymbol{D}^\star]$.
  Pairwise angle proportions are again preserved: if $\alpha_{ij}(t)$
  denotes the angle in the conditional embedding and
  $\alpha'_{ij}(t)$ the angle in sensitivity space, then
  $\cos\alpha'_{ij}(t) = \cos\alpha_{ij}(t)$, with embedding norms
  rescaled by the time-varying conformal factor~$c_2(t)$, i.e.\
  $\lVert \boldsymbol{\beta}_i(t)\rVert
    = c_2(t)\,\lVert \boldsymbol{m}^{\boldsymbol{D}^\star}_i(t)\rVert$.}
  \label{fig:enter-SecondConformalMap}
\end{figure}

\medskip
Across the two maps (unconditional\(\to\)conditional\(\to\)sensitivity), pairwise angles are preserved: co--movement patterns remain invariant while overall scales vary with \(t\). This ensures that angle--based diagnostics are powerful tests of model adequacy and that span/tangent--plane optimization remains well--posed and basis--invariant. Figures~\ref{fig:enter-1stconformalmap}--\ref{fig:enter-SecondConformalMap} depict the two conformal mappings from the unconditional mean--variance geometry to the conditional embedding and then to the sensitivity (beta) space, each preserving pairwise angle proportions up to a time--varying scalar.

\begin{figure}[htbp]
  \centering
  \begin{subfigure}[t]{0.4\textwidth}
    \centering
    \includegraphics[width=\linewidth]{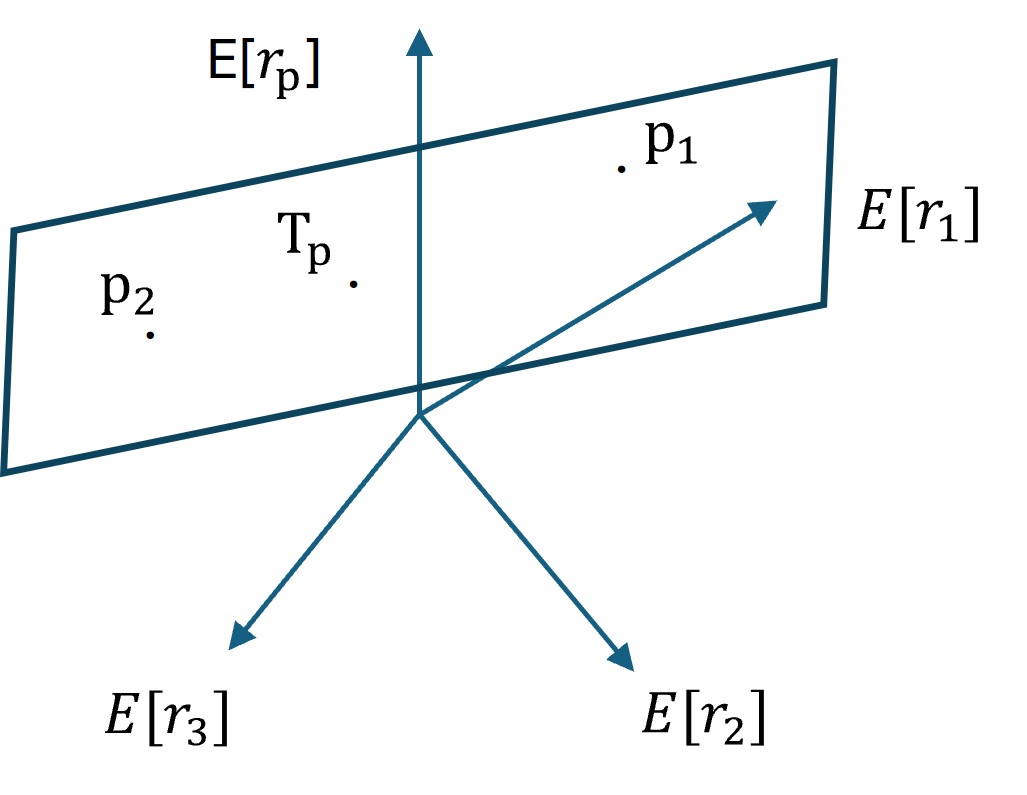}
    \caption{Mean--variance hyperplane with the tangent portfolio~$T_p$
    and two sample portfolios~$p_1$,~$p_2$ (unconditional case).
    The vectors $E[r_1]$, $E[r_2]$, $E[r_3]$ represent the
    unconditional expected returns of three assets; the hyperplane
    is the feasible mean--variance surface spanned by these exposures.}
    \label{fig:mv-uncond}
  \end{subfigure}\hfill
  \begin{subfigure}[t]{0.55\textwidth}
    \centering
    \includegraphics[width=\linewidth]{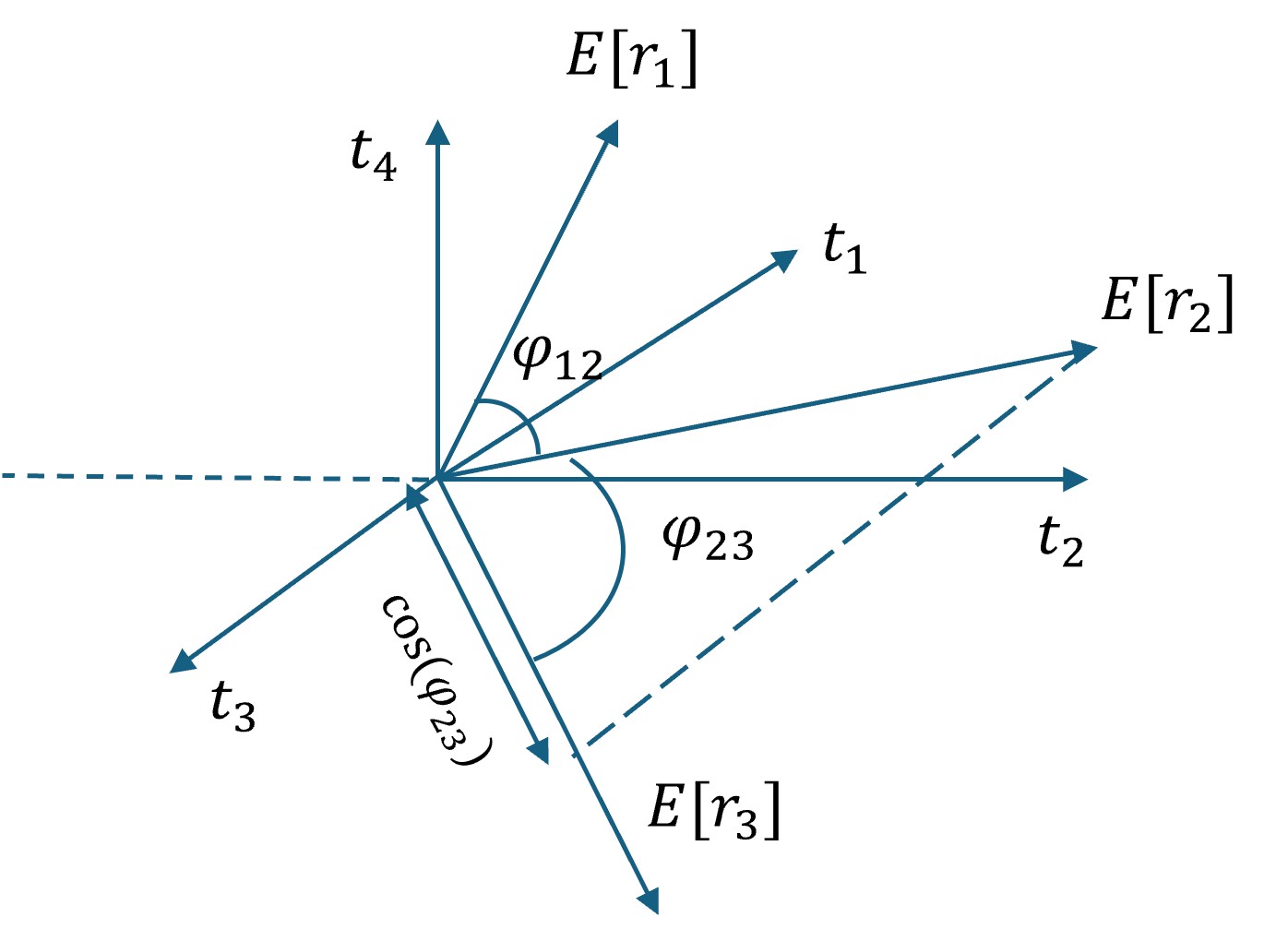}
    \caption{Mean--variance time embedding for the unconditional case.
    Each asset's expected-return vector $E[r_i]$ is plotted across time
    stamps $t_1,\ldots,t_4$. Pairwise angles $\varphi_{12}$,
    $\varphi_{23}$ between asset vectors encode co-movement; the cosine
    $\cos(\varphi_{23})$ equals the unconditional correlation between the
    embedding vectors of assets~2 and~3.}
    \label{fig:mv-cond}
  \end{subfigure}
  \caption{Unconditional mean--variance representations. Left: the
  static mean--variance hyperplane with tangent portfolio and asset
  expected returns $E[r_i]$. Right: the dynamic time-embedding view,
  where pairwise angles $\varphi_{ij}$ between centered return vectors
  capture co-movement patterns preserved by the conformal maps of
  Theorems~\ref{thm:conf1-rekey}--\ref{thm:conf2-rekey}.}
  \label{fig:uncond-vs-cond-embeddings}
\end{figure}

\smallskip
Minimizing $\mathcal G(\boldsymbol D)$ penalizes omitted confounders (inflated $\Delta_{ij}$) and avoids conditioning on colliders (which can induce spurious dependence), aligning with the causal Markov condition and Reichenbach’s Common Cause Principle.

\clearpage

\subsection{The Causal PDE--Control Model (CPCM) Meta--Class}
\label{sec:CPCM}

The results above motivate a unifying framework that generalizes portfolio models across stochastic drivers, filtering, and PDE control. We introduce the Causal PDE--Control Model (CPCM), which formalizes the interaction between drivers, beliefs, forward--backward PDEs, and control strategies. Classical specifications such as Markowitz, CAPM, APT, and Black--Litterman appear as limiting cases, while reinforcement learning and deep hedging can be interpreted as approximate variants lacking causal projection or pricing structure.

A CPCM begins with latent drivers \(\mathbf F_t\) that represent fundamental sources of variation. Because drivers are only partially observed, a filtering step produces a posterior \(\pi_t\) summarizing current beliefs based on observations \(\mathbf Y_t\). A forward equation (Fokker--Planck) describes how state densities evolve, while a backward equation (HJB) characterizes the dynamic return--risk trade--off. The two are linked through a control policy \(\boldsymbol{\theta}_t\) that determines weights and is projected onto a feasible set (the driver span). Together, filtering, forward evolution, backward control, and projection generate a portfolio path with instantaneous return \(p_t=\boldsymbol{\theta}_t^\top \mathbf r_t\) (cf.\ \ref{sec:drivers-assets}) and discounted value \(\tilde p_t\). Causal invariance (Figure~\ref{fig:symA}) formalizes that intervening on latent drivers induces driver-specific risk-neutral measures whose effective pricing kernel, under partial information, is the posterior mixture over those measures. Projection symmetry (Figure~\ref{fig:symB}) enforces that allocations are restricted to the instantaneous driver span, implemented by orthogonal projection of any tentative exposure onto the feasible subspace.

\begin{Definition}[Causal PDE--Control Model]
\label{def:cpcm}
A Causal PDE--Control Model is a tuple
\[
\mathfrak C=\big(\mathbf F_t,\ \pi_t,\ \rho(\tilde p,t\mid \mathbf F_t),\ u(\tilde p,t\mid \mathbf F_t),\ \boldsymbol{\theta}_t\big),
\]
consisting of:
\begin{enumerate} \renewcommand{\labelenumi}{(\roman{enumi})}
\item a driver process \(\mathbf F_t\in\mathbb R^m\) evolving as an It\^o SDE under \(\mathbb P\);
\item a filtering posterior \(\pi_t\) over \(\mathbf F_t\) generated by noisy observations \(\mathbf Y_t\), adapted to the observation filtration \(\mathcal F^{\mathbf Y}_t\);
\item a forward state density \(\rho(\tilde p,t\mid \mathbf F_t)\) of discounted portfolio value \(\tilde p_t\) solving a Fokker--Planck PDE conditional on \(\mathbf F_t\);\footnote{Under partial information we also use the posterior mixture \(\rho_\pi(\tilde p,t):=\int \rho(\tilde p,t\mid \mathbf f)\,\pi_t(d\mathbf f)\).}
\item a backward value function \(u(\tilde p,t\mid \mathbf F_t)\) solving a Hamilton--Jacobi--Bellman PDE;
\item an admissible control \(\boldsymbol{\theta}_t\), progressively measurable with respect to \(\mathcal F^{\mathbf Y}_t\), taking values in a feasible set \(\mathcal W\), and generating portfolio weights.
\end{enumerate}
\end{Definition}

This definition casts CPCMs as a meta-class in which probabilistic evolution, causal structure, and optimal control are inseparably linked. Pricing consistency is enforced via driver-conditional risk-neutral measures (and posterior mixtures), belief dynamics via filtering, and commonality via restrictions on exposures.

\subsubsection{Solver modules}
\label{subsec:cpcm-solvers}
Solving a CPCM requires three coupled components. The forward module evolves conditional densities under the adjoint generator of the driver dynamics, producing \(\rho(\tilde p,t\mid \mathbf F_t)\) (or \(\rho_\pi\)). The backward module solves the HJB for \(u\), determining optimal controls \(\boldsymbol{\theta}_t\). The filtering module updates \(\pi_t\) via Zakai or Kushner--Stratonovich SPDEs so that new observations \(\mathbf Y_t\) adjust beliefs consistently with the dynamics. These modules can be implemented by finite differences/elements in low dimensions, physics-informed neural networks in higher dimensions, and particle or ensemble Kalman filters for nonlinear filtering. Forward–backward PDE duality under partial information is summarized in Figure~\ref{fig:symC}.

\begin{figure}[t!]
\centering
\begin{tikzpicture}[
  >=Stealth, node distance=10mm,
  symbox/.style={draw, rounded corners, inner sep=2pt, minimum width=16mm, align=center},
  every node/.style={font=\footnotesize}
]

\node[symbox] (Z) {$Z_t$};
\node[symbox, right=18mm of Z] (F) {$\mathbf F_t$};
\node[symbox, right=18mm of F] (S) {$\mathbf S_t$};
\node[symbox, right=18mm of S] (p) {$p_t=\boldsymbol{\theta}_t^{\top}\mathbf r_t$};

\draw[-{Stealth}, thick] (Z) -- (F);
\draw[-{Stealth}, thick] (F) -- (S);
\draw[-{Stealth}, thick] (S) -- (p);

\node[symbox, below=12mm of F] (Y) {$\mathbf Y_t$};
\node[symbox, below=8mm of Y] (pi) {$\pi_t$};

\draw[-{Stealth}, thick] (Y) -- (pi);

\node[symbox, below=12mm of S, minimum width=40mm] (Qmix)
{$\mathbb Q^{\pi_t}=\int \mathbb Q^{\,\mathbf f}\,\pi_t(d\mathbf f)$};

\node[symbox, above=8mm of F, dashed, minimum width=22mm] (doF)
{$\mathrm{do}(\mathbf F_t=\mathbf f)$};

\node[symbox, above right=8mm and 10mm of S, dashed, minimum width=16mm] (Qf)
{$\mathbb Q^{\,\mathbf f}$};

\draw[dashed, -{Stealth}, thick] (doF) -- (F);
\draw[dashed, -{Stealth}, thick] (doF) -- (Qf);

\draw[-{Stealth}, thick] (pi) -| (Qmix);
\draw[dashed, -{Stealth}, thick] (Qf) |- (Qmix);

\draw[dashed, -{Stealth}, thick] (Qmix) -- (S);

\end{tikzpicture}

\caption{Causal invariance. The chain
$Z_t \to \mathbf{F}_t \to \mathbf{S}_t \to p_t$ represents
latent regimes, drivers, asset prices, and portfolio returns
$p_t = \boldsymbol{\theta}_t^\top \mathbf{r}_t$, respectively.
Noisy observations~$\mathbf{Y}_t$ yield the filtering
posterior~$\pi_t$ over~$\mathbf{F}_t$. Intervening via
$\mathrm{do}(\mathbf{F}_t = \mathbf{f})$ defines driver-specific
risk-neutral measures~$\mathbb{Q}^{\mathbf{f}}$. Under latent
drivers, the effective pricing measure is the posterior mixture
$\mathbb{Q}^{\pi_t}
  = \int \mathbb{Q}^{\mathbf{f}}\,\pi_t(d\mathbf{f})$,
which determines the pricing dynamics of~$\mathbf{S}_t$.
Solid arrows denote structural dependence; dashed elements denote
interventional and pricing constructs.}

\label{fig:symA}
\end{figure}

\begin{figure}[!t]
\centering
\resizebox{\columnwidth}{!}{%
\begin{tikzpicture}[
  >=Stealth, node distance=7mm,
  symbox/.style={draw, rounded corners, inner sep=2pt, align=center},
  every node/.style={font=\scriptsize}
]

\node[symbox, minimum width=0.56\linewidth] (span)
{$\mathcal V_t=\mathrm{Range}(\Sigma_t),\ \ \Sigma_t:=\boldsymbol \sigma(\mathbf F_t,t)$};

\node[symbox, left=18mm of span] (theta)
{$\boldsymbol{\theta}^\ast_t$};

\node[symbox, right=18mm of span] (thetap)
{$\Pi_{\mathcal V_t}\boldsymbol{\theta}^\ast_t$};

\draw[dashed, -{Stealth}, thick] (theta) -- (span);
\draw[-{Stealth}, thick] (span) -- (thetap);

\node[symbox, below=6mm of span, minimum width=0.70\linewidth] (phi)
{$\Sigma_t=U_t\Lambda_tU_t^\top,\qquad \mathbf z_t=\Lambda_t^{1/2}U_t^\top \mathbf r_t$};

\end{tikzpicture}%
}
\caption{Projection symmetry. The tentative optimal
weights~$\boldsymbol{\theta}^\ast_t$ are projected onto the
driver-implied covariance span
$\mathcal{V}_t = \mathrm{Range}(\boldsymbol{\Sigma}_t)$,
where $\boldsymbol{\Sigma}_t := \boldsymbol{\Sigma}(\mathbf{F}_t,t)
= \boldsymbol{\sigma}(\mathbf{F}_t,t)^\top
  \boldsymbol{\sigma}(\mathbf{F}_t,t)$
is the instantaneous conditional covariance, via the orthogonal
projection~$\Pi_{\mathcal{V}_t}\boldsymbol{\theta}^\ast_t$.
The spectral decomposition
$\boldsymbol{\Sigma}_t = U_t \Lambda_t U_t^\top$ induces
driver-aligned coordinates
$\mathbf{z}_t = \Lambda_t^{1/2} U_t^\top \mathbf{r}_t$,
in which only systematic risk directions are retained.}
\label{fig:symB}
\end{figure}

\begin{figure}[t!]
\centering
\resizebox{0.95\linewidth}{!}{%
\begin{tikzpicture}[
  >=Stealth, node distance=14mm,
  symbox/.style={draw, rounded corners, inner sep=2pt, align=center},
  every node/.style={font=\scriptsize}
]

\node[symbox, text width=0.44\linewidth] (FP)
{$\partial_t \rho(\tilde p,t\mid \mathbf f)=
-\partial_{\tilde p}\!\big[\mu_{\tilde p}(\mathbf f)\rho\big]
+\tfrac12\,\partial_{\tilde p\tilde p}\!\big[\sigma_{\tilde p}^2(\mathbf f)\rho\big]$};

\node[symbox, right=10mm of FP, text width=0.44\linewidth] (HJB)
{$\partial_t u+
\sup_{\boldsymbol{\theta}\in\mathcal W}\!\Big\{
\mu_{\tilde p}(\mathbf f,\boldsymbol{\theta})u_{\tilde p}
+\tfrac12\sigma_{\tilde p}^2(\mathbf f,\boldsymbol{\theta})u_{\tilde p\tilde p}
-r\,u\Big\}=0$};

\node[symbox, below=9mm of $(FP)!0.5!(HJB)$] (thetaStar)
{$\boldsymbol{\theta}_t^\ast(\mathbf f)$};

\draw[-{Stealth}, thick] (FP) -- (thetaStar);
\draw[-{Stealth}, thick] (HJB) -- (thetaStar);

\node[symbox, above=8mm of $(FP)!0.5!(HJB)$] (avg)
{$\displaystyle \int (\cdot)\,\pi_t(d\mathbf f)$};

\draw[-{Stealth}, thick] (avg) -- (FP);
\draw[-{Stealth}, thick] (avg) -- (HJB);

\end{tikzpicture}}
\caption{Forward--backward PDE duality under partial information.
The Fokker--Planck equation~(left) evolves the conditional density
$\rho(\tilde{p},t\mid\mathbf{f})$ of the discounted portfolio
value~$\tilde{p}_t$, with drift~$\mu_{\tilde{p}}(\mathbf{f})$ and
variance~$\sigma_{\tilde{p}}^2(\mathbf{f})$ determined by the driver
state~$\mathbf{f}$. The Hamilton--Jacobi--Bellman equation~(right)
solves for the value function~$u(\tilde{p},t\mid\mathbf{f})$ by
optimising the control~$\boldsymbol{\theta}_t$ over the admissible
set~$\mathcal{W}$, with $r$~the short rate. Together the two PDEs
yield the optimal feedback policy~$\boldsymbol{\theta}_t^\ast(\mathbf{f})$.
When drivers are latent, both equations are averaged against the
filtering posterior~$\pi_t$ via
$\int(\cdot)\,\pi_t(d\mathbf{f})$.}
\label{fig:symC}
\end{figure}

\subsubsection{Standing Assumptions and Well–Posedness}
\label{subsub:asumptionswell}

We impose the following standing assumptions ensuring well–posedness of the filtering and control problems.

\begin{Assumption}[State dynamics]
\label{A1}
The driver process $\mathbf F_t$ satisfies Lipschitz and linear-growth conditions ensuring existence and uniqueness of strong solutions.
\end{Assumption}

\begin{Assumption}[Observation process]
\label{A2}
The observation process $\mathbf Y_t$ has bounded second moments; the posterior $\pi_t$ exists and is adapted to $\mathcal{F}^{\mathbf Y}_t$.
\end{Assumption}

\begin{Assumption}[Admissible portfolios]
\label{A3}
The admissible set $\mathcal W$ is nonempty, convex, and compact, representing budget and leverage constraints.
\end{Assumption}

\begin{Remark}[Constraint hierarchy]
\label{rem:constraint-sets}
Portfolio weights $\boldsymbol\theta_t\in\mathbb{R}^n$ are subject to three
layers of constraints: (i)~\emph{integrability}, requiring progressive
measurability and
$\mathbb{E}\!\left[\int_0^T\boldsymbol\theta_t^\top
\boldsymbol \sigma(\mathbf F_t,t)\boldsymbol\theta_t\,\mathrm{d}t\right]<\infty$,
which defines the admissible set $\mathcal{A}$;
(ii)~\emph{pointwise feasibility}, requiring
$\boldsymbol\theta_t\in \mathcal W$ at each $t$, where $\mathcal W$ is the nonempty convex
compact set of Assumption~\ref{A3}, encoding leverage or short-sale limits;
and (iii)~the \emph{budget constraint}
$\mathbf{1}^\top\boldsymbol\theta_t=1$.
Define the budget- and leverage-constrained feasible set
$\mathcal{K}:=\{\boldsymbol\theta\in\mathbb R^n:\mathbf 1^\top\boldsymbol\theta=1,\;\boldsymbol\theta\in\mathcal W\}$.
In the continuous-time HJB~\eqref{eq:hjb}, the supremum is taken over
$\boldsymbol\theta\in \mathcal W$. In the manifold--constrained formulation
(§\ref{subsec:manifold-riskneutral}), weights are additionally projected
onto the estimated driver subspace $\mathcal{M}_\tau=\mathrm{span}(U_\tau)$
subject to $\boldsymbol\theta_\tau\in\mathcal{M}_\tau\cap\mathcal{K}$.
\end{Remark}

\begin{Assumption}[Utility regularity]
\label{A4}
The utility function $U$ is strictly concave, $C^1$, and satisfies the Inada conditions \citep{10.2307/2295809}.
\end{Assumption}

\begin{Assumption}[FP/HJB regularity]
\label{A5}
Coefficients in the forward Fokker–Planck equation and the associated Hamilton–Jacobi–Bellman equation are regular enough to ensure classical well–posedness and interchange of limits and expectations.
\end{Assumption}

Under Assumptions~\ref{A1}–\ref{A5}, the filtering problem, the forward Fokker–Planck equation, and the associated Hamilton–Jacobi–Bellman equation are well posed. In particular, these conditions ensure that the controlled state process is admissible and that the dynamic programming principle applies, providing the foundation for the verification arguments used in the portfolio control results developed later.

\begin{Remark}[Viscosity and weak–solution robustness]
\label{rem:A5prime}
Assumption~\ref{A5} can be relaxed to weaker regularity conditions without affecting the main results. Let $b(t,x)$ and $\sigma(t,x)$ denote the drift and diffusion of the controlled state on a Polish space $(\mathsf X,\mathcal B)$. Assume:

\begin{enumerate}
\item Linear growth and local Lipschitz continuity in $x$ (uniform in $t$) for $b$ and $\sigma$; the diffusion matrix $a=\sigma\sigma^\top$ is uniformly elliptic on compact subsets of $\mathsf X$.

\item Measurable control regularity: the running cost $L(t,x,u)$ is Borel in $(t,x,u)$, convex and lower semicontinuous in $u$, with at most quadratic growth; the terminal cost $G(x)$ is bounded from below and lower semicontinuous.

\item Admissibility and tightness: the set of admissible relaxed controls is tight under the canonical weak topology, and the controlled SDE admits a weak solution for every admissible control.
\end{enumerate}

Under these conditions, the HJB equation admits a bounded-from-below, lower-semicontinuous viscosity solution and the forward Fokker–Planck equation admits a weak (distributional) solution. The forward–backward pair solves the control problem via dynamic programming and comparison principles \citep{CrandallLions1992,FlemingSoner2006,Pham2009}. All verification arguments used later extend to viscosity solutions by replacing pointwise derivatives with super/sub-jet calculus and exploiting stability and uniqueness under comparison. In the empirical section, PINN/RL solvers remain consistent with viscosity notions via weak residual minimization. If one prefers a classical setting, Assumption~\ref{A5} implies the weaker conditions stated above; results established under the weaker assumptions simplify under additional smoothness.
\end{Remark}
\begin{Theorem}[Existence and uniqueness of optimal CPCM control]
\label{thm:existence-strong-main}
Under Assumptions \(\ref{A1}\)--\(\ref{A5}\) there exists a unique admissible control \(\boldsymbol{\theta}^\ast\in\mathcal{A}\), adapted to \(\mathcal{F}^{\mathbf Y}_t\), that maximizes \(\mathbb{E}\!\left[U(\tilde p_T)\mid \pi_0\right]\). Moreover: 
\begin{enumerate} \renewcommand{\labelenumi}{(\roman{enumi})}
\item \(\rho_\pi(\tilde p,t):=\int \rho(\tilde p,t\mid \mathbf f)\,\pi_t(d\mathbf f)\) and \(u(\tilde p,t):=\int u(\tilde p,t\mid \mathbf f)\,\pi_t(d\mathbf f)\) admit classical solutions to FP/HJB;
\item \(\boldsymbol{\theta}^\ast\) admits a measurable feedback \(\boldsymbol{\theta}^\ast_t=\vartheta^\ast(\tilde p_t,\pi_t)\);
\item the discounted wealth \(\tilde p_t\) is a martingale under the posterior mixture measure \(\mathbb Q^{\pi_t}\);
\item at rebalance dates \(\tau\), \(\boldsymbol{\theta}_\tau\in\mathcal{M}_\tau\cap\mathcal{K}\).
\end{enumerate}
\end{Theorem}
\noindent Proofs of Theorem~\ref{thm:existence-strong-main} and the posterior–mixture martingale property are provided in Appendix \ref{app:proofs}.

The SCM with common–cause drivers $\boldsymbol{D}^\star\subseteq\sigma(\mathbf F_t)$ fixes which shocks are economically admissible: conditional on $\mathbf F_t$, assets are independent idiosyncratically and all systematic co-movement flows through the driver span
\(
\mathcal{V}_t:=\mathrm{Range}\,\Sigma(\mathbf F_t,t),\qquad
\Sigma(\mathbf F_t,t)=\sigma(\mathbf F_t,t)^\top\sigma(\mathbf F_t,t).
\)
This causal restriction determines both (i) the feasible exposure set for portfolios (weights live in $\mathcal{V}_t$ or its local tangent space) and (ii) the pricing discipline: if 
\(
\boldsymbol{\mu}(\mathbf F_t,t)-r\mathbf 1\in \mathrm{Range}\,\Sigma(\mathbf F_t,t),
\)
then the Girsanov tilt defined on $\mathcal{V}_t$ yields, for each scenario $\mathbf f$, a measure $\mathbb Q^{\,\mathbf f}$ under which discounted prices are local martingales. With latent drivers, filtering provides a posterior $\pi_t$ and the posterior–integrated measure
\(
\mathbb Q^{\pi_t}=\int \mathbb Q^{\,\mathbf f}\,\pi_t(d\mathbf f),
\)
so valuation and control are carried out in the observable filtration. The forward–backward layer follows: the Fokker–Planck equation evolves observable laws consistent with $\boldsymbol{D}^\star$, and the HJB/BSDE supplies optimal policies restricted to $\mathcal{V}_t$. In short,
SCM $\Rightarrow$ causal span $\mathcal{V}_t$ $\Rightarrow$ risk–neutral discipline on $\mathcal{V}_t$ $\Rightarrow$ FP–HJB on the observable filtration.

\subsection{Main Results}

\subsubsection{Risk–Neutral Measures under Commonality}
\label{sec:risk-neutral}

The causal–structural framework described above extends classical martingale pricing to settings with common, partially observed drivers. We establish four foundational results: the existence of scenario–conditional risk–neutral measures, a filtered martingale property, conditions for replicability of contingent claims, and a criterion for causal completeness. Proofs are provided in Appendix \ref{app:proofs-main}.

\subsubsection{Existence of Conditional Risk--Neutral Measures}
\label{sec:existence-riskneutral}

\begin{Theorem}[Existence of Conditional Risk--Neutral Measures]
\label{thm:existence-riskneutral}
Work in the setup of Section \ref{sec:drivers-assets}, with drivers $\mathbf{F}_t$ evolving as in \eqref{eq:driver-sde} and asset prices $S^{(i)}_t$ following \eqref{eq:asset-sde1}, and let $r\ge 0$ be a bounded short rate. Suppose:

(i) for each $i$, $\boldsymbol{\sigma}_i(\cdot,\cdot)$ is locally Lipschitz with linear growth so that $\sigma(\mathbf{F}_t,t)$ is well defined and $\Sigma(\mathbf{F}_t,t)=\sigma(\mathbf{F}_t,t)^\top\sigma(\mathbf{F}_t,t)$ is finite a.s. (almost surely);

(ii) the drift vector $\boldsymbol{\mu}(\mathbf{F}_t,t)=(\mu_1(\mathbf{F}_t,t),\dots,\mu_n(\mathbf{F}_t,t))^\top$ is progressively measurable and of at most linear growth; and

(iii) the excess drift lies in the volatility span, equivalently in the covariance span, i.e.
\[
\boldsymbol{\mu}(\mathbf{F}_t,t)-r\,\mathbf{1}
\in
\mathrm{Range}\big(\sigma(\mathbf{F}_t,t)^\top\big)
=
\mathrm{Range}\big(\Sigma(\mathbf{F}_t,t)\big)
\quad \text{a.s.\ for a.e.\ } t\in[0,T].
\]

Here, a.s.\ means almost surely and a.e.\ means almost every. Then there exists a progressively measurable market price of risk $\boldsymbol{\lambda}_t\in\mathbb{R}^k$ satisfying
\[
\sigma(\mathbf{F}_t,t)^\top \boldsymbol{\lambda}_t
=
\boldsymbol{\mu}(\mathbf{F}_t,t)-r\,\mathbf{1}.
\]

If the Doléans--Dade exponential
\[
Z_t:=\mathcal{E}\!\Big(-\!\int_0^t \boldsymbol{\lambda}_s^\top\,\mathrm d\mathbf W_s\Big)
\]
is a true martingale (e.g., by Novikov or Kazamaki), then the equivalent measure $\mathbb Q^{\,\mathbf F}$ with
\[
\frac{\mathrm d\mathbb Q^{\,\mathbf F}}{\mathrm d\mathbb P}\Big|_{\mathcal F_t}=Z_t
\]
renders the discounted prices $\tilde S^{(i)}_t=e^{-rt}S^{(i)}_t$ local martingales. In particular, $\mathbb Q^{\,\mathbf F}$ is a driver--conditional equivalent martingale measure compatible with the diffusion risk structure induced by the drivers $\mathbf F_t$. Consequently, the model admits no instantaneous arbitrage relative to the driver-conditioned filtration.
\end{Theorem}

Proof of Theorem~\ref{thm:existence-riskneutral} is provided in Appendix~\ref{app:existence-Qf}.

\begin{Remark}[Interpretation and robustness of Assumption (iii)]
Assumption (iii) states that the excess drift lies in the volatility span (equivalently the covariance span). This condition guarantees the existence of a progressively measurable market price of diffusion risk $\boldsymbol{\lambda}_t$ satisfying
\[
\sigma(\mathbf{F}_t,t)^\top \boldsymbol{\lambda}_t
=
\boldsymbol{\mu}(\mathbf{F}_t,t)-r\mathbf 1 .
\]

In diffusion models this compatibility condition is precisely what permits a Girsanov change of measure producing an equivalent martingale measure and therefore corresponds to the standard no--instantaneous--arbitrage condition \citep{shreve2004stochastic,bjork2009arbitrage}. If the excess drift contains a component outside $\mathrm{Range}\big(\sigma(\mathbf{F}_t,t)^\top\big)$, then such a market price of risk may fail to exist and the equivalent martingale measure constructed in Theorem~\ref{thm:existence-riskneutral} need not exist. Economically this corresponds to a direction of expected return not associated with priced diffusion risk. For small violations one may decompose
\[
\boldsymbol{\mu}-r\mathbf 1
=
\Pi_{\mathrm{Range}\big(\sigma(\mathbf{F}_t,t)^\top\big)}(\boldsymbol{\mu}-r\mathbf 1)
+
\delta ,
\]
where $\Pi_{\mathrm{Range}\big(\sigma(\mathbf{F}_t,t)^\top\big)}$ denotes projection onto the volatility span and $\delta$ is orthogonal to that span. The pricing--control layer then applies to the projected component while $\delta$ represents an unspanned drift or model--misspecification term. Because expected returns evolve more slowly than individual observations, such deviations are typically small over economically relevant horizons, making Assumption (iii) empirically reasonable.
\end{Remark}

Intuitively, Assumption (iii) ensures that excess returns can be absorbed by a (possibly time--varying) price of diffusion risk compatible with the driver--conditioned covariance structure $\Sigma(\mathbf{F}_t,t)$, yielding a risk--neutral measure adapted to the evolving driver state.

\subsubsection{Filtered Martingale Property}
\label{sec:martingale}

\begin{Theorem}[Filtered Martingale Property]
\label{thm:filtered-martingale-representation}
Let $\pi_t$ denote the filtering posterior of the driver process $\mathbf{F}_t$ given observations in $\mathcal F^Y_t$. For any event $E\in\mathcal F$, define the posterior mixture measure
\[
\mathbb Q^{\pi_t}(E)
:=
\int \mathbb Q^{\,\mathbf f}(E)\,\pi_t(\mathrm d\mathbf f),
\]
where $\mathbb Q^{\,\mathbf f}$ is the driver–conditional risk–neutral measure from Theorem~\ref{thm:existence-riskneutral} associated with the state $\mathbf f$. Then, for any admissible strategy $\boldsymbol{\theta}\in\mathcal A$ (cf.\ \ref{sec:drivers-assets}) with discounted portfolio value $\tilde p_t$, one has
\[
\mathbb E_{\mathbb Q^{\pi_t}}\!\big[\tilde p_T\mid \mathcal F^Y_t\big]
=
\tilde p_t,
\]
so $\{\tilde p_t\}_{t\ge 0}$ is an $\mathcal F^Y_t$–martingale under $\mathbb Q^{\pi_t}$.
\end{Theorem}

\noindent
See Appendix~\ref{app:filtered-martingale} for the proof. The result promotes the driver–conditional martingale property to the observable filtration by integrating over posterior beliefs on $\mathbf{F}_t$.

\subsubsection{Replicability under Commonality}
\label{sec:replicable-claims}

\begin{Theorem}[Replicability under Commonality]
\label{thm:replicable-claims}
Let $\tilde{\mathbf S}_t$ be discounted prices and define the posterior–integrated covariance
\[
\Sigma^{\pi_t}(t)\ :=\ \int \Sigma(\mathbf f,t)\,\pi_t(\mathrm d\mathbf f)\ \in\ \mathbb R^{n\times n},
\quad \text{with }\ \Sigma(\mathbf f,t)=\sigma(\mathbf f,t)^\top\sigma(\mathbf f,t).
\]
A square–integrable claim $\Phi\in L^2(\mathcal F^Y_T)$ is exactly replicable using admissible strategies in $\mathcal A$ if and only if its martingale integrand $\varphi_t$ (in the \(\mathcal F^Y\)–martingale representation of the discounted claim) satisfies
\[
\varphi_t \ \in\ \mathrm{Range}\big(\Sigma^{\pi_t}(t)\big)\quad\text{for a.e.\ }t\in[0,T]\ \text{a.s.}
\]
\end{Theorem}

\noindent
The proof is in Appendix \ref{app:replicable-claims}. This theorem implies that exact hedging requires the claim's risk exposures to lie in the span generated by the driver–mediated covariance under posterior beliefs.

\begin{Theorem}[Causal completeness under partial information]
\label{thm:causal-completeness}
Under the setup of \ref{sec:drivers-assets}, let the (discounted) returns satisfy \eqref{eq:asset-sde1} with diffusion loading $\sigma(\mathbf{F}_t,t)\in\mathbb R^{n\times k}$ and observable filtration $\mathcal F^Y_t$. If $m\le n$ and standard no–arbitrage conditions hold, then the market is complete with respect to $\mathcal F^Y$ if and only if
\begin{equation}
\begin{aligned}
\operatorname{rank}\big(\sigma(\mathbf F_t,t)\big) = n 
&\quad \text{a.s.\ for a.e.\ } t,\\[3pt]
&\Longleftrightarrow\ 
\Sigma(\mathbf F_t,t)\ \text{is positive definite a.s.\ for a.e.\ } t.
\end{aligned}
\end{equation}
Equivalently, in posterior–integrated form, completeness holds if and only if
\[
\operatorname{rank}\big(\Sigma^{\pi_t}(t)\big)=n\ \text{ a.s.\ for a.e.\ }t,
\qquad \Sigma^{\pi_t}(t)=\int \Sigma(\mathbf f,t)\,\pi_t(\mathrm d\mathbf f).
\]
In the square case $n=k$, this reduces to $\det\big(\Sigma^{\pi_t}(t)\big)>0$ a.s.\ for a.e.\ $t$.
\end{Theorem}

\noindent
The proof is contained in Appendix \ref{app:causal-completeness}. Causal completeness means the driver manifold spans all systematic exposures under the observable information, so every square–integrable claim measurable w.r.t.\ $\mathcal F^Y_T$ can be replicated by portfolios in $\mathcal A$.

This section formalizes the structural properties that make CPCMs robust and interpretable. Building on the conditional risk–neutral measures and filtered martingale structure in Section~\ref{sec:risk-neutral}, manifold constraints implied by the Commonality Principle are shown to admit an information–geometric interpretation; causal semantics are expressed through structural causal models and preserved under partial observation; the martingale representation extends to the filtered, driver–mediated setting; and the projected formulation remains invariant under smooth reparametrizations of the driver subspace.

\begin{takeawaybox}
\textbf{What "risk–neutral" means.}
It is a property of the pricing measure, not investor preferences. 
Conditioned on driver state $\mathbf F_t=\mathbf f$, if the excess drift satisfies 
$\boldsymbol\mu(\mathbf f,t)-r\mathbf 1\in \mathrm{Range}\big(\Sigma(\mathbf f,t)\big)$,
there exists a market price of risk $\boldsymbol\lambda_t$ with
$\sigma(\mathbf f,t)^\top\boldsymbol\lambda_t=\boldsymbol\mu(\mathbf f,t)-r\mathbf 1$,
so discounted asset prices $\tilde S_t=e^{-rt}S_t$ are local martingales under $\mathbb Q^{\,\mathbf f}$.

\medskip
\textbf{From latent drivers to observables.}
With latent $\mathbf F_t$, integrate over the filtering posterior $\pi_t$ to form the posterior–mixture measure
$\mathbb Q^{\pi_t}=\int \mathbb Q^{\,\mathbf f}\,\pi_t(d\mathbf f)$.
Then, for any admissible self-financing strategy with value 
$p_t=\boldsymbol{\theta}_t^\top \mathbf S_t$, 
$\mathbb E_{\mathbb Q^{\pi_t}}[\tilde p_T\mid\mathcal F^Y_t]=\tilde p_t$,
so discounted portfolio values are $\mathcal F^Y_t$–(local) martingales. Risk–neutrality is therefore with respect to the portfolio/process under the observable filtration, not a particular optimizer.

\medskip
\textbf{Why the common–cause ($\boldsymbol{D}^\star$) architecture matters.}
Conditioning on the common drivers makes asset idiosyncrasies independent; all cross–sectional dependence flows through
$\Sigma(\mathbf F_t,t)=\sigma(\mathbf F_t,t)^\top\sigma(\mathbf F_t,t)$. 
"Controlling for $\boldsymbol{D}^\star$" means projecting portfolio exposures onto the driver span $\mathcal{V}_t=\mathrm{Range}\big(\Sigma(\mathbf F_t,t)\big)$.

\medskip
\textbf{Hedging/completeness.}
A claim is exactly replicable iff its martingale integrand lies in $\mathrm{Range}\big(\Sigma^{\pi_t}(t)\big)$, where
$\Sigma^{\pi_t}(t)=\int\Sigma(\mathbf f,t)\,\pi_t(d\mathbf f)$.
Causal completeness holds when $\operatorname{rank}\big(\Sigma^{\pi_t}(t)\big)=n$ 
(in the square case $n=k$, equivalently $\det\big(\Sigma^{\pi_t}(t)\big)>0$).
\end{takeawaybox}

\subsection{Projection–Divergence Duality}
\label{subsec:projection-duality}

Restricting portfolio weights to the span of common drivers admits two equivalent interpretations. Geometrically, it projects an unconstrained allocation onto the feasible driver subspace. Informationally, it selects, among all laws attainable within that subspace, the one closest to the ideal (but infeasible) target law under a suitable divergence. The constraint is therefore structural: portfolio risk is mediated only through persistent causal drivers. Let
\[
\mathcal{V}_t := \mathrm{Range}\big(\Sigma(\mathbf F_t,t)\big),
\qquad
\Sigma(\mathbf F_t,t)=\sigma(\mathbf F_t,t)^\top\sigma(\mathbf F_t,t)
\]
denote the causal driver span implied by the SCM. The duality below shows that geometric projection in \(\mathcal{V}_t\) coincides with minimal-distortion reweighting of the target law under a convex divergence. The following theorem formalizes this equivalence by showing that the geometric projection of portfolio weights onto the causal driver span coincides with the divergence projection of the induced portfolio law relative to the unconstrained optimum.

\begin{Theorem}[Projection–Divergence Duality]
\label{thm:projection-divergence}
Fix a driver realization \(\mathbf f\in\mathbb R^m\) and let \(\nu^\star\) denote the law induced by the unconstrained optimizer under \(\mathbb Q^{\,\mathbf f}\). Define
\[
\mathcal V(\mathbf f,t):=\mathrm{Range}\big(\Sigma(\mathbf f,t)\big),
\qquad
\mathcal M_{\mathrm{proj}}(\mathbf f)
:=
\Big\{
\mathcal L_{\mathbf f}^{\boldsymbol{\theta}}
:\;
\boldsymbol{\theta}\in \mathcal V(\mathbf f,t)\cap\mathcal W
\Big\},
\]
where \(\mathcal L_{\mathbf f}^{\boldsymbol{\theta}}\) denotes the law of the terminal payoff \(p_T(\boldsymbol{\theta})\) under \(\mathbb Q^{\,\mathbf f}\). Let \(D(\,\cdot\,\|\,\cdot\,)\) be any divergence on probability laws such that  

(i) \(D\) is lower semicontinuous in the weak topology on \(\mathcal P\);  

(ii) \(D(\,\cdot\,\|\,\nu^\star)\) is strictly convex on \(\mathcal M_{\mathrm{proj}}(\mathbf f)\).

Then the minimizers of the geometric projection problem and the divergence projection problem coincide. In particular, if
\[
\boldsymbol{\theta}^{\mathrm{proj}}
\in
\arg\min_{\boldsymbol{\theta}\in \mathcal V(\mathbf f,t)\cap\mathcal W}
\mathbb{E}^{\mathbb{Q}^f}\!\left[\Phi\!\bigl(p_T(\boldsymbol{\theta})\bigr)\right],
\]
then the induced law
\[
g^{\mathrm{proj}}=\mathcal L_{\mathbf f}^{\boldsymbol{\theta}^{\mathrm{proj}}}
\]
satisfies
\[
g^{\mathrm{proj}}
=
\arg\min_{g\in\mathcal M_{\mathrm{proj}}(\mathbf f)}
D\!\left(g\,\middle\|\,\nu^\star\right),
\]
and the minimizer is unique.
\end{Theorem}

\begin{Remark}
\label{rem:D-class}
The theorem includes, as special cases:  

(a) \(f\)-divergences \(D_\varphi\) with \(\varphi\) convex, lower semicontinuous (l.s.c.), \(\varphi(1)=0\), and coercive (strict convexity holds on \(\mathcal M_{\mathrm{proj}}(\mathbf f)\)); and  

(b) the quadratic Wasserstein distance \(W_2^2\) provided \(\mathcal M_{\mathrm{proj}}(\mathbf f)\) has uniformly bounded second moments (ensuring l.s.c.\ and strict convexity along feasible geodesics).

The proof combines KKT conditions for the geometric projection with the corresponding duality (Fenchel for \(f\)-divergences; Kantorovich for \(W_2\)). Full details are given in Appendix~\ref{appendix:duality-projection-divergence}.
\end{Remark}

The duality clarifies why CPCM projections remain stable when the unconstrained optimum is infeasible or poorly conditioned: the feasible portfolio is the least distorted element under \(D\) relative to the target law. Relative to entropy pooling\citep{Meucci2008}, which tilts a prior over scenarios, here the feasible set is fixed by the causal driver span rather than subjective priors. In practice, this acts as disciplined risk budgeting: weights concentrate on persistent directions, the divergence penalizes excursions off the manifold, and exposures remain interpretable across regimes with lower turnover and improved robustness.

More broadly, the result provides a geometric interpretation of the Commonality Principle. Restricting portfolio weights to the driver span removes directions of variation that are not supported by persistent causal drivers. The projection therefore acts as an information filter: among all attainable portfolios within the causal manifold, the selected allocation minimizes informational distortion relative to the unconstrained target law. In this sense the CPCM constraint does not merely reduce dimensionality; it identifies the most stable feasible representation of the optimal policy consistent with the causal structure.

\subsection{Manifold--Constrained Portfolio Optimization under Posterior--Integrated Risk--Neutrality}
\label{subsec:manifold-riskneutral}

Portfolio choice under CPCMs takes place in the causal driver manifold, extending classical martingale pricing into a lower-dimensional structure tied to common exposures. Asset returns $\mathbf r_t\in\mathbb{R}^n$ depend on drivers $\mathbf F_t\in\mathbb{R}^m$, which are only partially observed through noisy signals $\mathbf Y_t$. Investors therefore update beliefs $\pi_t$ about the latent driver state. These beliefs define the posterior--integrated risk--neutral measure $\mathbb Q^{\pi_t}$ characterized by

\[
\mathbb Q^{\pi_t}(E)
=
\int \mathbb Q^{\,\mathbf f}(E)\,\pi_t(\mathrm d\mathbf f),
\qquad
E\in\mathcal F,
\]
where $\mathbb Q^{\,\mathbf f}$ denotes the driver--conditional risk--neutral measure. Systematic dependence is restricted by the Commonality Principle. Let

\[
\Sigma(\mathbf F_t,t)
=
\sigma(\mathbf F_t,t)^\top\sigma(\mathbf F_t,t)
\]
denote the instantaneous covariance matrix implied by the diffusion structure of returns conditional on drivers. This object determines the structural directions of systematic risk generated by the driver process. By contrast, the conditional distributional covariance
\[
\mathrm{Cov}(\mathbf r_t\mid \mathbf F_t)
\]
describes the dispersion of returns conditional on the driver realization. While related in diffusion models, these quantities are conceptually distinct: the first is determined by the instantaneous diffusion structure of the model, whereas the second reflects the conditional distribution of returns. The causal driver span is therefore defined as
\[
\mathcal V_t
=
\mathrm{Range}\big(\Sigma(\mathbf F_t,t)\big)
\subset\mathbb R^n .
\]

The driver span admits a geometric interpretation (see Figure~\ref{fig:return-manifold}). If returns are generated by the structural relation
\[
\mathbf r_t=g(\mathbf F_t,\mathbf U_t),
\]
then the Jacobian
\[
B_t=\nabla_{\mathbf F}g(\mathbf F_t,\mathbf U_t)
\]
defines the tangent space of the return manifold
\[
\mathcal M=\{g(\mathbf f,\mathbf u)\}.
\]
Under the diffusion representation
\[
\mathrm d\mathbf r_t
=
\boldsymbol\mu(\mathbf F_t,t)\,\mathrm dt
+
\sigma(\mathbf F_t,t)\,\mathrm d\mathbf W_t,
\]
the covariance matrix satisfies
\[
\Sigma(\mathbf F_t,t)
=
\sigma(\mathbf F_t,t)^\top\sigma(\mathbf F_t,t)
=
B_t\,\Gamma_t\Gamma_t^\top B_t^\top
\]
for some factor--volatility matrix $\Gamma_t$. Consequently
\[
\mathcal V_t
=
\mathrm{Range}\big(\Sigma(\mathbf F_t,t)\big)
=
\mathrm{Range}(B_t)
=
T_{\mathbf r_t}\mathcal M ,
\]
so the CPCM driver span coincides with the tangent space of the return manifold induced by the drivers. A spectral decomposition
\[
\Sigma(\mathbf F_t,t)|_{\mathcal V_t}
=
U_t\Lambda_tU_t^\top
\]
induces the linear map
\[
\Psi_t:\mathbb{R}^n\to\mathbb{R}^m,
\qquad
\Psi_t x=\Lambda_t^{1/2}U_t^\top x ,
\]
which transports returns into coordinates that preserve the covariance metric. Exposures expressed in
\[
\mathbf z_t=\Psi_t\mathbf r_t
\]
reflect only causal directions, filtering out spurious correlations and anchoring risk premia to persistent drivers. At a rebalance time $\tau$, the manifold is estimated from a recent window $W_\tau$. The average Jacobian with respect to drivers is

\[
B_\tau
=
\mathbb E_{t\in W_\tau}
\big[
\nabla_{\mathbf F}g(\mathbf F_t,\mathbf U_t)
\big],
\qquad
B_\tau\in\mathbb R^{n\times m}.
\]

A thin singular value decomposition
\[
B_\tau
=
U_\tau\Lambda_\tau V_\tau^\top
\]
identifies the active driver subspace
\[
\mathcal M_\tau
=
\mathrm{span}(U_\tau).
\]

To avoid artificial jumps between windows, $U_\tau$ is aligned with the previous basis $U_{\tau^-}$ via orthogonal Procrustes rotation, ensuring smooth manifold transport. Portfolio tilts in driver space are determined by the quadratic utility problem
\[
\boldsymbol\phi_\tau
=
\arg\max_{\boldsymbol\phi\in\mathbb R^m}
\left(
\boldsymbol\mu_{F,\tau}^\top\boldsymbol\phi
-
\frac{\gamma}{2}
\boldsymbol\phi^\top
\boldsymbol{\Sigma}_{F,\tau}
\boldsymbol\phi
\right),
\]
with solution
\[
\boldsymbol\phi_\tau
=
\gamma^{-1}\boldsymbol{\Sigma}_{F,\tau}^{-1}\boldsymbol\mu_{F,\tau}.
\]
Mapping back to the return space yields raw portfolio weights
\[
\tilde{\boldsymbol\theta}_\tau
=
B_\tau\boldsymbol\phi_\tau .
\]
These weights are projected onto the feasible set
\[
\{\boldsymbol\theta\in\mathbb R^n:
\mathbf 1^\top\boldsymbol\theta=1\},
\qquad
\boldsymbol\theta_\tau^{\mathrm{man}}
=
\Pi_{\mathcal{M}_\tau\cap\mathcal{K}}
\big(\tilde{\boldsymbol\theta}_\tau\big),
\]
ensuring both budget balance and alignment with the driver manifold. The allocation $\boldsymbol\theta_\tau^{\mathrm{man}}$ determines the exposure direction but not its scale. To calibrate the scale, a one--dimensional HJB control problem is solved on the recent portfolio path,
\[
\partial_t u+
\max_{\vartheta\in\mathbb R}
\left\{
\mu_t\,\vartheta\, u_x
+
\frac12\sigma_t^2\,\vartheta^2\, u_{xx}
\right\}
=0,
\qquad
u(T,x)=-e^{-\gamma x}.
\]
This produces a tail control $\vartheta_{\mathrm{tail}}$ and clipped scale
\[
s_\tau
=
\operatorname{clip}(1+\vartheta_{\mathrm{tail}},
s_{\min},s_{\max}).
\]
The scaled allocation becomes
\[
\boldsymbol\theta_\tau^{\mathrm{PDE}}
=
s_\tau\,\boldsymbol\theta_\tau^{\mathrm{man}},
\]
while the unscaled allocation remains
\[
\boldsymbol\theta_\tau^{\mathrm{noPDE}}
=
\boldsymbol\theta_\tau^{\mathrm{man}}.
\]
A convex blend
\[
\boldsymbol\theta_\tau^{(\omega)}
=
\Pi_{\mathcal{M}_\tau\cap\mathcal{K}}
\Big[
(1-\omega)\,\boldsymbol\theta_\tau^{\mathrm{noPDE}}
+
\omega\,\boldsymbol\theta_\tau^{\mathrm{PDE}}
\Big],
\qquad
\omega\in[0,1],
\]
traces the return--structure frontier between raw manifold alignment and PDE--guided scaling. Structural coherence is monitored through a martingale--defect proxy computed from the portfolio return process
\[
p_t
=
\boldsymbol{\theta}_t^\top\mathbf r_t .
\]
Fitting the regression
\[
p_t=a+b\,p_{t-1}+\varepsilon_t
\]
on rolling windows and averaging $|a+b\,p_t|$ yields a diagnostic of pricing consistency. Low defect values indicate that discounted portfolios behave as martingales under $\mathbb Q^{\pi_t}$, whereas persistent deviations signal structural misspecification or unstable manifold estimates.

\begin{figure}[t]
\centering
\begin{tikzpicture}[
  >=Latex,
  node distance=1.8cm,
  box/.style={rectangle, draw, rounded corners=3pt, minimum width=18mm,
              minimum height=10mm, align=center, font=\small},
  lbl/.style={font=\footnotesize, text=black!70},
  arr/.style={->, thick},
  dasharr/.style={->, dashed, thin, black!60}
]

\node[box, fill=gray!8] (F) {$\mathbf{F}_t$};
\node[lbl, above=1pt of F] {Drivers};

\node[box, fill=blue!6, right=3.5cm of F] (r) {$\mathbf{r}_t$};
\node[lbl, above=1pt of r] {Returns};

\draw[arr] (F) -- node[above, lbl] {structural map $g$} (r);

\node[box, fill=orange!8, above=2.2cm of r] (M)
  {$\mathcal{M}=\{g(\mathbf{f},\mathbf{u})\}$};
\node[lbl, above=1pt of M] {Return manifold};

\draw[dasharr] (r) -- node[right, lbl, align=left] {image of $g$} (M);

\node[box, fill=green!6, below=2.2cm of F] (B)
  {$B_t = \nabla_{\mathbf{F}}\,g$};
\node[lbl, below=1pt of B] {Jacobian};

\node[box, fill=green!6, below=2.2cm of r] (V)
  {$\mathcal{V}_t = \mathrm{Range}\!\left(\Sigma(\mathbf{F}_t,t)\right)$};
\node[lbl, below=1pt of V] {Driver span};

\draw[arr] (F) -- node[left, lbl] {differentiate} (B);
\draw[arr] (r) -- node[right, lbl] {covariance} (V);
\draw[arr] (B) -- node[above, lbl] {$\mathrm{Range}(B_t) = \mathcal{V}_t$} (V);

\draw[dasharr, bend left=20] (V.north east)
  to node[right, lbl, align=left] {tangent space\\$T_{\mathbf{r}_t}\!\mathcal{M}$} (M.south east);

\end{tikzpicture}
\caption{Return manifold induced by the structural mapping from drivers
to returns. The Jacobian $B_t=\nabla_{\mathbf{F}}g$ defines the tangent
space of the manifold $\mathcal{M}$, which coincides with the covariance
span $\mathcal{V}_t=\mathrm{Range}(\Sigma(\mathbf{F}_t,t))$, where
$\Sigma(\mathbf{F}_t,t)=\sigma(\mathbf{F}_t,t)^\top\sigma(\mathbf{F}_t,t)$.
Portfolios constrained to $\mathcal{V}_t$ have exposure only to
systematic driver risk.}
\label{fig:return-manifold}
\end{figure}

\subsection{Causal Semantics with Correlated Layers}
\label{subsec:scm-correlated}

Structural Causal Models (SCMs) provide semantics for CPCMs by encoding interventions on drivers and tracing their propagation to asset returns and portfolios. The causal chain
\[
Z_t \;\to\; \mathbf F_t \;\to\; \mathbf r_t \;\to\; p_t
\]
formalizes the Commonality Principle: latent regimes $Z_t$ influence drivers $\mathbf F_t$, which determine returns $\mathbf r_t$, and portfolios aggregate these into
\[
p_t=\boldsymbol{\theta}_t^\top \mathbf r_t .
\]
Exogenous shocks are assumed to be independent across layers in the baseline, meaning that the structural disturbances
\[
(\varepsilon^Z_t,\varepsilon^F_t,\varepsilon^r_t)
\]
factorize. This block–triangular structure rules out unmeasured backdoor paths and implies that interventional and observational conditionals coincide along the chain. 

To study departures from independence, let the Brownian innovations driving $\mathbf F_t$ and $\mathbf r_t$ be instantaneously correlated,
\[
\mathbb E\!\left[\mathrm d\mathbf W^{r}_t\,\mathrm d\mathbf W^{F\top}_t\right]
=
\Xi\,\mathrm dt,
\qquad
\Xi\in\mathbb R^{n\times m}.
\]
In the generator for the value function $u(p,t\mid \mathbf f)$ this introduces a cross–variation term
\[
u_{p\mathbf f}^\top
\boldsymbol{\Sigma}_F^{1/2}
\Xi^\top
\boldsymbol{\Sigma}_\varepsilon^{1/2}
\boldsymbol\theta,
\]
so the Hamilton–Jacobi–Bellman first–order condition becomes
\[
\mathbf f\,u_p
+
\boldsymbol{\Sigma}_\varepsilon\,u_{pp}\,\boldsymbol\theta
+
\boldsymbol{\Sigma}_\varepsilon^{1/2}
\Xi
\boldsymbol{\Sigma}_F^{1/2}\,
u_{p\mathbf f}
=0 .
\]

The optimal driver tilt therefore decomposes into a myopic and a hedging component,
\[
\boldsymbol\theta^\ast(\mathbf f,t)
=
\frac{u_p}{-u_{pp}}
\,\boldsymbol{\Sigma}_\varepsilon^{-1}\mathbf f
-
\frac{1}{u_{pp}}
\,\boldsymbol{\Sigma}_\varepsilon^{-1/2}
\Xi
\boldsymbol{\Sigma}_F^{1/2}\,
u_{p\mathbf f}.
\]

Evaluating the Hamiltonian at $\boldsymbol\theta^\ast$ yields the running reward
\begin{align}
\mathcal R(\mathbf f,u_p,u_{pp},u_{p\mathbf f})
=
\frac{1}{2}\frac{1}{-u_{pp}}
\Big[
&u_p^2\,\mathbf f^\top\boldsymbol{\Sigma}_\varepsilon^{-1}\mathbf f \nonumber\\
&+2u_p\,\mathbf f^\top\boldsymbol{\Sigma}_\varepsilon^{-1/2}\Xi\boldsymbol{\Sigma}_F^{1/2}u_{p\mathbf f} \nonumber\\
&+u_{p\mathbf f}^\top
\boldsymbol{\Sigma}_F^{1/2}
\Xi^\top
\boldsymbol{\Sigma}_\varepsilon^{-1}
\Xi
\boldsymbol{\Sigma}_F^{1/2}
u_{p\mathbf f}
\Big],
\end{align}

which reduces to
\[
\frac{1}{2}\frac{u_p^2}{-u_{pp}}
\,\mathbf f^\top
\boldsymbol{\Sigma}_\varepsilon^{-1}
\mathbf f
\]
when $\Xi=0$. Thus, with correlated exogenous shocks, the optimal value accrues not only from the driver's Mahalanobis strength but also from hedging gains against shared shocks.

\begin{Theorem}[Partial Identifiability of Filtered Counterfactuals]
\label{thm:partial-identifiability}

If returns are conditionally independent given $\mathbf F_t$ and the posterior $\pi_t$ is regular, then the filtered counterfactual distribution
\[
\bar\rho(p,t)
=
\int
\rho(p,t\mid
\mathrm{do}(\mathbf F_t=\mathbf f))
\,\pi_t(\mathrm d\mathbf f)
\]
is uniquely determined from the observable filtration $\mathcal F^Y_t$ up to posterior support. Moreover, for any alternative posterior $\pi_t'$,
\[
W_2(\bar\rho,\bar\rho')
\le
L\,W_2(\pi_t,\pi_t'),
\]
for some Lipschitz constant $L$ depending on the driver–density mapping.
\end{Theorem}

Proof is provided in Appendix~\ref{appendix:counterfactual-identifiability}. Independence across layers is therefore relevant for both identification and control. When $\Xi\neq0$ and this dependence is modeled, coherent semantics are restored either by explicit augmentation with a confounder $U_t$ or by retaining the cross term in the diffusion and using the adjusted $\boldsymbol\theta^\ast$ and $\mathcal R$ above.

\subsection{Generalized Martingale Representation}
\label{subsec:generalized-martingale}

We now extend the martingale representation to the filtered setting.

\begin{Theorem}[Generalized martingale representation]
\label{thm:generalized-martingale}
Under the Commonality Principle and the SCM structure, with the posterior $\pi_t$ evolving according to the Zakai SPDE associated with the observation filtration $\mathcal F^Y_t$, any claim $\Phi\in L^2(\mathcal F^Y_T)$ admits the representation
\[
\Phi
=
\mathbb E_{\mathbb Q^{\pi_t}}\!\big[\Phi \mid \mathcal F^Y_t\big]
+
\int_0^T
\varphi_t^\top\,\mathrm d M_t^{\mathcal F^Y},
\]
where $M_t^{\mathcal F^Y}$ denotes the $\mathcal F^Y$–innovation martingale and $\mathbb Q^{\pi_t}$ is the posterior–integrated risk–neutral measure defined earlier.
\end{Theorem}

\noindent
The proof is provided in Appendix~\ref{appendix:generalized-martingale}. This result guarantees that all attainable claims can be priced consistently under incomplete information. For practitioners, it ensures that hedging strategies remain valid even when state variables are only partially observed, so valuation and risk management remain coherent under filtering.

\subsection{Extensions to the Projected Framework}
\label{subsec:extensions}

CPCMs admit two complementary projection modalities. 
In-span projections act directly in the (approximately) fixed linear span of identified common drivers, matching the static factor-style viewpoint used earlier. 
Tangential (manifold) projections, by contrast, operate on a dynamic low-dimensional manifold traced by the driver-implied risk geometry, where local tangent spaces evolve smoothly over time. 
The present subsection focuses on the dynamic-manifold case and develops two additional properties needed for stability and interpretability when the driver geometry is itself time-varying: a conformal (bounded-distortion) map into driver coordinates, and a smooth transport of the driver subspace across re-estimation windows. 
These results extend the earlier in-span statements by ensuring that local reparametrizations neither distort risk metrics nor induce artificial jumps in exposures. 
Formal statements are provided here, with complete proofs in Appendix~\ref{app:extensions-proofs}.

\subsubsection{Conformal transport induced by commonality}

At each time $t$, let
\[
\boldsymbol \sigma(\mathbf F_t,t)
=
\boldsymbol \sigma(\mathbf F_t,t)^\top\boldsymbol \sigma(\mathbf F_t,t)
\]
denote the instantaneous covariance matrix implied by the diffusion structure of returns conditional on the driver state. 
Assume $\mathrm{rank}(\boldsymbol \sigma(\mathbf F_t,t))=m<n$ and define the driver span
\[
\mathcal V_t
=
\mathrm{Range}\big(\boldsymbol \sigma(\mathbf F_t,t)\big).
\]

Let the spectral decomposition on this subspace be
\[
\boldsymbol \sigma(\mathbf F_t,t)\big|_{\mathcal V_t}
=
U_t\Lambda_tU_t^\top .
\]

Define the linear transport map
\[
\Psi_t:\mathbb{R}^n\to\mathbb{R}^m,
\qquad
\Psi_t x=\Lambda_t^{1/2}U_t^\top x .
\]

\begin{Theorem}[Conformal transport and causal invariance]
\label{thm:conformal-transport}
Let
\[
\mathbf z_t=\Psi_t\mathbf r_t .
\]
For any $\mathbf u,\mathbf v\in\mathcal V_t$,
\[
\langle \Psi_t \mathbf u,\Psi_t \mathbf v\rangle
=
\mathbf u^\top \boldsymbol \sigma(\mathbf F_t,t)\, \mathbf v .
\]
Hence $\Psi_t$ is an isometry between $(\mathcal V_t,\langle\cdot,\cdot\rangle_{\Sigma})$ and $(\mathbb R^m,\langle\cdot,\cdot\rangle)$. More generally it is a $K_t$--quasi--conformal map with distortion
\[
K_t
=
\sqrt{\frac{\lambda_{\max}(\Lambda_t)}{\lambda_{\min}(\Lambda_t)}}\,.
\]
If $\boldsymbol \sigma(\mathbf F_t,t)|_{\mathcal V_t}=c_t\, I_m$ (isotropy), then $K_t=1$ and $\Psi_t$ is strictly conformal. Conditional independence of asset returns given $\mathbf F_t$ is preserved in the coordinates $\mathbf z_t$, so causal invariance carries through to the reduced representation.
\end{Theorem}

See proof in Appendix~\ref{app:proof-conformal-transport}. The map $\Psi_t$ provides coordinates in which covariances in the driver-implied subspace are measured without distortion: angles are preserved and lengths are scaled uniformly within eigenspaces. Economically, shocks measured in $\mathbf z_t$ align with causal driver directions rather than with spurious cross-asset correlations, so estimated premia are metric-consistent and robust to reparametrization of $\mathcal V_t$.

\subsubsection{Smooth evolution of the driver subspace}

Because $\mathcal V_t$ is re-estimated over rolling windows, stability requires preventing spurious jumps. Let
\[
\{\widehat{\mathcal V}_{t_k}\}\subset \mathrm{Gr}(m,n)
\]
be the sequence of estimated driver subspaces, where $\mathrm{Gr}(m,n)$ denotes the Grassmann manifold of $m$-dimensional subspaces of $\mathbb R^n$. Smoothness in the Grassmann metric ensures time-consistent projections.

\begin{Theorem}[Continuity under smooth subspace transport]
\label{thm:subspace-continuity}
If
\[
d_{\mathrm{Gr}}(\widehat{\mathcal V}_{t_{k+1}},\widehat{\mathcal V}_{t_k})\to 0
\quad\text{as}\quad
t_{k+1}-t_k\to 0,
\]
then for orthonormal bases transported via Procrustes alignment, Stiefel interpolation, or Grassmann geodesics, the transformed coordinates
\[
z_{t_k}
=
\Psi_{t_k}\mathbf r_{t_k}
\]
satisfy
\[
\lim_{t_{k+1}-t_k\to 0}
\|z_{t_{k+1}}-z_{t_k}\|
=
0
\quad\text{in probability}.
\]
\end{Theorem}

See proof in Appendix~\ref{app:proof-subspace-continuity}.

\begin{Corollary}[Pricing and control invariance]
\label{cor:pricing-filtering-consistency}
Let the filtering posterior $\pi_t$ evolve according to the Zakai or Kushner–Stratonovich SPDE. Then the forward Fokker–Planck and backward HJB equations expressed in $(z_t,p_t)$ remain invariant under orthonormal reparametrizations of $\mathcal V_t$ and continuous deformations of its path. Optimal controls $\boldsymbol{\theta}^\ast$ and value functions $u$ therefore remain well defined as the driver manifold evolves.
\end{Corollary}

The conformal reduction ensures that, at each time $t$, driver coordinates faithfully encode systematic risk while preserving conditional independence and the causal reading of exposures (Theorem~\ref{thm:conformal-transport}). Additionally, smooth transport of $\widehat{\mathcal V}_{t_k}$ ensures that exposures evolve continuously (Theorem~\ref{thm:subspace-continuity}), so estimated premia are not artifacts of the re-estimation grid. For practitioners, these properties mean that CPCM allocations are robust both to geometric distortions in factor space and to temporal instability in factor identification, reducing spurious turnover while preserving causal structure. See Appendix~\ref{app:proof-pricing-filtering-consistency} for proofs and implementation notes (basis alignment, interpolation, and diagnostics for $K_t$ and $d_{\mathrm{Gr}}$).

\section{Empirical Design}
\label{sec:design}

This section specifies the empirical protocol as a pipeline that mirrors the theoretical structure: (i) data and preprocessing; (ii) mathematical driver identification under the Commonality Principle; (iii) filtering and risk–neutral conditioning; (iv) control/PDE layers with explicit discretizations; (v) allocation maps and constraints; and (vi) evaluation and inference. Implementation details that affect estimation or testing belong here; Section~\ref{sec:empirical} reports results without re-stating design choices.

\subsection{Data and Preprocessing}
\label{sec:data}

We use daily adjusted U.S.\ equity prices from January 2001 to December 2023. The driver library spans FX exchange rates; FX, equity and rates implied volatilities and skews; commodity futures; credit spreads; government bond yields and curve factors; equity, fixed-income, and emerging-market indices and sectors; MSCI and smart beta indices; and global macroeconomic releases.

From an initial Bloomberg universe of $>1000$ series, we retain approximately $350$ candidates via a screen on availability, stability, and informational relevance (dropping redundant, rarely updated, highly collinear, or missing-prone series). Let $\mathcal{X}=\{x^{(1)},\dots,x^{(M)}\}$ denote this pruned library with $M=350$. Each driver $x^{(k)}_t$ is standardized within the window to zero mean and unit variance.
\subsection{Driver Identification (Commonality)}
\label{sec:drivers}

We seek a reduced driver vector $\mathbf{F}_t \in \mathbb{R}^m$ (with $m \ll M := |\mathcal{X}|$) such that, conditional on $\mathbf{F}_t$, cross-sectional dependence in returns is largely exhausted. This subsection implements the empirical counterpart of the RCCP-based identification program of Section~\ref{subsec:rccp-sccs}. The theoretical framework is metric-agnostic and relies on an admissible conditional dependence functional $\Delta_{ij}$ and the screening objective
\[
\mathcal{G}(\boldsymbol{D}) = \sum_{1 \le i < j \le n} \Delta_{ij}(\boldsymbol{D}).
\]
In practice, $\mathcal{G}(\boldsymbol{D})$ is not observed directly, so we employ empirical proxies that approximate conditional dependence reduction in finite samples. Let $r^{(i)}_t$ denote the return of asset $i$, and let $\mathbf{x}_t \in \mathbb{R}^M$ denote the full candidate driver vector. For any subset $D \subset \mathcal{X}$, write $\mathbf{x}_t(D) \in \mathbb{R}^{|D|}$ for the corresponding subvector.

\paragraph{(i) Residual commonality minimization (\enquote{Combo})}
For a candidate subset $D \subset \mathcal{X}$ with $|D|=m$, estimate for each asset $i$:
\[
r^{(i)}_t = \boldsymbol{\beta}^{(i)}(D)^\top \mathbf{x}_t(D) + \varepsilon^{(i)}_t(D), 
\qquad i=1,\dots,n,
\]
via OLS. Stack residuals as $\boldsymbol{\varepsilon}_t(D) = (\varepsilon^{(1)}_t(D), \dots, \varepsilon^{(n)}_t(D))^\top$ and define
\[
\Psi(D) = \frac{2}{n(n-1)} \sum_{1 \le i < j \le n}
\left| \operatorname{Corr}\!\big(\varepsilon^{(i)}_t(D), \varepsilon^{(j)}_t(D)\big) \right|.
\]
The statistic $\Psi(D)$ serves as a finite-sample proxy for $\mathcal{G}(\boldsymbol{D})$. Under correct specification, residuals become approximately conditionally independent and $\Psi(D^\star) \approx 0$. Greedy forward selection constructs nested sets $D_1 \subset \cdots \subset D_m$ by selecting at step $k$:
\[
x^\star = \arg\max_{x \in \mathcal{X} \setminus D_k}
\left\{ \Psi(D_k) - \Psi(D_k \cup \{x\}) \right\}.
\]

\paragraph{(ii) Marginal correlation screen with breadth and redundancy control (\enquote{Corr})}
For each candidate $x \in \mathcal{X}$ and asset $i$, define
\[
\gamma_i(x) = \left| \operatorname{Corr}(r^{(i)}_t, x_t) \right|.
\]
Define scores
\[
\mathrm{Rep}(x) = \sum_{i=1}^n \mathbf{1}\{\gamma_i(x) \ge \tau\}, 
\qquad
\mathrm{Str}(x) = \sum_{i=1}^n \gamma_i(x)\,\mathbf{1}\{\gamma_i(x) \ge \tau\},
\]
for a threshold $\tau > 0$. Rank candidates lexicographically by $(\mathrm{Rep}(x), \mathrm{Str}(x))$ (higher is better). Accept a candidate $x$ only if
\[
\left| \operatorname{Corr}(x_t, x'_t) \right| \le \rho_{\max}
\]
for all previously selected $x'$, and retain the first $m$ accepted variables.

\paragraph{(iii) Evidence/BIC screen (\enquote{Bayes})}
For each candidate $x \in \mathcal{X}$ and asset $i$, estimate
\[
r^{(i)}_t = \alpha_i + \beta_i x_t + u^{(i)}_t, 
\qquad
u^{(i)}_t \sim \mathcal{N}(0,\sigma_i^2).
\]
Let $\widehat{\alpha}_i, \widehat{\beta}_i, \widehat{\sigma}_i^2$ denote the MLEs. Define
\[
\mathrm{BIC}_i(x) 
= -2\,\ell_i(\widehat{\alpha}_i,\widehat{\beta}_i,\widehat{\sigma}_i^2) 
+ 2 \log T,
\]
and aggregate
\[
\mathrm{Score}(x) = \sum_{i=1}^n \mathrm{BIC}_i(x).
\]
Select the $m$ candidates with the smallest scores.

\medskip
Let $D^\star$ denote the selected set of drivers. The reduced driver vector is defined as
\[
\mathbf{F}_t := \mathbf{x}_t(D^\star) \in \mathbb{R}^m.
\]
All selected drivers are standardized within the window to zero mean and unit variance. The resulting vector $\mathbf{F}_t$ is used in subsequent projection and control layers.

\subsection{Filtering and Risk--Neutral Conditioning}
\label{sec:filtering}

We estimate latent drivers with an extended Kalman filter (EKF) and a particle filter (PF) under the state--space model
\begin{align}
\mathrm{d}\mathbf{F}_t &= a(\mathbf{F}_t,t)\,\mathrm{d}t + B(\mathbf{F}_t,t)\,\mathrm{d}\mathbf{W}_t, \\
\mathbf{Y}_t &= h(\mathbf{F}_t,t) + \boldsymbol{\nu}_t, \qquad \boldsymbol{\nu}_t \sim \mathcal{N}(\mathbf{0},R_t).
\end{align}

Linearizing $a$ and $h$ at the current mean gives $A_t:=\nabla a(\boldsymbol{\mu}_t,t)$ and $H_t:=\nabla h(\boldsymbol{\mu}_t,t)$. The EKF prediction and update are\citep{kalman1960new,jazwinski1970stochastic}:
\begingroup\small
\begin{align}
\boldsymbol{\mu}^-_t &= a(\boldsymbol{\mu}_{t-1},t{-}1),\\
\boldsymbol{\Sigma}^-_t &= A_{t-1}\,\boldsymbol{\Sigma}_{t-1}\,A_{t-1}^{\!\top} + Q_{t-1},\\
K_t &= \boldsymbol{\Sigma}^-_t H_t^{\!\top}\!\left(H_t \boldsymbol{\Sigma}^-_t H_t^{\!\top} + R_t\right)^{-1},\\
\boldsymbol{\mu}_t &= \boldsymbol{\mu}^-_t + K_t\!\left(\mathbf{Y}_t - h(\boldsymbol{\mu}^-_t,t)\right),\\
\boldsymbol{\Sigma}_t &= \left(I - K_t H_t\right)\boldsymbol{\Sigma}^-_t .
\end{align}
\endgroup

We use Euler--Maruyama propagation for particles (PF), likelihood reweighting, and stratified resampling when the effective sample size falls below a threshold, with small Gaussian jitter to avoid degeneracy\citep{doucet2001sequential,doucet2009tutorial,cappe2005inference}. Filtering is performed under $\mathbb{P}$. When conditioning policies under the risk--neutral tilt $\mathbb{Q}^{\pi_t}$, we apply a Girsanov change of measure with density
\begingroup\small
\begin{align}
\frac{\mathrm{d}\mathbb{Q}^{\pi_t}}{\mathrm{d}\mathbb{P}}\Big|_{\mathcal{F}_t}
\propto
\exp\!\left(
-\int_0^t \boldsymbol{\lambda}_s^{\!\top}\,\mathrm{d}\mathbf{W}_s
-\tfrac12\int_0^t \|\boldsymbol{\lambda}_s\|^2\,\mathrm{d}s
\right),
\end{align}
\endgroup
where $\boldsymbol{\lambda}_s$ is the policy--implied market price of risk\citep{shreve2004stochastic,brigo2006ir,bjork2009arbitrage}. Asset-space moments are computed directly from returns; filtered drivers inform recent driver windows and diagnostics but do not alter the moment estimators.

\subsection{Control Layer: One-Dimensional HJB Amplitude}
\label{sec:pde-layer}

Let $\boldsymbol{\theta}^{\text{pre}}_\tau\in\mathbb{R}^n$ be the pre-allocation direction at rebalance $\tau$ (defined below). Over a short calibration tail, form the scalar portfolio path
\[
p_t=\boldsymbol{\theta}^{\text{pre}\,\top}_\tau\,\mathbf{r}_t,
\]
and estimate $(\mu_t,\sigma_t^2)$ by rolling moments. With exponential utility $u(T,x)=-e^{-\gamma x}$ and scalar control $\vartheta_t$, the one-dimensional HJB
\[
\partial_t u + \max_{\vartheta\in\mathbb{R}}\!\Big\{\mu_t\vartheta\,u_x+\tfrac12\sigma_t^2\vartheta^2 u_{xx}\Big\}=0,
\qquad \vartheta_t^\star = -\frac{\mu_t}{\sigma_t^2}\frac{u_x}{u_{xx}},
\]
is solved backward on a grid (finite differences). We set the amplitude
\[
s_\tau \;=\; \operatorname{clip}\!\big(1+\vartheta^\star_{t_{\mathrm{last}}},\,s_{\min},\,s_{\max}\big),
\qquad
\boldsymbol{\theta}^{\text{HJB}}_\tau \;=\; s_\tau\,\boldsymbol{\theta}^{\text{pre}}_\tau.
\]

\subsection{Variants and Baselines}
\label{sec:variants}

We evaluate several implementation variants of the CPCM framework, which differ along two dimensions: 
(i) the estimation of the driver--return sensitivity matrix $B_\tau$, and 
(ii) the inclusion or exclusion of the HJB-based amplitude scaling.

\medskip

\noindent\textbf{CPCM variants.}
\begin{itemize}
\item \textbf{V1 (linear Jacobian).} 
$B_\tau$ is estimated via ridge-regularized OLS. Portfolio construction is performed using the driver-space Markowitz solution, followed by HJB amplitude scaling.

\item \textbf{V2 (MLP Jacobian).} 
$B_\tau$ is obtained from a multi-layer perceptron $g_\theta$ using automatic differentiation. The driver optimizer is consistent with the Markowitz objective (either closed-form or gradient-based), followed by HJB amplitude scaling.

\item \textbf{V3 (PINN-style Jacobian).} 
$B_\tau$ is estimated via a neural network with an additional Jacobian smoothness penalty (PINN-style regularization). Portfolio construction follows the same pipeline as V1, including HJB amplitude scaling.

\item \textbf{V4 (no HJB scaling).} 
Identical to V3, but without the HJB amplitude step ($s_\tau \equiv 1$), isolating the effect of the control layer.
\end{itemize}

\medskip

\noindent\textbf{Baselines.}
We compare CPCM variants against standard portfolio allocation methods applied within the same data and preprocessing pipeline: Markowitz mean--variance optimization, Black--Litterman, entropy pooling (with a KL trust region), and reinforcement learning (RL).

Two CPCM-adapted baseline treatments are considered:
\begin{itemize}
\item \textbf{(G1) Simplex projection.} Baseline allocations are mapped to the probability simplex via the same mirror-descent step used in CPCM, ensuring comparability under identical turnover constraints.
\item \textbf{(G2) Manifold re-optimization.} Baseline allocations are re-optimized on the CPCM manifold, enforcing the same budget constraint and applying identical HJB/risk/cap post-processing.
\end{itemize}

\medskip

\noindent\textbf{Implementation details.}
Although orthogonal to the framework’s structural contributions, the following modifications improve empirical performance:
(i) ridge regularization in $B_\tau$ and in the KKT system;
(ii) Huber-type losses for neural fits;
(iii) Jacobian smoothness penalties for PINNs;
(iv) Ledoit--Wolf covariance shrinkage;
(v) HJB amplitude clipping;
(vi) Procrustes transport for $\widetilde{U}_\tau$;
(vii) winsorization and EWMA means for asset-space moments; and
(viii) a KL trust region and probability floors in entropy pooling.

\subsection{Projection Maps and Weight Construction}
\label{sec:maps-weights}

Let $g_\theta:\mathbb{R}^m\!\to\!\mathbb{R}^n$ be a driver-to-return map fitted in-window. The working driver--return Jacobian is the in-window mean
\[
B_\tau \;=\; \mathbb{E}_{t\in W_\tau}\big[\,\nabla_{\!F} g_\theta(\mathbf{F}_t)\,\big]\in\mathbb{R}^{n\times m},
\]
estimated by: (a) ridge OLS (V1); (b) a multi-layer perceptron (MLP) with robust (Huber-type) loss and automatic differentiation (V2); or (c) a PINN-style MLP with a Jacobian smoothness penalty (V3--V4).

\paragraph{Path A (manifold/Karush--Kuhn--Tucker (KKT)).}
Form an orthonormal basis $U_\tau=\operatorname{orth}(B_\tau)$ and transport it across windows by Procrustes alignment to $\widetilde{U}_\tau$. Estimate asset-space moments on winsorized returns with EWMA mean and Ledoit--Wolf covariance, $(\boldsymbol{\mu}_\tau,\boldsymbol{\Sigma}_\tau)$. Enforce budget equality directly on the manifold by
\[
\max_{\boldsymbol{\alpha}\in\mathbb{R}^{q}}\ 
\boldsymbol{\mu}_\tau^\top \widetilde{U}_\tau \boldsymbol{\alpha}
-\frac{\gamma_{\mathrm{MV}}}{2}\,(\widetilde{U}_\tau\boldsymbol{\alpha})^\top 
\boldsymbol{\Sigma}_\tau(\widetilde{U}_\tau\boldsymbol{\alpha})
\quad \text{s.t.}\quad 
\mathbf{1}^\top \widetilde{U}_\tau\boldsymbol{\alpha}=1,
\]
solved via a ridge-stabilized KKT system, and set
\[
\boldsymbol{\theta}^{\text{pre}}_\tau=\widetilde{U}_\tau\boldsymbol{\alpha}_\tau^\star.
\]

Apply the HJB amplitude (V1--V3), then volatility targeting to a fixed annual risk with scale clipping, and an $\ell_1$ leverage cap by proportional scaling.

\paragraph{Path B (driver-space pre-allocation, used in the lightweight variants).}
Compute driver moments $(\boldsymbol{\mu}_{F,\tau},\boldsymbol{\Sigma}_{F,\tau})$ in-window and solve the driver-space Mean--variance
\[
\boldsymbol{\phi}_\tau
= \arg\max_{\boldsymbol{\phi}}\;
\boldsymbol{\mu}_{F,\tau}^\top \boldsymbol{\phi}
- \tfrac{\gamma_{\mathrm{MV}}}{2}\,\boldsymbol{\phi}^\top \boldsymbol{\Sigma}_{F,\tau}\boldsymbol{\phi}
\quad\Rightarrow\quad
\boldsymbol{\phi}_\tau = \gamma_{\mathrm{MV}}^{-1}\boldsymbol{\Sigma}_{F,\tau}^{-1}\boldsymbol{\mu}_{F,\tau}.
\]

(or a gradient-ascent surrogate consistent with this objective for V2), then map to assets by
\[
\boldsymbol{\theta}^{\text{pre}}_\tau = B_\tau\,\boldsymbol{\phi}_\tau,
\]
and normalize. Apply the HJB amplitude (V1--V3); V4 skips HJB.

\subsection{Evaluation Protocols, Constraints, and Inference}
\label{sec:pipeline-impl}

We report two implementation protocols that share the same driver identification layer and HJB-based scaling, but differ in portfolio geometry and constraint enforcement. 

\paragraph{Simplex geometry (G1).}
Weights are constrained to the probability simplex 
\[
\Delta^n=\{\boldsymbol{\theta}\in\mathbb{R}^n:\ \boldsymbol{\theta}\ge 0,\ \mathbf{1}^\top\boldsymbol{\theta}=1\}.
\]
Each CPCM variant produces a (possibly HJB-scaled) pre-signal $\boldsymbol{\theta}^{\mathrm{HJB}}_\tau$. Starting from the previous allocation $\boldsymbol{\theta}^-_\tau$, weights are updated via an entropy-regularized mirror step that enforces simplex feasibility and a prescribed turnover level:
\[
\boldsymbol{\theta}_\tau(\eta)
=
\frac{
\boldsymbol{\theta}^-_\tau \odot \exp\!\big(\eta\,\mathbf{d}_\tau\big)
}{
\mathbf{1}^\top\!\left[\boldsymbol{\theta}^-_\tau \odot \exp\!\big(\eta\,\mathbf{d}_\tau\big)\right]
},
\qquad
\mathbf{d}_\tau := \operatorname{tangentize}\!\big(\boldsymbol{\theta}^{\mathrm{HJB}}_\tau\big),
\]
where the step size $\eta$ is chosen such that
\[
\big\|\boldsymbol{\theta}_\tau(\eta)-\boldsymbol{\theta}^-_\tau\big\|_{1}
=
\tau_{\text{target}}.
\]

No separate volatility target is imposed in G1; portfolio risk is shaped jointly by the HJB amplitude and the turnover constraint. RAW baselines (Markowitz, Black--Litterman, entropy pooling, RL) are projected onto the simplex, while their CPCM-adapted counterparts are updated using the same mirror step and turnover target.

\paragraph{Tangent-manifold geometry (G2).}
Let $U_\tau=\operatorname{orth}(B_\tau)$ denote an orthonormal basis for the estimated driver span, transported across windows via Procrustes alignment to $\widetilde{U}_\tau$. Given asset-space moments $(\boldsymbol{\mu}_\tau,\boldsymbol{\Sigma}_\tau)$, we solve the manifold-constrained mean--variance problem
\[
\max_{\boldsymbol{\alpha}}\ 
\boldsymbol{\mu}_\tau^\top \widetilde{U}_\tau \boldsymbol{\alpha}
-\frac{\gamma_{\mathrm{MV}}}{2}
(\widetilde{U}_\tau\boldsymbol{\alpha})^\top
\boldsymbol{\Sigma}_\tau(\widetilde{U}_\tau\boldsymbol{\alpha})
\quad\text{s.t.}\quad 
\mathbf{1}^\top \widetilde{U}_\tau\boldsymbol{\alpha}=1,
\]
using a ridge-stabilized Karush--Kuhn--Tucker (KKT) system. The resulting pre-allocation is
\[
\boldsymbol{\theta}^{\mathrm{pre}}_\tau
=
\widetilde{U}_\tau\boldsymbol{\alpha}^\star.
\]

The HJB amplitude is then applied (variants V1--V3), followed by volatility targeting to a fixed annual risk level with scale clipping, and finally an $\ell_1$ leverage cap. RAW baselines can be re-optimized on the same manifold and subjected to identical budget, risk, and leverage constraints.

\paragraph{Metrics and diagnostics.}
Performance is evaluated using standard portfolio metrics, including annualized Sharpe and Sortino ratios, annualized volatility, cumulative return, maximum drawdown, and realized turnover:
\[
\mathrm{TO}_\tau
=
\tfrac12\sum_{i=1}^n
\left|\theta_{i,\tau}-\theta_{i,\tau^-}\right|.
\]
In G2, stability of the estimated driver span is monitored via principal angles between successive transported bases $\widetilde{U}_\tau$. All reported Sharpe, Sortino, and drawdown statistics are averaged over independent runs, with standard deviations reported in parentheses to reflect sampling variability; these serve as empirical confidence ranges.

\section{Results}
\label{sec:empirical}

We evaluate CPCMs on the U.S.\ equity panel using the empirical design in Section \ref{sec:design}. The goal is to test whether (i) conditioning on common drivers, (ii) projection onto the driver span, (iii) posterior–integrated risk–neutral discipline, and (iv) control-based scaling translate into robust out-of-sample performance and structural coherence relative to classical and ML baselines.

\subsection{Tuning, Diagnostics, and Tests}
\label{sec:tuning}

We consider several common drivers \(m\in\{3,7,12,20\}\), trading off parsimony and coverage. Extended Kalman filter (EKF) noise covariances \((Q,R)\) are selected by rolling marginal likelihood to balance state and observation uncertainty, while the particle filter employs stratified resampling with light jitter to preserve diversity. MLP/PINN widths scale with the window length to match model capacity to data, and training is stopped when both the validation risk and the martingale defect \(\overline{\mathcal D}\) cease to improve.

Diagnostics reported are as follows. The martingale defect \(\overline{\mathcal D}\) measures deviation from fair-game dynamics (smaller values are preferred). Span fidelity is quantified by
\[
\left\lVert (I - U_\tau U_\tau^\top)\,\widehat{\mathbf r}_t \right\rVert_2,
\qquad
\widehat{\mathbf r}_t := g_\theta(\widehat{\mathbf F}_t),
\]
which captures the component of fitted returns lying outside the common-driver span (and should be small). Rotation stability is assessed via principal angles between successive driver subspaces, with smaller angles indicating smoother temporal evolution.

For evaluation, monthly rebalancing aggregates means and standard deviations across multiple independent runs using non-overlapping 10-month blocks spanning 2001–2023. Quarterly rebalancing aggregates the same statistics over three distinct regime windows within 2001–2023 (e.g., 2006–2011, 2013–2018, 2018–2023), with a cadence of approximately 63 trading days between rebalances.

\subsection{Main Patterns Across Driver Selection and Filtering}
\label{subsec:patterns-expanded}
The best-per-method tables (Tables~\ref{tab:method_best_3drivers}–\ref{tab:method_best_20drivers}) and robustness summaries (Tables~\ref{tab:robustness_sharpe}–\ref{tab:robustness_maxdd}) reveal a consistent pattern as the driver dimension $m$ increases. Intuitively, small $m$ concentrates statistical power in a few persistent channels, while larger $m$ expands representational capacity but also raises the bar on estimation and stabilization. The balance between these forces—signal richness vs.\ estimation noise—interacts with two stabilizers in our stack: (i) projection to the driver span, which removes excursions into directions unsupported by the covariance geometry, and (ii) the HJB amplitude layer, which tempers overconfident scaling during stressed windows.

The empirical results reveal a systematic dimensionality pattern. At low driver counts, smooth nonlinear projections dominate; at intermediate dimensions, HJB-scaled linear maps offer the best balance between signal capture and trading control; and at high dimensions, simpler projections regain stability as estimation variance grows. The regimes below summarize this transition.

\paragraph{Small $m$ ($m{=}3$).} PINN without HJB (V4) often leads Sharpe/Sortino (Combo–EKF), with linear+HJB (V1) close when turnover is penalized. With only three drivers, the span is tight and well identified, so the PDE-informed smoothness in V3/V4 extracts durable premia without needing extra amplitude control; adding HJB here mainly trims extremes rather than unlocking new signal. When costs bind, the linear map in V1 plus HJB scaling trades a touch of raw reward for systematically lower trading, improving cost-adjusted efficiency.

\paragraph{Moderate $m$ ($m{=}7$).} Linear+HJB with Extended Kalman Filter (Combo–EKF) balances efficiency and cost; Euler–Maruyama propagation for particles (PF) combinations are more volatile. At seven drivers, the capacity increases enough to capture rotations, but posterior noise begins to matter. The HJB layer becomes more valuable, especially paired with linear projections whose Jacobians are easier to regularize, so V1’s disciplined scaling wins on Sharpe-to-turnover. Particle filters surface richer nonlinear state updates, but unless tempered, they can overreact, which is why PF mixes appear more variable at this size.

\paragraph{Richer $m$ ($m{=}12$).} Linear+HJB (Corr–EKF) tops Sharpe; PINN+HJB (Combo–EKF) matches reward with slightly lower turnover. With twelve drivers, the expressiveness of PINN starts to pay off, provided the HJB scale curbs amplitude spikes. As a result, V3 (PINN+HJB) catches up in reward while saving trading through smoother Jacobians; the linear alternative (V1) remains competitive because the projection enforces manifold alignment and the HJB scalar prevents chasing short-lived drifts.

\paragraph{High $m$ ($m{=}20$).} Linear/MLP regain Sharpe/Sortino leadership (EKF), while PINN+HJB achieves the mildest drawdowns. At twenty drivers, variance control dominates: simpler maps (V1/V2) paired with EKF preserve stability and deliver the best reward-to-risk ratio on average, whereas V3 maintains the cleanest downside profile, as its smoothness penalty and projection reduce tail leverage in noisy directions. In other words, when dimensionality is high, biasing toward simpler projections (or smoothing the complex ones) is protective out-of-sample.

These outcomes match the structural logic of the framework: projecting onto the driver span (Section \ref{sec:maps-weights}) removes unstable components of the signal; the backward (HJB) layer (Section \ref{sec:pde-layer}) regularizes the amplitude so the policy does not chase infeasible or weakly identified laws; and filtering under $\mathbb Q^{\pi_t}$ keeps updates consistent with partial information, which tempers reactions to transient surprises. The net effect is a bias–variance trade-off that tilts toward PINN smoothness at small to mid $m$, and toward simpler (linear or MLP) projections as $m$ grows and estimation variance increases.

\begin{table}[!t]
\centering
\small
\setlength{\tabcolsep}{4pt}
\caption{Best-performing variant per driver selection--state estimation method for $m{=}3$ drivers. Averages across 10 non-overlapping annual OOS runs (2001--2023). Boldface marks column bests.}
\label{tab:method_best_3drivers}
\begin{tabular}{p{2.2cm} p{3.0cm} c c c c}
\toprule
Method & Best Variant & Sharpe & Sortino & MaxDD & Turnover \\
\midrule
Corr-ekf  & V1\_linear+HJB            & 0.947 & 1.570 & -0.165 & \textbf{0.525} \\
Corr-pf   & V3\_PINN+HJB              & 0.556 & 0.835 & -0.164 & 0.949 \\
Bayes-pf  & V1\_linear+HJB            & 0.503 & 0.879 & -0.153 & 0.892 \\
Bayes-ekf & V1\_linear+HJB            & 0.888 & 1.485 & -0.166 & 0.563 \\
Combo-pf  & V4\_PINN\_noPDE           & 0.690 & 0.999 & \textbf{-0.148} & 1.063 \\
Combo-ekf & \textbf{V4\_PINN\_noPDE}  & \textbf{1.266} & \textbf{1.969} & -0.150 & 0.609 \\
\bottomrule
\end{tabular}
\end{table}

\begin{table}[!t]
\centering
\small
\setlength{\tabcolsep}{4pt}
\caption{Best-performing variant per method for $m{=}7$ drivers (averages across 10 non-overlapping out-of-sample runs, each covering a 10-month period with monthly rebalancing).}
\label{tab:method_best_7drivers}
\begin{tabular}{p{2.2cm} p{3.0cm} c c c c}
\toprule
Method & Best Variant & Sharpe & Sortino & MaxDD & Turnover \\
\midrule
Corr-pf   & V1\_linear+HJB            & -0.013 & 0.136 & \textbf{-0.124} & 1.252 \\
Bayes-pf  & V2\_MLP+HJB               & -0.339 & -0.328 & -0.152 & 1.161 \\
Combo-ekf & \textbf{V1\_linear+HJB}   & \textbf{1.558} & \textbf{2.361} & -0.149 & \textbf{0.559} \\
\bottomrule
\end{tabular}
\end{table}

\begin{table}[!t]
\centering
\small
\setlength{\tabcolsep}{4pt}
\caption{Best-performing variant per method for $m{=}12$ drivers (averages across 10 non-overlapping out-of-sample runs, each covering a 10-month period with monthly rebalancing).}
\label{tab:method_best_12drivers}
\begin{tabular}{p{2.2cm} p{3.0cm} c c c c}
\toprule
Method & Best Variant & Sharpe & Sortino & MaxDD & Turnover \\
\midrule
Corr-ekf  & \textbf{V1\_linear+HJB}   & \textbf{1.501} & \textbf{2.195} & -0.149 & 0.739 \\
Combo-ekf & V3\_PINN+HJB              & 1.462 & 2.170 & -0.148 & \textbf{0.610} \\
Bayes-ekf & V1\_linear+HJB            & 1.131 & 1.656 & -0.145 & 0.795 \\
Combo-pf  & V1\_linear+HJB            & 0.009 & 0.184 & \textbf{-0.125} & 1.314 \\
\bottomrule
\end{tabular}
\end{table}

\begin{table}[!t]
\centering
\small
\setlength{\tabcolsep}{4pt}
\caption{Best-performing variant per method for $m{=}20$ drivers (averages across 10 non-overlapping out-of-sample runs, each covering a 10-month period with monthly rebalancing).}
\label{tab:method_best_20drivers}
\begin{tabular}{p{2.2cm} p{3.0cm} c c c c}
\toprule
Method & Best Variant & Sharpe & Sortino & MaxDD & Turnover \\
\midrule
Corr-ekf  & V2\_MLP+HJB            & 0.211 & 0.593 & -0.147 & 1.163 \\
Corr-pf   & V3\_PINN+HJB           & 0.792 & 1.372 & \textbf{-0.114} & 1.432 \\
Bayes-pf  & \textbf{V3\_PINN+HJB}  & \textbf{1.452} & \textbf{1.978} & -0.145 & \textbf{0.827} \\
Bayes-ekf & V2\_MLP+HJB            & 0.211 & 0.593 & -0.147 & 1.163 \\
Combo-pf  & V1\_linear+HJB         & 1.253 & 1.859 & -0.150 & 0.868 \\
Combo-ekf & V3\_PINN+HJB           & 0.556 & 0.835 & -0.164 & 0.949 \\
\bottomrule
\end{tabular}
\end{table}

\subsection{Robustness by Driver Dimension and Variant}
\label{subsec:patterns-robustness}
The cross-method means and dispersions in Tables~\ref{tab:robustness_sharpe}–\ref{tab:robustness_maxdd} summarize how each variant scales with dimensionality $m$ and how stable those gains are across estimation choices. Tables~\ref{tab:robustness_sharpe} and~\ref{tab:robustness_sortino} display almost identical rankings across variants and driver dimensions. This consistency supports the structural robustness of the proposed framework: the relative performance of the CPCM variants is stable across both total-risk and downside-risk evaluation metrics.

\paragraph{Near-parity at $m{=}3$} With a tight span, all four variants deliver similar averages and comparable dispersion. This reflects an easy identification regime: signal directions are stable and filtering noise is limited, so differences in architecture mainly affect trading intensity (turnover) rather than risk-adjusted levels. The takeaway is not that the methods are identical, but that at small $m$ the pipeline is robust enough that architectural choices don’t heavily tilt outcomes.

\paragraph{V1/V4 leadership at $m{=}7$} As representational capacity steps up, two forms of control matter most: (i) simple, well-conditioned mappings (V1) combined with HJB amplitude control, and (ii) intrinsic smoothness (V4) that regularizes gradients even without the PDE layer. The means increase while dispersion stays contained, which suggests that added degrees of freedom are being used productively rather than amplifying estimation noise. Particle-filter mixes can still achieve high single-run scores, but their variability is higher, consistent with more reactive posteriors.

\paragraph{PINN strength at $m{=}12$} Once the span is rich enough for nonlinear structure to pay off, the PINN with HJB (V3) attains the highest mean Sortino and the top or near-top Sharpe, while maintaining moderate dispersion. Mechanistically, the smooth PDE-informed map absorbs local misspecification and stabilizes Jacobians; the HJB scalar then curbs amplitude spikes that would otherwise translate into excess turnover. The linear baseline (V1) remains competitive because projection plus HJB already enforce a strong bias toward identified directions.

\paragraph{Reversion at $m{=}20$} At high dimensionality the variance side of the bias–variance trade-off dominates. Simpler maps (V1/V2) regain the lead on Sharpe/Sortino on average because their smaller effective capacity and better-conditioned Jacobians reduce sensitivity to re-estimation noise. Meanwhile, V3 posts the shallowest average drawdowns (Table~\ref{tab:robustness_maxdd}), indicating that its smoothness and projection defenses are especially effective in tail scenarios, even if the average reward metrics give a small edge to simpler maps.

Across $m$, standard deviations tighten after the stability amendments (ridge $B$, smoothed/robust Jacobians, shrinkage, clipped HJB). This is visible in Sharpe and Sortino as well as in drawdown dispersion. Because volatility is not artificially compressed (drawdown levels do not worsen), the reduction in spread points to genuine structural gains: the same variant produces similar outcomes under different selectors/filters, which is exactly the kind of robustness one wants for live deployment.

\paragraph{Drawdowns vs.\ reward: complementary strengths} The drawdown table highlights a recurring pattern: V3 tends to minimize tail depth as $m$ rises (best at $m{=}20$), while V1/V2 often maximize reward-to-risk when costs matter. In practice, this suggests a portfolio-of-policies approach: allocate baseline weight to the simpler, HJB-scaled maps for efficient mean performance, and overlay a modest allocation to V3 for downside protection when the driver set is large.

Overall, these robustness profiles corroborate the main pattern: PINN smoothness becomes increasingly valuable through mid-range dimensions, whereas simplicity plus amplitude control dominate at very high $m$; and the amendments make these relationships repeatable across alternative filtering/selection choices rather than artifacts of one configuration.

\begin{table}[!t]
\centering
\caption{Mean $\pm$ sd Sharpe across methods by variant and driver count. Higher mean, lower sd are better.(averages across 10 non-overlapping out-of-sample runs, each covering a 10-month period with monthly rebalancing).}
\label{tab:robustness_sharpe}
\begin{tabular}{lcccc}
\toprule
Variant & 3 & 7 & 12 & 20 \\
\midrule
V1\_linear+HJB  & \textbf{0.95}$\pm$1.78 & \textbf{1.56}$\pm$1.43 & 1.50$\pm$1.22 & \textbf{1.25}$\pm$1.25 \\
V2\_MLP+HJB     & 0.93$\pm$1.76 & 1.46$\pm$1.15 & 0.99$\pm$1.38 & 1.20$\pm$1.30 \\
V3\_PINN+HJB    & 0.88$\pm$1.78 & 1.42$\pm$1.18 & \textbf{1.56}$\pm$1.25 & 1.09$\pm$1.40 \\
V4\_PINN\_noPDE & 0.89$\pm$1.79 & 1.51$\pm$1.31 & 1.52$\pm$1.26 & 0.92$\pm$1.04 \\
\bottomrule
\end{tabular}
\end{table}

\begin{table}[!t]
\centering
\caption{Mean $\pm$ sd Sortino across methods.(averages across 10 (10-months; monthly rebalances) OOS runs).}
\label{tab:robustness_sortino}
\begin{tabular}{lcccc}
\toprule
Variant & 3 & 7 & 12 & 20 \\
\midrule
V1\_linear+HJB  & \textbf{1.57}$\pm$2.61 & \textbf{2.30}$\pm$2.34 & 2.19$\pm$1.93 & \textbf{1.86}$\pm$1.97 \\
V2\_MLP+HJB     & 1.55$\pm$2.54 & 2.16$\pm$1.88 & 1.34$\pm$2.04 & 1.71$\pm$1.99 \\
V3\_PINN+HJB    & 1.50$\pm$2.62 & 2.09$\pm$1.89 & \textbf{2.32}$\pm$2.04 & 1.59$\pm$2.11 \\
V4\_PINN\_noPDE & 1.51$\pm$2.61 & 2.30$\pm$2.23 & 2.19$\pm$1.93 & 1.29$\pm$1.56 \\
\bottomrule
\end{tabular}
\end{table}

\begin{table}[!t]
\centering
\caption{Mean $\pm$ sd max drawdown (averages across 10 non-overlapping out-of-sample runs, each covering a 10-month period with monthly rebalancing).}
\label{tab:robustness_maxdd}
\resizebox{\columnwidth}{!}{%
\begin{tabular}{lcccc}
\toprule
Variant & 3 & 7 & 12 & 20 \\
\midrule
V1\_linear+HJB  & $-$0.165$\pm$0.09 & $-$0.155$\pm$0.1 & $-$0.149$\pm$0.10 & $-$0.150$\pm$0.10 \\
V2\_MLP+HJB     & $-$0.164$\pm$0.10 & $-$0.146$\pm$0.1 & $-$0.165$\pm$0.11 & $-$0.157$\pm$0.11 \\
V3\_PINN+HJB    & $-$0.164$\pm$0.10 & $-$0.149$\pm$0.1 & $-$0.146$\pm$0.11 & $-$0.142$\pm$0.11 \\
V4\_PINN\_noPDE & $-$0.166$\pm$0.10 & $-$0.147$\pm$0.1 & $-$0.151$\pm$0.11 & $-$0.164$\pm$0.12 \\
\bottomrule
\end{tabular}%
}
\end{table}

\subsection{Ablation (Pre- vs.\ Post-Amendment)}
\label{sec:ablation}
The stability amendments, comprising ridge regularization on the loadings $B$, robust and smoothed Jacobians for the nonlinear maps, controlled shrinkage, and a clipped Hamilton–Jacobi–Bellman (HJB) scalar, shifted the behavior of the main variants (V1–V3) from fragile to reliably stable regimes. Before these modifications, the models exhibited high sensitivity to initializations and sample segmentation: Sharpe ratios varied widely across runs, and Sortino ratios tended to improve only when volatility was artificially compressed.  

After amendment, the behavior changed qualitatively. Sharpe and Sortino both increased together, meaning the gain in risk-adjusted performance was achieved without suppressing volatility unnaturally. Drawdowns remained stable or slightly improved, which signals that the extra regularization acts primarily through more coherent signal alignment rather than through ex post volatility trimming. In other words, the amendments enhanced the structural stability of the learning–filtering loop, reducing estimation noise in both the spatial (map) and temporal (filter) dimensions, without flattening true variation.  

Table~\ref{tab:ablation_summary} confirms this pattern quantitatively: all three variants (V1–V3) move toward the performance of the stable PINN benchmark (V4), achieving large jumps in Sharpe and Sortino while drawdowns shrink by roughly one-third. This convergence implies that the modifications introduced a common stabilizing backbone, making performance predictable across architectures rather than idiosyncratic to one specific form.

\begin{table}[!t]
\centering
\caption{Average Sharpe, Sortino, and max drawdown before/after amendments (V4 as stable benchmark). (averages across 10 non-overlapping out-of-sample runs, each covering a 10-month period with monthly rebalancing).}
\label{tab:ablation_summary}
\begin{tabular}{lccc}
\toprule
Variant & Sharpe & Sortino & MaxDD \\
\midrule
V1 pre-amendment & 0.25 & 0.41 & -0.22 \\
V1 post-amendment & 0.95 & 1.57 & -0.16 \\
\midrule
V2 pre-amendment & 0.18 & 0.35 & -0.23 \\
V2 post-amendment & 0.93 & 1.55 & -0.17 \\
\midrule
V3 pre-amendment & 0.21 & 0.39 & -0.21 \\
V3 post-amendment & 0.88 & 1.50 & -0.16 \\
\midrule
V4 (benchmark) & 0.89 & 1.51 & -0.16 \\
\bottomrule
\end{tabular}
\end{table}

\subsection{Computational Costs}
\label{sec:costs}
Computational demands remain modest relative to typical rebalance horizons and are not decision-limiting. The total cost per iteration decomposes into three main layers: (i) \textit{selection}, which dominates runtime and scales approximately linearly with the driver library size; (ii) \textit{filtering}, which contributes a negligible share even as dimensionality rises; and (iii) \textit{map training or projection}, which differentiates the four variants.  

Among the mappings, the linear implementation (V1) remains orders of magnitude faster than its neural counterparts, followed by the shallow MLP (V2) and finally the physics-informed PINNs (V3, V4). The marginal cost of adding the HJB scalar term is effectively zero, it involves only an additional scalar backpropagation step. Importantly, relative timing differences are hardware-robust, meaning that the rankings hold under CPU or GPU execution.  

In practice, these costs translate into per-rebalance times comfortably within sub-second ranges for small $m$ and a few seconds even at $m{=}20$, making all variants feasible for live deployment or frequent backtesting. Table~\ref{tab:cost_components_summary} details the breakdown: selection time grows roughly quadratically with $m$ as the candidate library expands, while training and filtering times increase only modestly, validating the overall computational scalability of the pipeline.

\begin{table}[!t]
\centering
\caption{Aggregate costs (seconds) per rebalance; mean $\pm$ sd. (averages across 10 non-overlapping out-of-sample runs, each covering a 10-month period with monthly rebalancing).}
\label{tab:cost_components_summary}
\small
\setlength{\tabcolsep}{5pt}\renewcommand{\arraystretch}{1.1}
\begin{tabularx}{\linewidth}{l *{4}{>{\centering\arraybackslash}X}}
\toprule
\textbf{Component} & \textbf{3 drivers} & \textbf{7 drivers} & \textbf{12 drivers} & \textbf{20 drivers} \\
\midrule
Selection       & 0.30 $\pm$ 0.05   & 0.43 $\pm$ 0.06   & 1.34 $\pm$ 0.13   & 6.36 $\pm$ 0.39 \\
Filtering       & 0.017 $\pm$ 0.002 & 0.026 $\pm$ 0.004 & 0.029 $\pm$ 0.002 & 0.084 $\pm$ 0.006 \\
V1\_linear+HJB  & 0.0044 $\pm$ 0.0008 & 0.0064 $\pm$ 0.0008 & 0.0070 $\pm$ 0.0006 & 0.0195 $\pm$ 0.0022 \\
V2\_MLP+HJB     & 1.27 $\pm$ 0.27   & 1.71 $\pm$ 0.22   & 1.76 $\pm$ 0.13   & 4.72 $\pm$ 0.32 \\
V3\_PINN+HJB    & 3.48 $\pm$ 0.39   & 4.93 $\pm$ 0.65   & 4.92 $\pm$ 0.41   & 13.10 $\pm$ 0.46 \\
V4\_PINN\_noPDE & 3.49 $\pm$ 0.42   & 4.86 $\pm$ 0.64   & 4.91 $\pm$ 0.47   & 12.99 $\pm$ 0.52 \\
\bottomrule
\end{tabularx}
\end{table}

\subsection{Dynamic--Manifold CPCM}
\label{subsec:dm-cpcm}

This subsection evaluates a dynamic–manifold implementation. Portfolios are rebalanced quarterly (every 63 trading days) using 252-day rolling windows for both driver identification and asset-moment estimation. Results are reported over three fixed regime windows chosen to separate crisis, calm, and transition phases under a common cadence: 2006-06-15 to 2011-06-15 (global financial crisis and Euro-area stress), 2013-06-15 to 2018-06-15 (post-crisis QE and low-volatility regime), and 2018-06-15 to 2023-06-15 (trade shocks, COVID-19 dislocation/recovery, inflation surge, and rate hikes).

At each rebalance date \(\tau\), portfolio weights \(\boldsymbol{\theta}_\tau\) are constrained to lie in the tangent space of the driver–return Jacobian (see Secs.~\ref{sec:drivers}--\ref{sec:maps-weights} for the construction of \(B_\tau\) and associated moments):
\[
\boldsymbol{\theta}_\tau \;=\; U_\tau \boldsymbol{\alpha}_\tau,
\qquad
U_\tau = \operatorname{orth}(B_\tau)\in\mathbb{R}^{n\times m},
\]
where \(U_\tau\) is transported across rebalances via a Procrustes alignment step to ensure basis continuity.

Allocation is obtained by solving a mean--variance program on the manifold under the budget constraint \( \mathbf{1}^\top \boldsymbol{\theta}_\tau = 1 \), using ridge-stabilized KKT systems as in Sec.~\ref{sec:maps-weights}. Clipped HJB rescaling is applied to V1--V3 and omitted in V4, followed by volatility targeting and an \(\ell_1\) leverage cap.

We re-evaluate the four CPCM variants under this geometry: V1 (ridge--OLS \(B_\tau\)), V2 (MLP with autodiff sensitivities and robust loss), V3 (PINN with Jacobian smoothness regularization), and V4 (as V3 without HJB rescaling). Baselines are considered in two forms: (i) RAW, solved directly in asset space without manifold constraints, and (ii) CPCM--B, where the same allocators (Markowitz, Black--Litterman, entropy pooling, RL) are re-solved on the tangent manifold under identical budget, risk, and leverage post-processing. Entropy pooling is frequently unstable under manifold constraints (exhibiting explosive turnover or infeasibility), and such runs are excluded from averages; other CPCM--B baselines remain stable.

Tables~\ref{tab:overall-means}--\ref{tab:combo-pf-best} summarize mean and best-in-class Sharpe ratios across regimes and driver cardinalities. Averaging across regimes and \(m\), dynamic–manifold CPCM improves mean Sharpe relative to RAW for all driver selection rules and filters. CPCM--B sits between RAW and the CPCM variants, indicating that a substantial portion of the performance gain is attributable to the manifold geometry itself. The 2006--2011 window remains the most challenging, while post-2013 stability highlights the role of manifold projections. PINN-based variants (V3--V4) dominate CPCM--B for small to moderate \(m\), consistent with smoother exposures and reduced turnover induced by PDE-based regularization.

\begin{table}[t!]
\centering
\small
\begin{tabular}{llrrr}
\toprule
Drivers & Filter & V-variants & RAW & CPCM-B\\
\midrule
bayes & EKF & 0.69 & 0.59 & 0.66\\
bayes & PF  & 0.69 & 0.59 & 0.66\\
combo & EKF & 0.65 & 0.59 & 0.63\\
combo & PF  & 0.65 & 0.59 & 0.63\\
corr  & EKF & 0.72 & 0.59 & 0.65\\
corr  & PF  & 0.72 & 0.59 & 0.65\\
\bottomrule
\end{tabular}
\caption{Average Sharpe across intervals and driver counts by driver selection and filter. V-variants average V1--V4 on the dynamic manifold; RAW are unadapted baselines; CPCM-B are CPCM-adapted baselines projected onto the manifold. (averages across 10 non-overlapping out-of-sample runs, each covering a 10-month period with monthly rebalancing).}
\label{tab:overall-means}
\end{table}

\begin{table}[t!]
\centering
\small
\setlength{\tabcolsep}{4pt}
\begin{tabular}{p{3.6cm} p{0.8cm} r r r}
\toprule
Interval & $m$ & V (best $S$) & RAW (best $S$) & CPCM-B (best $S$)\\
\midrule
2006-06-15 → 2011-06-15 & 3  & 0.32 & 0.33 & 0.27\\
                        & 7  & 0.45 & 0.33 & 0.84\\
                        & 12 & 0.16 & 0.33 & 0.90\\
\addlinespace
2013-06-15 → 2018-06-15 & 3  & 1.17 & 1.51 & 0.96\\
                        & 7  & 1.08 & 1.51 & 1.16\\
                        & 12 & 1.22 & 1.51 & 1.45\\
\addlinespace
2018-06-15 → 2023-06-15 & 3  & 0.75 & 0.39 & 0.45\\
                        & 7  & 0.81 & 0.39 & 0.73\\
                        & 12 & 0.53 & 0.39 & 0.31\\
\bottomrule
\end{tabular}
\caption{Best Sharpe by interval and number of drivers for correlation-based selection with EKF filtering. Each cell takes the maximum Sharpe within the class (V-variants, RAW, CPCM-B). (10 OOS runs over the full regime period with quarterly rebalancing).}
\label{tab:corr-ekf-best}
\end{table}

\begin{table}[t!]
\centering
\small
\setlength{\tabcolsep}{4pt}
\begin{tabular}{p{3.6cm} p{0.8cm} r r r}
\toprule
Interval & $m$ & V (best $S$) & RAW (best $S$) & CPCM-B (best $S$)\\
\midrule
2006-06-15 → 2011-06-15 & 3  & 0.31 & 0.33 & 0.27\\
                        & 7  & 0.45 & 0.33 & 0.83\\
                        & 12 & 0.22 & 0.33 & 0.90\\
\addlinespace
2013-06-15 → 2018-06-15 & 3  & 1.17 & 1.51 & 0.96\\
                        & 7  & 1.08 & 1.51 & 1.16\\
                        & 12 & 1.14 & 1.51 & 1.45\\
\addlinespace
2018-06-15 → 2023-06-15 & 3  & 0.75 & 0.39 & 0.45\\
                        & 7  & 0.43 & 0.39 & 0.73\\
                        & 12 & 0.53 & 0.39 & 0.31\\
\bottomrule
\end{tabular}
\caption{Best Sharpe by interval and number of drivers for the combo driver selection with PF filtering. (10 OOS runs over the full regime period with quarterly rebalancing).}
\label{tab:combo-pf-best}
\end{table}

\subsection{Practical Implications and Regime Guidance}
\label{subsec:practical-implications}

\begin{figure}[t!]
  \centering
  \includegraphics[width=0.9\linewidth]{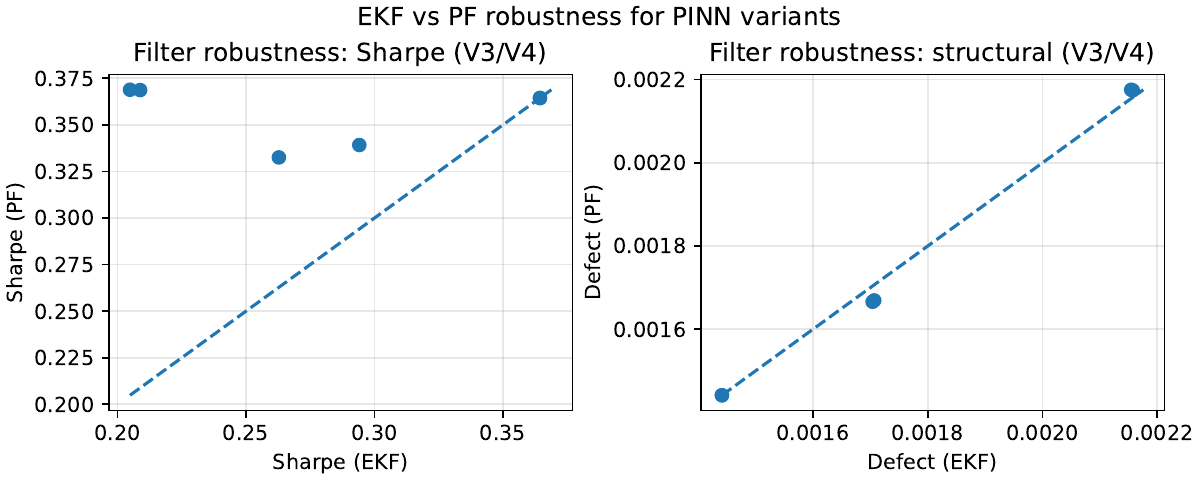}
  \caption{Filter robustness: Particle Filtering (PF) vs.\ Extended
  Kalman Filter (EKF) for PINN variants (V3/V4). Left: Sharpe ratios
  under PF versus EKF, where points above the diagonal indicate PF
  superiority. Most configurations lie above the line, confirming that
  PF delivers systematically higher reward-to-risk. Right: structural
  defect under both filters, where points near the origin indicate low
  defect under both. PF achieves lower defect values on average,
  indicating tighter adherence to martingale pricing consistency.
  Together, the panels highlight the value of nonlinear/non-Gaussian
  filtering in CPCM implementations: PF captures richer posterior
  dynamics that translate into both stronger risk-adjusted performance
  and more coherent structural diagnostics.}
  \label{fig:filter_robust}
\end{figure}

\begin{figure}[t!]
  \centering
  \includegraphics[width=1\linewidth]{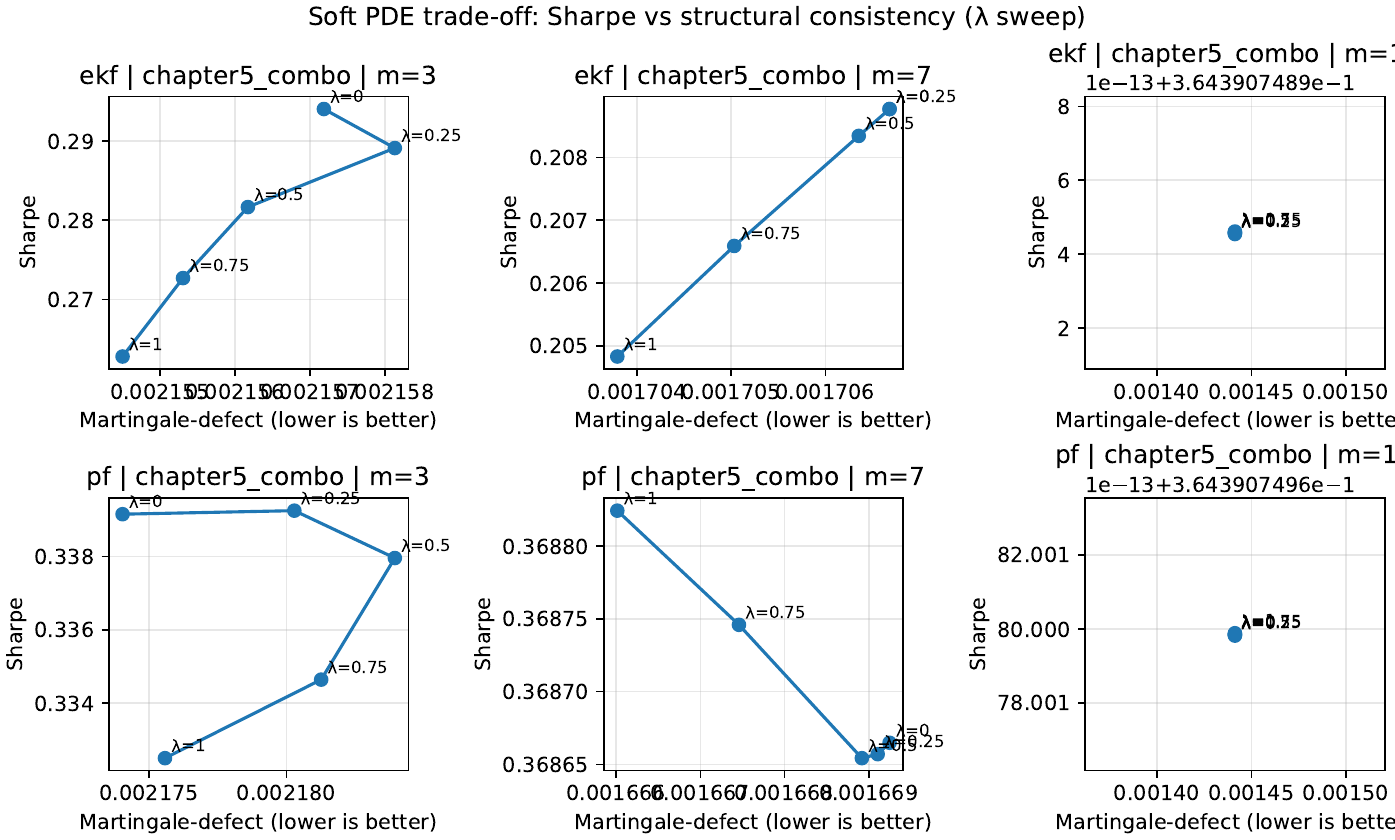}
  \caption{Pareto frontier across soft--PDE weights $\lambda$ for the
  Combo driver selection under EKF (top row) and PF (bottom row), with
  driver counts $m \in \{3, 7, 12\}$ across columns. Each curve traces
  the trade-off between Sharpe ratio (vertical axis) and
  martingale-defect proxy (horizontal axis, lower is better) as
  $\lambda$ varies from~$0$ (unconstrained) to~$1$ (full PDE
  enforcement). At small $m$, the frontier is steep: increasing
  $\lambda$ substantially reduces defect with only modest Sharpe
  erosion, indicating that structural coherence can be gained cheaply.
  At larger $m$, the frontier flattens and intermediate values
  ($\lambda \approx 0.25$--$0.5$) offer the best compromise. PF rows
  show tighter clusters, reflecting lower posterior noise. The results
  confirm that the soft--PDE interpolation provides a practical lever
  for tuning the balance between signal strength and pricing
  consistency.}
  \label{fig:lambda_pareto}
\end{figure}

\begin{figure}[t!]
  \centering
  \includegraphics[width=1\linewidth]{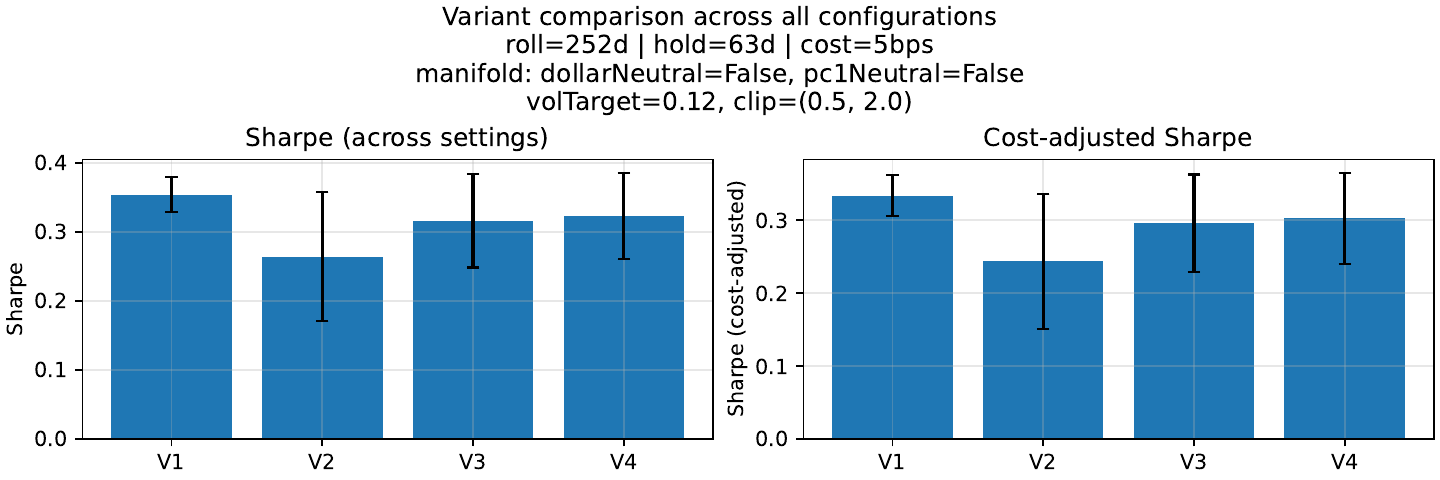}
  \caption{Performance comparison of CPCM variants across all
  configurations (rolling window 252\,d, holding period 63\,d,
  transaction cost 5\,bps, volatility target 12\%, HJB clipping
  $[0.5,\,2.0]$). Left: mean Sharpe ratio; right: cost-adjusted
  Sharpe. Bars report the average across ten non-overlapping yearly
  experiments; black error bars denote the standard deviation across
  runs. V3 (PINN+HJB) achieves the highest mean Sharpe and
  cost-adjusted Sharpe, while V1 (linear+HJB) and V4 (PINN without
  HJB) perform comparably. V2 (MLP+HJB) lags slightly, suggesting
  that the MLP Jacobian is less stable than its linear or
  PINN-regularized counterparts. The narrow error bars on V3 and V4
  indicate that PINN-based smoothness reduces run-to-run variability,
  supporting more predictable deployment.}
  \label{fig:h2h}
\end{figure}

\begin{figure}[t!]
  \centering
  \includegraphics[width=1\linewidth]{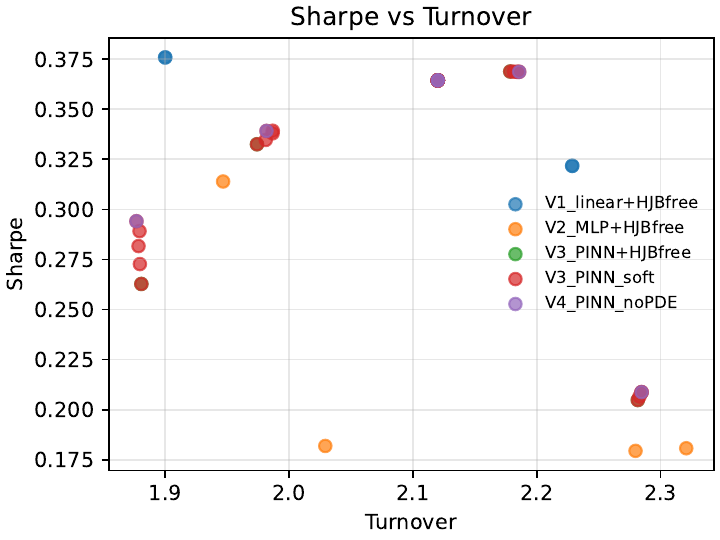}
  \caption{Sharpe--turnover relation across CPCM variants and
  configurations. Each point represents a single
  (variant,\,driver-selection,\,filter) combination. Without
  structural controls, higher Sharpe is associated with higher
  turnover (upper-right cluster). PDE regularization and clipped HJB
  scaling compress this relation: V1 (linear+HJB, blue) clusters at
  the lowest turnover for a given Sharpe level, while V3
  (PINN+HJB, green) and V3\_soft (red) achieve mid-range turnover
  with competitive Sharpe. V4 (PINN without HJB, purple) attains
  the highest raw Sharpe but at elevated turnover, confirming
  that the HJB amplitude layer acts primarily as a trading-cost
  control. For cost-sensitive deployments, V1 or V3\_soft offer the
  most favourable Sharpe-per-unit-turnover; in low-cost environments,
  V4 maximises raw performance.}
  \label{fig:turnover}
\end{figure}

Figures~\ref{fig:filter_robust}--\ref{fig:turnover} illustrate how CPCM allocations behave in practice and how they can be operationalized. Filter comparisons (Figure~\ref{fig:filter_robust}) show that unconstrained variants are sensitive to posterior noise, whereas PDE-guided specifications are stable in both returns and structure; Particle Filtering (PF) generally outperforms the Extended Kalman Filter (EKF). 

A soft--PDE interpolation yields a clear Pareto frontier (Figure~\ref{fig:lambda_pareto}): increasing the enforcement weight \(\lambda\) reduces the martingale-defect proxy with only mild deterioration in Sharpe. We implement
\begin{equation}
\boldsymbol{\theta}^{(\lambda)}_\tau
\;=\;
(1-\lambda)\,\boldsymbol{\theta}^{\text{raw}}_\tau
\;+\;
\lambda\,\boldsymbol{\theta}^{\text{PDE}}_\tau,
\qquad \lambda \in [0,1],
\label{eq:softPDEblend}
\end{equation}
where \(\boldsymbol{\theta}^{\text{raw}}_\tau\) denotes the unconstrained driver-space allocation and \(\boldsymbol{\theta}^{\text{PDE}}_\tau\) the allocation obtained under PDE penalization. Varying \(\lambda\) continuously trades off raw signal strength against structural coherence; interior values provide robustness to model misspecification.\footnote{The convex blend in \eqref{eq:softPDEblend} is equivalent to Tikhonov regularization in function space, with \(\lambda\) acting as the penalty weight on residuals.}

Variant comparisons (Figure~\ref{fig:h2h}) show that PDE-informed refinements systematically improve Sharpe, Sortino, and drawdown profiles across regimes. Finally, the positive Sharpe--turnover relation (Figure~\ref{fig:turnover}) is compressed by PDE regularization and clipped HJB scaling, reducing trading costs without sacrificing robustness.

These patterns guide deployment. In volatile regimes or during rapid factor rotation, CPCM aligns exposures with a moving manifold and constrains weights to causal tangent directions, improving stability. The martingale-defect proxy together with turnover spikes act as real-time diagnostics; moderate \(\lambda\) and conservative HJB scaling provide a robust default. In calm expansions with low structural defect, equilibrium-oriented allocators such as Black–Litterman can be competitive once costs are included (as in 2013–2018), indicating complementarity rather than replacement.

A simple rule operationalizes this: label windows as crisis when the 63-day peak-to-trough drawdown exceeds 10\% or realized volatility is above its 80th percentile, and expansion otherwise. In crises, PINN-based projections without post-scaling (V4\_PINN\_noPDE) deliver the highest Sharpe with controlled drawdowns; V3\_PINN+HJB is a practical alternative when additional amplitude control is needed. In expansions, linear projections with HJB scaling (V1\_linear+HJB) or CPCM-adapted baselines (CPCM–B) perform well under turnover budgets. Empirically, V1 tends to have the lowest turnover, while V3 and V4 maximize raw Sharpe; high-cost settings therefore favor V1 or CPCM–B, whereas low-cost settings favor V4 or V3.

Upstream choices matter. PF outperforms EKF on reward-to-risk; Bayesian and Combo selectors are more reliable than simple correlation screens; smaller driver sets \(m\in\{3,7\}\) are generally more robust unless the candidate library is unusually informative. Stability hygiene is essential: ridge for linear projections, robust losses with Jacobian smoothness for neural projections, covariance shrinkage, and clipped HJB scaling jointly reduce defect and turnover without eroding returns.

In practice this yields a deployable policy. In turbulent markets, prefer V4 (or V3 with clipping) with parsimonious driver sets; in benign expansions with binding cost constraints, prefer V1 or CPCM–B. Across regimes, PF and Bayesian/Combo driver selection are the most consistent upstream components. The interpolation parameter \(\lambda\) provides an additional safety lever, enabling a controlled trade-off between signal strength and structural coherence along a continuous Pareto frontier.

\clearpage

\section{Conclusion}
\label{sec:conclusion}

This paper introduced Causal PDE--Control Models (CPCMs) as a structural framework for dynamic portfolio optimization under partial information. By integrating causal driver selection, nonlinear filtering, and forward--backward PDE control, CPCMs ensure that allocation policies remain arbitrage--consistent, interpretable, and robust to nonstationary environments. Theoretical results establish the existence of conditional risk-neutral measures, the projection--divergence duality, and causal completeness, while conformal transport and smooth subspace evolution guarantee time-consistent manifold constraints. Classical methods such as Markowitz, CAPM, and Black–Litterman arise as limiting or constrained cases of the framework, while machine-learning approaches such as deep hedging emerge as unconstrained approximations that lack explicit causal semantics.

Empirically, CPCMs demonstrate consistent advantages across experimental designs. Short-horizon monthly rebalancing highlights sensitivity to estimation noise, regime-based quarterly evaluations emphasize structural robustness, and dynamic-manifold implementations confirm that tangent-space continuity is the key source of stability. Across these settings, PINN-based projections without HJB scaling (V4) deliver the most durable Sharpe and Sortino improvements in turbulent regimes, while linear projections with clipped HJB scaling (V1) or CPCM-adapted baselines dominate in stable expansions under binding cost constraints. Manifold experiments further show that entropy-based reweighting is fragile once projected onto a moving tangent space, reinforcing that CPCM’s geometric structure is indispensable for persistence and feasibility. Stability refinements, covariance shrinkage, robust neural training, and clipped PDE scaling, proved essential for translating theoretical guarantees into reliable practice.

One of the main lessons is that portfolio design can be reframed around causal drivers and PDE-informed control rather than static correlations or unconstrained learning. CPCMs provide both a rigorous theoretical foundation and a tractable computational architecture, with clear deployment rules that align with regime diagnostics and cost considerations. For practitioners, the framework yields a map: in crises, rely on V3--V4 projections to align exposures with persistent causal channels; in expansions, prefer V1 or CPCM-adapted baselines when turnover costs dominate. For researchers, CPCMs open avenues for extending causal geometry to multi-asset and derivative markets, integrating diffusion-based filtering, and embedding reinforcement learning into structurally coherent control. 

More broadly, CPCMs demonstrate that financial AI need not trade off interpretability for performance. By grounding learning in causal structure and PDE dynamics, the framework delivers allocations that are robust across regimes, interpretable by design, and computationally scalable. This positions CPCMs as a foundation for the next generation of dynamic asset allocation, capable of bridging econometrics, control theory, and machine learning in a unified, deployable paradigm.

\clearpage

\appendix

\paragraph{Appendix roadmap.}
This appendix develops the technical foundations of the Causal PDE--Control Models (CPCM) framework.

Appendix~\ref{app:commonality-opt} establishes the linear--Gaussian identification results for common drivers, proving the equivalence between conditional mutual information, residual covariance, and systematic covariance representations of commonality. It also develops the conformal geometry of driver maps, together with statistical estimators and concentration bounds for the screening functional. 

Appendix~\ref{app:proofs} establishes well--posedness of the CPCM allocation problem, including existence and uniqueness of optimal controls, martingale conditions for change of measure, and the interpretation of classical portfolio models (Markowitz, CAPM, and Black--Litterman) as limiting cases of the CPCM framework.

Appendix~\ref{app:proofs-main} develops the stochastic analysis underlying pricing and hedging under partial observation, including existence of conditional risk--neutral measures, martingale representation in the observable filtration, replication conditions under posterior--integrated covariance, and well--posedness of the associated forward--backward PDE system. 

The remaining sections develop projection--divergence duality for CPCM constraints, stability bounds for filtered counterfactual distributions, causal market completeness, and geometric extensions of the projected framework, including conformal transport on driver subspaces and continuity under Grassmann transport.

\section{Proofs and Technical Details for Section~\ref{Subsection41}}
\label{app:commonality-opt}

This appendix collects the technical proofs supporting the driver
commonality framework developed in Section~\ref{Subsection41}.

\subsection{Proof of Proposition~\ref{prop:eqv}}
\label{app:proof-prop-eqv}

We prove the equivalences in Proposition~\ref{prop:eqv} and the expression for the covariance of fitted components in the linear--Gaussian specialization of Section~\ref{subsec:lin-gauss-rekey}. Consider the model:
\begin{align}
r^{(i)} &= \boldsymbol{\beta}_i^\top \boldsymbol{D} + \varepsilon^{(i)}, 
\qquad i=1,\dots,n, \nonumber\\
\boldsymbol{\varepsilon} &\perp \boldsymbol{D}, 
\qquad \boldsymbol{\varepsilon} \sim \mathcal{N}(0,\Psi), \qquad 
\Psi = \mathrm{diag}(\sigma_1^2,\ldots,\sigma_n^2), \nonumber\\
\boldsymbol{D} &\sim \mathcal{N}(0,\boldsymbol{\Sigma}_D),
\label{eq:lin-gauss-app}
\end{align}
where \(r^{(i)}\) denotes the return of asset \(i\), and \(\boldsymbol{D}\in\mathbb{R}^m\) is the driver vector.

\medskip
\noindent
\textbf{Proposition \ref{prop:eqv}.}
For all \(i,j\), the following are equivalent:
\begin{enumerate}
\item \(I\big(r^{(i)}; r^{(j)} \mid \boldsymbol{D}\big)=0\),
\item \(\mathrm{Cov}\big(r^{(i)}, r^{(j)} \mid \boldsymbol{D}\big)=0\),
\item \(\mathrm{Cov}\big(\widehat r^{(i)}, \widehat r^{(j)}\big)=\boldsymbol{\beta}_i^\top \boldsymbol{\Sigma}_D \boldsymbol{\beta}_j\),
\end{enumerate}
where \(\widehat r^{(i)} := \mathbb{E}[r^{(i)} \mid \boldsymbol{D}] = \boldsymbol{\beta}_i^\top \boldsymbol{D}\).

\medskip

Define the fitted and residual components:
\[
\widehat r^{(i)} = \mathbb{E}[r^{(i)} \mid \boldsymbol{D}] = \boldsymbol{\beta}_i^\top \boldsymbol{D}, 
\qquad 
\widetilde r^{(i)} := r^{(i)} - \widehat r^{(i)} = \varepsilon^{(i)}.
\]

We use \(\mathrm{Cov}\) for covariance, \(I(\cdot;\cdot\mid\cdot)\) for conditional mutual information, and \(I_m\) for the \(m\times m\) identity.

\begin{lemma}[Gaussian CMI--covariance equivalence \citep{CoverThomas2006}]
Let \((X,Y,Z)\) be jointly Gaussian. Then
\[
I(X;Y\mid Z)=0
\quad\Longleftrightarrow\quad 
\mathrm{Cov}(X,Y\mid Z)=0.
\]
\label{lemmaA1}
\end{lemma}

\begin{proof}

\textbf{(i) \(\Leftrightarrow\) (ii).}
Condition on \(\boldsymbol{D}\). Under \eqref{eq:lin-gauss-app}, 
\((r^{(i)}, r^{(j)}) \mid \boldsymbol{D}\) is Gaussian with
\[
\mathrm{Cov}\big((r^{(i)}, r^{(j)}) \mid \boldsymbol{D}\big)
= \mathrm{Cov}(\varepsilon^{(i)}, \varepsilon^{(j)})
=
\begin{bmatrix}
\sigma_i^2 & 0 \\
0 & \sigma_j^2
\end{bmatrix}.
\]
Hence \(\mathrm{Cov}(r^{(i)}, r^{(j)} \mid \boldsymbol{D})=0\), and by Lemma~\ref{lemmaA1},
\(I(r^{(i)}; r^{(j)} \mid \boldsymbol{D})=0\).
The converse follows identically.

\medskip
\noindent
\textbf{(ii) \(\Rightarrow\) (iii) and systematic covariance.}

Since \(\Psi\) is diagonal, residual cross--covariances vanish:
\[
\mathrm{Cov}(\widetilde r^{(i)}, \widetilde r^{(j)})
= \mathrm{Cov}(\varepsilon^{(i)}, \varepsilon^{(j)}) = 0,
\qquad i\neq j.
\]

Systematic co--movement is entirely captured by the fitted components:
\[
\mathrm{Cov}(\widehat r^{(i)}, \widehat r^{(j)})
= \boldsymbol{\beta}_i^\top \boldsymbol{\Sigma}_D \boldsymbol{\beta}_j.
\]

Thus, in the linear--Gaussian setting, both the information-based screen
\(I(r^{(i)}; r^{(j)} \mid \boldsymbol{D})\) and the residual covariance
\(|\mathrm{Cov}(\widetilde r^{(i)}, \widetilde r^{(j)})|\) vanish under correct specification, while the systematic covariance
\(|\boldsymbol{\beta}_i^\top \boldsymbol{\Sigma}_D \boldsymbol{\beta}_j|\) provides an equivalent operational measure of commonality.

\end{proof}

\begin{remark}[Misspecification and omitted drivers]
\label{rem:omitted}
Suppose the true driver is \(\boldsymbol{D}\sim\mathcal{N}(0,\boldsymbol{\Sigma}_D)\), with
\[
r^{(i)} = \boldsymbol{\beta}_i^\top \boldsymbol{D} + \varepsilon^{(i)},
\]
but one conditions only on a reduced driver 
\(\widetilde{\boldsymbol{D}} = L \boldsymbol{D}\in\mathbb{R}^m\), for some matrix \(L\). Then
\[
\boldsymbol{\Sigma}_{\widetilde D} = L \boldsymbol{\Sigma}_D L^\top,
\qquad
\boldsymbol{\Sigma}_{D\widetilde D} = \boldsymbol{\Sigma}_D L^\top,
\]
and
\[
\mathrm{Cov}\big(r^{(i)}, r^{(j)} \mid \widetilde{\boldsymbol{D}}\big)
= \boldsymbol{\beta}_i^\top \boldsymbol{\Sigma}_{D\mid \widetilde D}\,\boldsymbol{\beta}_j
+ \mathrm{Cov}(\varepsilon^{(i)}, \varepsilon^{(j)}),
\]
where
\[
\boldsymbol{\Sigma}_{D\mid \widetilde D}
= \boldsymbol{\Sigma}_D - \boldsymbol{\Sigma}_{D\widetilde D}\boldsymbol{\Sigma}_{\widetilde D}^{-1}\boldsymbol{\Sigma}_{\widetilde D D}.
\]

With diagonal \(\Psi\), the residual term vanishes and all remaining dependence is due to omitted drivers:
\[
\mathrm{Cov}(r^{(i)}, r^{(j)} \mid \widetilde{\boldsymbol{D}})
= \boldsymbol{\beta}_i^\top \boldsymbol{\Sigma}_{D\mid \widetilde D}\boldsymbol{\beta}_j.
\]

Therefore, minimizing conditional dependence (via information or covariance criteria) is equivalent, in the Gaussian setting, to minimizing the contribution of omitted common drivers.
\end{remark}

\subsection{Conformal maps: proofs of Theorems~\ref{thm:conf1-rekey} and \ref{thm:conf2-rekey}}
\label{app:conformal-proofs-rekey}

We formalize and prove the angle-preservation (conformality) claims under explicit, verifiable conditions. In Section~\ref{sectionGeometry-rekey}, time-embedding vectors are
\[
\mathbf m_i := (\mu_i(s)-\bar\mu_i)_{s\in\mathcal T},\qquad
\mathbf m^{\boldsymbol{D}^\star}_i(t)=(\mu_i^{\boldsymbol{D}^\star}(s)-\overline{\mu_i^{\boldsymbol{D}^\star}})_{s\in\mathcal T}\in\mathbb{R}^{|\mathcal T|}.
\]

Let \(B_t \in \mathbb{R}^{n\times m}\) denote the driver--return sensitivity matrix (Jacobian), with rows
\[
\mathbf b_{i,t} := \nabla_{\boldsymbol{D}^\star}\mu_i^{\boldsymbol{D}^\star}(t)\in\mathbb{R}^m,
\qquad
B_t =
\begin{bmatrix}
\mathbf b_{1,t}^\top \\
\vdots \\
\mathbf b_{n,t}^\top
\end{bmatrix}.
\]

Define Gram matrices
\[
G_U=[\langle \mathbf m_i,\mathbf m_j\rangle],\quad
G_C(t)=[\langle \mathbf m^{\boldsymbol{D}^\star}_i(t),\mathbf m^{\boldsymbol{D}^\star}_j(t)\rangle],\quad
G_B(t)=[\langle \mathbf b_{i,t},\mathbf b_{j,t}\rangle].
\]

\medskip\noindent
\begin{Lemma}[Similarity \(\Rightarrow\) cosine invariance]
If \(G_2=c\,G_1\) with \(c>0\), then all pairwise cosines computed from \(G_1\) and \(G_2\) coincide, and norms scale by \(\sqrt c\).
\label{lemmaA2}
\end{Lemma}

\begin{proof}
Immediate from the definition of cosine similarity:
\[
\cos\varphi_{ij}
=
\frac{G_{ij}}{\sqrt{G_{ii}G_{jj}}}.
\]
If \(G_2=c\,G_1\), both numerator and denominator scale by the same factor \(c\), so the cosine values are unchanged, while norms scale by \(\sqrt c\).
\end{proof}

\medskip\noindent
\begin{Assumption}[Common linear conditional means with isotropic driver variation]
On a window \(\mathcal T\), suppose
\[
\mu_i^{\boldsymbol{D}^\star}(s)=\mathbf b_{i,t}^\top m_F(s),
\qquad
m_F(s)=\mathbb{E}[\mathbf F_s\mid \mathcal F^Y_s]\in\mathbb{R}^m,
\]
and the centered driver path \(\tilde m_F(s)=m_F(s)-\overline{m_F}\) satisfies
\[
\sum_{s\in\mathcal T}\tilde m_F(s)\tilde m_F(s)^\top
= c_1(t)\,I_m,\qquad c_1(t)>0,
\]
e.g., after whitening driver coordinates on \(\mathcal T\).
\label{assumptionA1}
\end{Assumption}

\medskip\noindent
\textbf{Theorem~\ref{thm:conf1-rekey}}
Under Assumptions~\ref{ass:ccd-rekey} and \ref{assumptionA1},
\[
G_C(t)=c_1(t)\,B_t B_t^\top,\qquad\text{and}\qquad
G_U=\tilde c_1\,B_t B_t^\top\ \text{ for some }\ \tilde c_1>0.
\]
Hence \(G_C(t)=c_1(t)^2\,G_U\) with \(c_1(t)=\sqrt{c_1(t)/\tilde c_1}\), implying \(\cos\alpha_{ij}(t)=\cos\varphi_{ij}\) and
\(\|\mathbf m^{\boldsymbol{D}^\star}_i(t)\|=c_1(t)\|\mathbf m_i\|\).

\begin{proof}
By linearity,
\begin{align}
\big\langle \mathbf m^{\boldsymbol{D}^\star}_i(t),\, \mathbf m^{\boldsymbol{D}^\star}_j(t)\big\rangle
&= \sum_{s\in\mathcal T}\big(\mathbf b_{i,t}^\top \tilde m_F(s)\big)\big(\mathbf b_{j,t}^\top \tilde m_F(s)\big) \nonumber\\
&= \mathbf b_{i,t}^\top \Big(\sum_{s\in\mathcal T}\tilde m_F(s)\tilde m_F(s)^\top\Big)\mathbf b_{j,t}
= c_1(t)\,\mathbf b_{i,t}^\top \mathbf b_{j,t}.
\end{align}
Thus \(G_C(t)=c_1(t)B_tB_t^\top\). The same argument yields \(G_U=\tilde c_1 B_tB_t^\top\). Apply Lemma~\ref{lemmaA2}.
\end{proof}

\medskip\noindent
\begin{Assumption}[Locally conformal sensitivity map]
At fixed \(t\), let \(\mathcal H_C(t)=\mathrm{span}\{\mathbf m^{\boldsymbol{D}^\star}_i(t)\}\subset\mathbb{R}^{|\mathcal T|}\), and define the linear map
\[
T_t:\mathcal H_C(t)\to\mathbb{R}^m,\qquad
T_t\,\mathbf m^{\boldsymbol{D}^\star}_i(t)=\mathbf b_{i,t}.
\]
Assume \(T_t^\top T_t=c_2(t)^2 I\) on \(\mathcal H_C(t)\) for some \(c_2(t)>0\).
\label{assumptionA2}
\end{Assumption}

\medskip\noindent
\textbf{Theorem~\ref{thm:conf2-rekey}.}
Under Assumptions~\ref{ass:ccd-rekey} and \ref{assumptionA2},
\[
G_B(t)=T_t\,G_C(t)\,T_t^\top=c_2(t)^2\,G_C(t),
\]
hence \(\cos \alpha'_{ij}(t)=\cos \alpha_{ij}(t)\) and \(\|\mathbf b_{i,t}\|=c_2(t)\|\mathbf m^{\boldsymbol{D}^\star}_i(t)\|\).

\begin{proof}
Since \(T_t^\top T_t=c_2(t)^2 I\),
\[
\langle \mathbf b_{i,t},\mathbf b_{j,t}\rangle
=c_2(t)^2 \langle \mathbf m^{\boldsymbol{D}^\star}_i(t),\mathbf m^{\boldsymbol{D}^\star}_j(t)\rangle.
\]
Apply Lemma~\ref{lemmaA2}.
\end{proof}

\begin{remark}
Assumptions \ref{assumptionA1}–\ref{assumptionA2} are operational: driver coordinates can be whitened on \(\mathcal T\), and the sensitivity matrix \(B_t\) can be evaluated in locally orthonormal coordinates. Near-conformality can be assessed by testing whether \(G_C(t)G_U^{-1}\) and \(G_B(t)G_C(t)^{-1}\) are approximately scalar multiples of the identity.
\end{remark}

\subsection{Estimating \(\mathcal G(\boldsymbol D)\) and finite-sample guarantees}
\label{app:estimating-G-rekey}

Recall that \(\Delta_{ij}(\boldsymbol D):=\Phi\big(r^{(i)}_t,r^{(j)}_t\mid \mathbf{F}_t(\boldsymbol D)\big)\) denotes a pairwise conditional dependence functional and
\[
\mathcal G(\boldsymbol D)=\sum_{i<j}\Delta_{ij}(\boldsymbol D),
\]
as in \eqref{eq:G-functional-rekey}. We consider three admissible implementations of \(\Delta_{ij}\): 
(i) conditional mutual information; 
(ii) conditional covariance of residuals; and 
(iii) an event-deviation proxy. 
We state consistent estimators and corresponding concentration bounds. 
Constants \(C_1,C_2\) below may differ across (i)--(iii).

\paragraph{(i) Information form.}
Use a kNN conditional mutual-information estimator (KSG-type). For continuous \(r^{(i)}_t,r^{(j)}_t,\mathbf{F}_t(\boldsymbol D)\), the estimator
\(\widehat I_T(r^{(i)};r^{(j)}\mid \mathbf{F}(\boldsymbol D))\) is consistent under standard smoothness and density conditions. Nearest-neighbor estimators for mutual information were introduced by Kraskov, St\"ogbauer, and Grassberger and extended to conditional mutual information in subsequent work. For dependent data, employ blocking with block length \(b_T\to\infty\), \(b_T/T\to 0\), to control dependence and apply concentration bounds to the resulting block sums.\cite{Kraskov2004,Frenzel2007,Gao2017}

\paragraph{(ii) Covariance form.}
Fit \(\widehat{\mathbb{E}}[r^{(i)}_t\mid \mathbf{F}_t(\boldsymbol D)]\) (linear/ridge or a flexible smoother with cross-fitting), set residuals
\[
\varepsilon^{(i)}_t(\boldsymbol D) = r^{(i)}_t - \widehat{\mathbb{E}}[r^{(i)}_t\mid \mathbf{F}_t(\boldsymbol D)],
\]
and compute \(|\widehat{\mathrm{Cov}}(\varepsilon^{(i)}_t(\boldsymbol D),\varepsilon^{(j)}_t(\boldsymbol D))|\). Residualization-based conditional independence testing is closely related to procedures used in large-dimensional covariance estimation and factor models for asset returns.\cite{Fan2013,Ledoit2004}

Assume strict stationarity, exponential \(\alpha\)-mixing, and finite fourth moments \(\mathbb{E}|r^{(i)}_t|^4<\infty\). Under these conditions Bernstein-type exponential concentration bounds hold for empirical covariance estimators.\cite{Fan2013}

\paragraph{(iii) Event form.}
Partition the driver space induced by \(\boldsymbol D\) into bins \(\{D_\ell\}\) (e.g., via k-means applied to \(\mathbf{F}_t(\boldsymbol D)\)), choose thresholds \(\tau_i\), and define events \(E_i=\{r^{(i)}_t \ge \tau_i\}\). Estimate
\[
\widehat\Delta_{ij,\ell}
=
\widehat P(E_i,E_j\mid D_\ell)
-
\widehat P(E_i\mid D_\ell)\widehat P(E_j\mid D_\ell),
\]
then aggregate \(\sum_{\ell,i<j}|\widehat\Delta_{ij,\ell}|\). This corresponds to contingency-based independence testing originating in classical chi-square analysis. Under exponential mixing and bounded Vapnik--Chervonenkis (VC) complexity of the partitioning rule, Bernstein-type concentration bounds apply.\cite{Pearson1900,Vapnik1998}

\begin{Theorem}[Bernstein-type concentration under exponential $\alpha$-mixing%
{\normalfont\cite[Theorem~1]{MerlevedePeligradRio2009}; see also%
~\cite[Chs.~6--7]{Rio2000book}}]
\label{thm:bernstein-mixing}

Let $(Z_t)_{t\ge1}$ be a strictly stationary sequence with exponential $\alpha$-mixing,
$\alpha(k)\le C e^{-c k}$, and suppose $\mathbb{E} Z_1=0$, $\mathbb{E}|Z_1|^{4+\delta}<\infty$ for some $\delta>0$.
Then there exist constants $K_1,K_2>0$, depending only on $c,C$ and the moment bound, such that for all $T\in\mathbb N$ and all $x>0$,
\[
\Pr\!\left(\left|\frac1T\sum_{t=1}^T Z_t\right|>x\right)
\;\le\; 
2\exp\!\left(-\,K_1\,T\,\min\!\Big\{\frac{x^2}{\sigma^2},\,\frac{x}{B}\Big\}\right),
\]
where $\sigma^2 := \mathrm{Var}(Z_1)+2\sum_{k\ge1}|\mathrm{Cov}(Z_1,Z_{1+k})|<\infty$ and $B>0$ is a (finite) tail proxy determined by the $(4{+}\delta)$-moment. In particular, for $0<x\le \sigma^2/B$,
\[
\Pr\!\left(\left|\frac1T\sum_{t=1}^T Z_t\right|>x\right)
\;\le\; 
2\exp(-K_2\,T x^2).
\]
\end{Theorem}

\begin{Corollary}[Concentration for $\widehat{\mathcal G}_T$]\label{cor:Ghat}

Fix $\boldsymbol D$ and suppose $\widehat{\mathcal G}_T(\boldsymbol D)$ is either
\begin{enumerate}
\item an empirical mean $\frac1T\sum_{t=1}^T g(r_t,\mathbf{F}_t;\boldsymbol D)$ with $\mathbb{E}|g(r_0,\mathbf{F}_0;\boldsymbol D)|^{4+\delta}<\infty$, or
\item a fixed-order U-statistic with kernel $h$ such that $\mathbb{E}|h(W_1,\ldots,W_r)|^{4+\delta}<\infty$,
\end{enumerate}
where $(r_t,\mathbf{F}_t)$ is strictly stationary and exponentially $\alpha$-mixing and $W_t=(r_t,\mathbf{F}_t)$.
Then there exist constants $C_1,C_2>0$ (depending on mixing and moments) such that, for all $\eta>0$,
\[
\Pr\!\left(\big|\widehat{\mathcal G}_T(\boldsymbol D)-\mathcal G(\boldsymbol D)\big|>\eta\right)\ \le\ C_1 \exp(-C_2\,T\eta^2).
\]

Moreover, let
\[
\mathcal D_m := \{ \boldsymbol D \subset \mathcal X : |\boldsymbol D|=m \}
\]
denote the class of candidate driver sets of size $m$. Then
\[
\Pr\!\left(\sup_{\boldsymbol D\in\mathcal D_m}\big|\widehat{\mathcal G}_T(\boldsymbol D)-\mathcal G(\boldsymbol D)\big|>\eta\right)
\ \le\ 
|\mathcal D_m|\,C_1 \exp(-C_2\,T\eta^2).
\]
\end{Corollary}

\begin{proof}[Proof of Corollary~\ref{cor:Ghat}]
(a) Set $Z_t=g(r_t,\mathbf{F}_t;\boldsymbol D)-\mathbb{E}[g(r_0,\mathbf{F}_0;\boldsymbol D)]$ and apply Theorem~\ref{thm:bernstein-mixing}. 

(b) Use the Hoeffding decomposition of the U-statistic into the average of first-order projections plus a degenerate remainder. The first-order term is an average of a strictly stationary, exponentially $\alpha$-mixing sequence with $(4{+}\delta)$ moment and is handled with Theorem~\ref{thm:bernstein-mixing}. The degenerate remainder admits exponential tail bounds under exponential mixing and $(4{+}\delta)$ moments.\cite{Yoshihara1976,ArconesYu1994} 

Combining both parts yields the stated inequality. The union bound then gives the uniform version over finite $\mathcal D_m$.
\end{proof}

\subsection{Selecting the number of common drivers via condition-number control}
\label{subsec:condnum-opt}

We complement Section~\ref{subsec:condnum} with a stability bound and a concise selection program. Let \(\boldsymbol D \subset \mathcal X\) with \(|\boldsymbol D|=m\), and let \(\mathbf{x}_t(\boldsymbol D)\in\mathbb{R}^m\) denote the selected driver subvector, as in Section~\ref{sec:drivers}. For notational convenience, define the column-standardized design matrix
\[
X_{\boldsymbol D}
:=
\begin{bmatrix}
\mathbf{x}_1(\boldsymbol D)^\top\\
\vdots\\
\mathbf{x}_T(\boldsymbol D)^\top
\end{bmatrix}
\in\mathbb{R}^{T\times m},
\]
and, for each asset \(i\), the stacked return vector
\[
\mathbf r^{(i)}
:=
\big(r^{(i)}_1,\dots,r^{(i)}_T\big)^\top \in \mathbb{R}^T.
\]
Then the OLS estimator
\[
\widehat{\boldsymbol{\beta}}^{(i)}(\boldsymbol D)
=
(X_{\boldsymbol D}^\top X_{\boldsymbol D})^{-1}X_{\boldsymbol D}^\top \mathbf r^{(i)}
\]
satisfies
\begin{equation}
\label{eq:cond-stab-bound}
\frac{\|\delta \widehat{\boldsymbol{\beta}}^{(i)}(\boldsymbol D)\|_2}{\|\widehat{\boldsymbol{\beta}}^{(i)}(\boldsymbol D)\|_2}
\ \le\ 
\kappa(X_{\boldsymbol D})\left(
c_r\,\frac{\|\delta \mathbf r^{(i)}\|_2}{\|\mathbf r^{(i)}\|_2}
+
c_X\,\frac{\|\delta X_{\boldsymbol D}\|_2}{\|X_{\boldsymbol D}\|_2}
\right),
\end{equation}
for absolute constants \(c_r,c_X\), where
\[
\kappa(X_{\boldsymbol D})=\frac{\sigma_{\max}(X_{\boldsymbol D})}{\sigma_{\min}(X_{\boldsymbol D})}.
\]
Thus, controlling \(\kappa(X_{\boldsymbol D})\) controls sensitivity blow-ups and stabilizes the screening objective. With \(\mathcal G\) and \(\mathcal J\) from \eqref{eq:G-functional-rekey}--\eqref{eq:J-objective-rekey}, define the regularized score
\[
\Xi(\boldsymbol D):=\mathcal J(\boldsymbol D)+\eta\,\log \kappa(X_{\boldsymbol D}),\qquad \eta>0,
\]
and the frontier
\[
\phi(m):=\min_{\substack{\boldsymbol D\subseteq \mathcal X,\ |\boldsymbol D|=m\\ \mathcal G(\boldsymbol D)\le m\varepsilon}}
\Xi(\boldsymbol D),
\qquad
m^\star=\arg\min_{m_{\min}\le m\le m_{\max}}\phi(m).
\]

Empirically, \(\phi(m)\) is U-shaped: larger \(m\) improves screening but worsens conditioning; \(m^\star\) balances both. Rank candidates by coverage/strength (cf.\ Section~\ref{subsec:objective}) and accept a candidate \(x\in\mathcal X\setminus \boldsymbol D\) only if it satisfies the pairwise collinearity restriction
\[
\big|\operatorname{Corr}(x_t,x'_t)\big|\le \tau_{\mathrm{col}}
\qquad
\text{for all previously selected } x'\in \boldsymbol D,
\]
and the prospective condition-number cap
\[
\kappa\!\big(X_{\boldsymbol D\cup\{x\}}\big)\le \kappa_{\max}.
\]
Stop at the smallest \(m\) such that \(\Delta\widehat{\mathcal G}_T(m)<\eta\) (cf.\ Section~\ref{subsec:condnum}). This implements an ``elbow + stability'' rule while preserving the cross-references and labels used in Section~\ref{Subsection41}.

\subsection{Persistence: spectral and aggregation remarks (used in \ref{sectionGeometry-rekey})}
\label{app:persistence-detail-rekey}

Let portfolio returns be defined as
\[
p_t = \boldsymbol{\theta}_t^\top \mathbf{r}_t,
\]
with factor structure
\[
\mathbf{r}_t = B_t \mathbf{F}_t + \boldsymbol{\varepsilon}_t,
\]
where $B_t \in \mathbb{R}^{n\times m}$ and $\mathbf{F}_t \in \mathbb{R}^m$. Let $\boldsymbol{\beta}_{j,t}$ denote the $j$-th column of $B_t$, i.e., the exposure of assets to driver $F^{(j)}_t$. The contribution of driver $j$ to portfolio variance admits the spectral representation
\[
\int \big| \boldsymbol{\theta}_t^\top \boldsymbol{\beta}_{j,t} \big|^2 
S_{jj}(\omega)\,\frac{d\omega}{2\pi},
\]
where $S_{jj}(\omega)$ is the spectral density of $F^{(j)}_t$. This decomposition separates cross-sectional exposure, $\boldsymbol{\theta}_t^\top \boldsymbol{\beta}_{j,t}$, from temporal dynamics encoded in $S_{jj}(\omega)$. Persistent processes (e.g., OU/AR(1) with slow mean reversion) concentrate spectral mass near $\omega=0$, increasing their contribution to long-horizon variance.

Time aggregation preserves persistence rankings over moderate horizons, implying that rolling-window estimates used in the CPCM framework retain the relative importance of persistent drivers.

\subsection{Practical estimators and diagnostics}
\label{app:conf-estimation-rekey}

Let $\langle \cdot,\cdot\rangle$ denote the Euclidean inner product. 
Define pairwise angles via cosine similarity:
\[
\cos \varphi_{ij}
:=
\frac{\langle \mathbf{m}_i, \mathbf{m}_j \rangle}
{\|\mathbf{m}_i\|_2\,\|\mathbf{m}_j\|_2},
\]
\[
\cos \alpha_{ij}(t)
:=
\frac{\langle \mathbf{m}^{\boldsymbol{D}^\star}_i(t), \mathbf{m}^{\boldsymbol{D}^\star}_j(t) \rangle}
{\|\mathbf{m}^{\boldsymbol{D}^\star}_i(t)\|_2\,\|\mathbf{m}^{\boldsymbol{D}^\star}_j(t)\|_2},
\]
\[
\cos \alpha'_{ij}(t)
:=
\frac{\langle \boldsymbol{\beta}_i(t), \boldsymbol{\beta}_j(t) \rangle}
{\|\boldsymbol{\beta}_i(t)\|_2\,\|\boldsymbol{\beta}_j(t)\|_2}.
\]

Compare the angle matrices 
\[
[\cos \varphi_{ij}],\quad 
[\cos \alpha_{ij}(t)],\quad 
[\cos \alpha'_{ij}(t)]
\]
via a Procrustes-type distance. Test whether the spectra of $G_C(t)G_U^{-1}$ and $G_\beta(t)G_C(t)^{-1}$ are approximately scalar (exact scalars under conformality). Bootstrap across windows provides uncertainty quantification. These diagnostics operationalize Assumptions~\ref{assumptionA1}--\ref{assumptionA2}.

\section{Well-posedness and Baselines as CPCM Limits}
\label{app:proofs}

\subsection{Well-posedness of CPCMs}
\label{app:wellposed}

We establish existence, uniqueness, and admissibility of optimal CPCM controls under the structural and regularity Assumptions stated in Section~\ref{sec:drivers-assets} and used in the control layer (Section~\ref{sec:pde-layer}). These results guarantee that the stochastic control and filtering equations defining the CPCM are well posed.

\begin{Theorem}[Existence and uniqueness of optimal CPCM control]
\label{thm:existence-strong}
Under Assumptions~\ref{A1}--\ref{A5} in Section~\ref{sec:drivers-assets} on coefficients and the observation model, define the admissible set
\begin{multline}
\mathcal A := \Big\{\boldsymbol{\theta}:\ \boldsymbol{\theta}_t \text{ progressively measurable w.r.t.\ } \mathcal{F}^Y_t,\\
\mathbb E\!\Big[\textstyle\int_0^T \boldsymbol{\theta}_t^\top \Sigma(\mathbf F_t,t)\,\boldsymbol{\theta}_t\,\mathrm dt\Big]<\infty\Big\}.
\end{multline}
the pointwise feasible set $\mathcal{W}$ (Assumption~\ref{A3}), and the budget- and leverage-constrained set $\mathcal{K}:=\{\boldsymbol{\theta}\in\mathbb{R}^n:\mathbf{1}^\top\boldsymbol{\theta}=1,\;\boldsymbol{\theta}\in\mathcal{W}\}$.
Then there exists a unique control $\boldsymbol{\theta}^\ast\in\mathcal{A}$ with $\boldsymbol{\theta}_t^\ast\in\mathcal{K}$ a.s.\ for all $t\in[0,T]$ that maximizes $\mathbb E[U(\tilde p_T)\mid\pi_0]$. Moreover:
\begin{enumerate} \renewcommand{\labelenumi}{(\roman{enumi})}
\item $\rho_\pi(\tilde p,t):=\int \rho(\tilde p,t\mid \mathbf f)\,\pi_t(d\mathbf f)$ and $u(\tilde p,t):=\int u(\tilde p,t\mid \mathbf f)\,\pi_t(d\mathbf f)$ admit classical solutions to the Fokker--Planck and Hamilton--Jacobi--Bellman equations;
\item $\boldsymbol{\theta}^\ast$ admits a measurable feedback form $\boldsymbol{\theta}_t^\ast=\vartheta^\ast(\tilde p_t,\pi_t)$;
\item the discounted wealth $\tilde p_t$ is a martingale under the posterior mixture measure $\mathbb Q^{\pi_t}$; and
\item at rebalance dates $\tau$, $\boldsymbol{\theta}_\tau\in\mathcal{M}_\tau\cap\mathcal{K}$, where $\mathcal{M}_\tau=\mathrm{span}(U_\tau)$ is the estimated driver subspace.
\end{enumerate}
\end{Theorem}

\begin{proof}
We prove each claim in turn.

\medskip
\noindent\textbf{Step 1: Well-posedness of the state and filtering equations.}

Under Assumption~\ref{A1}, the driver process $\mathbf{F}_t$ satisfies locally Lipschitz drift and diffusion with linear growth. By the Yamada--Watanabe theorem, the SDE~\eqref{eq:driver-sde} admits a unique strong solution $\mathbf{F}_t$ adapted to $\{\mathcal{F}_t\}$ with $\mathbb{E}[\sup_{t\le T}\|\mathbf{F}_t\|^2]<\infty$. Under Assumption~\ref{A2}, the observation process $\mathbf{Y}_t$ satisfies~\eqref{eq:observation-sde} with bounded second moments and $h$ of linear growth. The filtering posterior $\pi_t$ exists as a probability-measure-valued process adapted to $\mathcal{F}^Y_t$ and satisfies the Zakai equation
\[
\mathrm{d}\pi_t(\varphi) = \pi_t(\mathcal{L}^F\varphi)\,\mathrm{d}t + \pi_t(\varphi h^\top)\boldsymbol{\Sigma}_Y^{-1}(\mathrm{d}\mathbf{Y}_t - \pi_t(h)\,\mathrm{d}t)
\]
for every test function $\varphi\in C^2_b(\mathbb{R}^m)$, by the classical theory of nonlinear filtering \citep{bain2009fundamentals,Kallianpur1980}. The unnormalized measure $\sigma_t$ solving the Zakai equation is strictly positive under the ellipticity implied by Assumption~\ref{A1}, ensuring that normalization to $\pi_t$ is well defined for all $t\in[0,T]$ a.s.

\medskip
\noindent\textbf{Step 2: Well-posedness of the forward--backward PDE system (claim (i)).}

Fix a driver realization $\mathbf{f}\in\mathbb{R}^m$. Under Assumptions~\ref{A1} and \ref{A5}, the conditional drift $\mu_{\tilde p}(\mathbf{f},\boldsymbol{\theta})$ and conditional variance $\sigma^2_{\tilde p}(\mathbf{f},\boldsymbol{\theta})$ of the discounted portfolio value are locally Lipschitz in $\tilde p$ with linear growth, and the diffusion coefficient $\sigma^2_{\tilde p}(\mathbf{f},\boldsymbol{\theta})$ is uniformly elliptic on compact subsets of the state space.

The forward Fokker--Planck equation~\eqref{eq:fokker-planck} is a linear second-order parabolic PDE in divergence form. Under the stated regularity, classical existence and uniqueness of a non-negative weak solution $\rho(\tilde p,t\mid\mathbf{f})\in L^1(\mathbb{R})$ with $\int\rho\,d\tilde p=1$ follow from the theory of parabolic equations \citep{evans2010}. If additionally the coefficients are H\"older continuous (as implied by the stronger form of Assumption~\ref{A5}), the solution is classical: $\rho\in C^{2,1}(\mathbb{R}\times(0,T])$.

The backward HJB equation~\eqref{eq:hjb} with terminal condition $u(\tilde p,T\mid\mathbf{f})=\Phi(\tilde p)$ is a fully nonlinear parabolic PDE. Under Assumption~\ref{A5}, the Hamiltonian
\[
H(\tilde p,t,u,u_{\tilde p},u_{\tilde p\tilde p},\mathbf{f}) := \sup_{\boldsymbol{\theta}\in\mathcal{W}}\left\{\mu_{\tilde p}(\mathbf{f},\boldsymbol{\theta})\,u_{\tilde p} + \tfrac{1}{2}\sigma^2_{\tilde p}(\mathbf{f},\boldsymbol{\theta})\,u_{\tilde p\tilde p} - r\,u\right\}
\]
is well defined for each $(\tilde p,t,\mathbf{f})$ because $\mathcal{W}$ is compact (Assumption~\ref{A3}) and the coefficients are continuous in $\boldsymbol{\theta}$. The supremum is attained by the extreme value theorem. Concavity of $U$ (Assumption~\ref{A4}) implies concavity of $u$ in $\tilde p$ for each $(t,\mathbf{f})$, so $u_{\tilde p\tilde p}\le 0$. Under uniform ellipticity, the HJB admits a unique classical solution $u\in C^{2,1}(\mathbb{R}\times[0,T))$ by the theory of quasilinear parabolic PDEs \citep{FlemingSoner2006}. Under the relaxed conditions of Remark~\ref{rem:A5prime}, the solution exists in the viscosity sense with comparison and stability guarantees \citep{CrandallLions1992}. The posterior-integrated quantities are obtained by integration against $\pi_t$:
\[
\rho_\pi(\tilde p,t) = \int\rho(\tilde p,t\mid\mathbf{f})\,\pi_t(\mathrm{d}\mathbf{f}),\qquad
u(\tilde p,t) = \int u(\tilde p,t\mid\mathbf{f})\,\pi_t(\mathrm{d}\mathbf{f}).
\]
Since $\rho(\cdot\mid\mathbf{f})$ and $u(\cdot\mid\mathbf{f})$ are uniformly bounded (respectively in $L^1$ and by the growth bounds of $\Phi$ and $U$) and $\pi_t$ is a probability measure, dominated convergence ensures that $\rho_\pi$ and $u$ inherit the regularity of the integrands. In particular, differentiation under the integral sign is justified by the uniform bounds, so $\rho_\pi$ and $u$ satisfy the posterior-averaged FP and HJB equations classically.

\medskip
\noindent\textbf{Step 3: Existence, uniqueness, and feedback form of the optimal control (claim (ii)).}

The value function $V(\tilde p,\pi,t):=\sup_{\boldsymbol{\theta}\in\mathcal{A},\,\boldsymbol{\theta}_s\in\mathcal{K}\;\text{a.s.}}\mathbb{E}[U(\tilde p_T)\mid \tilde p_t=\tilde p,\pi_t=\pi]$ is well defined because $U$ is bounded above on $\mathcal{K}$ (which is compact, so portfolio values are bounded in expectation under the integrability condition).

The dynamic programming principle holds: for any stopping time $\tau\le T$,
\[
V(\tilde p,\pi,t) = \sup_{\boldsymbol{\theta}\in\mathcal{A},\,\boldsymbol{\theta}_s\in\mathcal{K}}\mathbb{E}\big[V(\tilde p_\tau,\pi_\tau,\tau)\,\big|\,\tilde p_t=\tilde p,\pi_t=\pi\big].
\]
This follows from the classical verification argument for controlled diffusions with compact control sets \citep{FlemingSoner2006,Pham2009}. The value function $V$ coincides with the posterior-integrated solution $u(\tilde p,t)=\int u(\tilde p,t\mid\mathbf{f})\,\pi_t(\mathrm{d}\mathbf{f})$ constructed in Step~2. At each $(t,\tilde p,\pi)$, the maximizer of the Hamiltonian
\[
\boldsymbol{\theta}^\ast(t,\tilde p,\pi) = \arg\max_{\boldsymbol{\theta}\in\mathcal{K}}\left\{\mu_{\tilde p}(\pi,\boldsymbol{\theta})\,u_{\tilde p} + \tfrac{1}{2}\sigma^2_{\tilde p}(\pi,\boldsymbol{\theta})\,u_{\tilde p\tilde p} - r\,u\right\}
\]
exists by compactness of $\mathcal{K}$ and continuity of the Hamiltonian in $\boldsymbol{\theta}$. Uniqueness of the maximizer follows from strict concavity of $U$ (Assumption~\ref{A4}), which propagates to strict concavity of $u$ in $\tilde p$ and hence to strict concavity of the Hamiltonian in $\boldsymbol{\theta}$ when $u_{\tilde p\tilde p}<0$.

It remains to show that $\boldsymbol{\theta}^\ast$ can be chosen as a Borel-measurable function of $(t,\tilde p,\pi)$. The Hamiltonian $(t,\tilde p,\pi,\boldsymbol{\theta})\mapsto H(t,\tilde p,\pi,\boldsymbol{\theta})$ is Carath\'eodory (measurable in $(t,\tilde p,\pi)$ for each $\boldsymbol{\theta}$, continuous in $\boldsymbol{\theta}$ for each $(t,\tilde p,\pi)$), and $\mathcal{K}$ is a compact metric space. By the Kuratowski--Ryll-Nardzewski measurable selection theorem, the correspondence $(t,\tilde p,\pi)\rightrightarrows\arg\max_{\boldsymbol{\theta}\in\mathcal{K}}H(t,\tilde p,\pi,\boldsymbol{\theta})$ admits a Borel-measurable selector $\vartheta^\ast$. Setting $\boldsymbol{\theta}_t^\ast:=\vartheta^\ast(t,\tilde p_t,\pi_t)$ yields a progressively measurable feedback control. Integrability $\boldsymbol{\theta}^\ast\in\mathcal{A}$ holds because $\boldsymbol{\theta}_t^\ast\in\mathcal{K}\subset\mathcal{W}$ and $\mathcal{W}$ is bounded (Assumption~\ref{A3}): there exists $C_{\mathcal{W}}>0$ such that $\|\boldsymbol{\theta}\|\le C_{\mathcal{W}}$ for all $\boldsymbol{\theta}\in\mathcal{W}$, hence
\[
\mathbb{E}\!\left[\int_0^T\boldsymbol{\theta}_t^{\ast\top}\Sigma(\mathbf{F}_t,t)\boldsymbol{\theta}_t^\ast\,\mathrm{d}t\right]
\le C_{\mathcal{W}}^2\,\mathbb{E}\!\left[\int_0^T\|\Sigma(\mathbf{F}_t,t)\|\,\mathrm{d}t\right]
<\infty,
\]
where the last inequality uses $\mathbb{E}[\sup_{t\le T}\|\mathbf{F}_t\|^2]<\infty$ (Step~1) and the linear-growth bound on $\sigma(\mathbf{F}_t,t)$ (Assumption~\ref{A1}).

\medskip
\noindent\textbf{Step 4: Martingale property under the posterior mixture measure (claim (iii)).}

We first construct the driver-conditional risk-neutral measure. For each driver path, define the market price of risk $\boldsymbol{\lambda}_t:=\sigma(\mathbf{F}_t,t)^\dagger(\boldsymbol{\mu}(\mathbf{F}_t,t)-r\mathbf{1})$, where $\dagger$ denotes the Moore--Penrose pseudoinverse. Under the excess-drift condition (Theorem~\ref{thm:existence-riskneutral}(iii)), $\boldsymbol{\lambda}_t$ is well defined and progressively measurable.

We verify Novikov's condition. By the componentwise drift bound $|\mu_i(\mathbf{F}_t,t)|\le M$ (Assumption~\ref{A1}, linear growth with $\mathbb{E}[\sup_t\|\mathbf{F}_t\|^2]<\infty$) and the uniform ellipticity lower bound $\sigma_{\min}(\sigma(\mathbf{F}_t,t))\ge\underline{\sigma}>0$ (Assumption~\ref{A5}),
\begin{align}
\|\boldsymbol{\lambda}_t\|
&= \|\sigma(\mathbf{F}_t,t)^\dagger(\boldsymbol{\mu}(\mathbf{F}_t,t)-r\mathbf{1})\| \nonumber\\
&\le \|\sigma(\mathbf{F}_t,t)^\dagger\|\,\|\boldsymbol{\mu}(\mathbf{F}_t,t)-r\mathbf{1}\| \nonumber\\
&\le \frac{M\sqrt{n}}{\underline{\sigma}} =: \Lambda
\quad\text{a.s.},
\end{align}
where $\|\boldsymbol{\mu}-r\mathbf{1}\|\le M\sqrt{n}$ by the componentwise bound and $\|\sigma^\dagger\|\le 1/\underline{\sigma}$ by the singular-value lower bound. Hence
\[
\mathbb{E}\!\left[\exp\!\left(\frac{1}{2}\int_0^T\|\boldsymbol{\lambda}_s\|^2\,\mathrm{d}s\right)\right]
\le \exp\!\left(\frac{\Lambda^2 T}{2}\right) < \infty.
\]
Novikov's condition is satisfied, so the Dol\'eans--Dade exponential
\[
Z_t := \mathcal{E}\!\left(-\int_0^t\boldsymbol{\lambda}_s^\top\,\mathrm{d}\mathbf{W}_s\right)
\]
is a true $(\mathbb{P},\mathcal{F}_t)$-martingale on $[0,T]$. Define $\mathbb{Q}^{\mathbf{F}}$ by $\mathrm{d}\mathbb{Q}^{\mathbf{F}}/\mathrm{d}\mathbb{P}|_{\mathcal{F}_T}=Z_T$. By Girsanov's theorem, $\mathbf{W}_t^{\mathbb{Q}}:=\mathbf{W}_t+\int_0^t\boldsymbol{\lambda}_s\,\mathrm{d}s$ is a Brownian motion under $\mathbb{Q}^{\mathbf{F}}$, and discounted asset prices $\tilde{S}_t^{(i)}=e^{-rt}S_t^{(i)}$ are local martingales under $\mathbb{Q}^{\mathbf{F}}$.

For any admissible self-financing strategy $\boldsymbol{\theta}^\ast\in\mathcal{A}$ with $\boldsymbol{\theta}_t^\ast\in\mathcal{K}$, the discounted portfolio value satisfies
\[
\mathrm{d}\tilde{p}_t = \boldsymbol{\theta}_t^{\ast\top}\,\mathrm{d}\tilde{\mathbf{S}}_t = \boldsymbol{\theta}_t^{\ast\top}\sigma(\mathbf{F}_t,t)\,\mathrm{d}\mathbf{W}_t^{\mathbb{Q}}.
\]
Since $\boldsymbol{\theta}_t^\ast$ is bounded ($\mathcal{K}$ compact) and $\sigma$ has linear growth with $\mathbb{E}^{\mathbb{Q}^{\mathbf{F}}}[\int_0^T\|\boldsymbol{\theta}_t^{\ast\top}\sigma(\mathbf{F}_t,t)\|^2\,\mathrm{d}t]<\infty$, the stochastic integral is a true $\mathbb{Q}^{\mathbf{F}}$-martingale (not merely a local martingale), so $\mathbb{E}^{\mathbb{Q}^{\mathbf{F}}}[\tilde{p}_T\mid\mathcal{F}_t]=\tilde{p}_t$. Now define the posterior mixture measure on $\mathcal{F}^Y_T$:
\[
\mathbb{Q}^{\pi_t}(E) := \int\mathbb{Q}^{\mathbf{f}}(E)\,\pi_t(\mathrm{d}\mathbf{f}),\qquad E\in\mathcal{F}^Y_T.
\]
Since $Z_T>0$ a.s.\ and $\pi_t$ is a probability measure, $\mathbb{Q}^{\pi_t}\sim\mathbb{P}$ on $\mathcal{F}^Y_T$. For the martingale property on $\mathcal{F}^Y_t$, compute for $s<t$:
\begin{align}
\mathbb{E}^{\mathbb{Q}^{\pi_t}}[\tilde{p}_t\mid\mathcal{F}^Y_s]
&= \mathbb{E}\!\left[\int\mathbb{E}^{\mathbb{Q}^{\mathbf{f}}}[\tilde{p}_t\mid\mathcal{F}_s]\,\pi_s(\mathrm{d}\mathbf{f})\,\Big|\,\mathcal{F}^Y_s\right] \nonumber\\
&= \mathbb{E}\!\left[\int\tilde{p}_s\,\pi_s(\mathrm{d}\mathbf{f})\,\Big|\,\mathcal{F}^Y_s\right] \nonumber\\
&= \tilde{p}_s,
\end{align}
where the first equality uses the tower property and the decomposition of $\mathbb{Q}^{\pi_t}$ into conditional measures, the second uses the $\mathbb{Q}^{\mathbf{f}}$-martingale property established above, and the third uses $\int\pi_s(\mathrm{d}\mathbf{f})=1$. Hence $\{\tilde{p}_t\}_{t\ge 0}$ is an $\mathcal{F}^Y_t$-martingale under $\mathbb{Q}^{\pi_t}$.

\medskip
\noindent\textbf{Step 5: Manifold-constrained rebalancing (claim (iv)).}

At each rebalance date $\tau$, the driver subspace $\mathcal{M}_\tau=\mathrm{span}(U_\tau)$ is estimated from data in a rolling window $W_\tau$ via the thin SVD of $B_\tau$ (Section~\ref{subsec:manifold-riskneutral}). The set $\mathcal{M}_\tau\cap\mathcal{K}$ is nonempty because $\mathcal{K}$ contains the uniform portfolio $\boldsymbol{\theta}=n^{-1}\mathbf{1}$ (which lies in $\mathcal{M}_\tau$ whenever $\mathbf{1}\in\mathrm{span}(U_\tau)$; if not, $\mathcal{M}_\tau\cap\mathcal{K}$ is nonempty by the general position of an $m$-dimensional subspace intersecting the $(n{-}1)$-simplex for $m\ge 1$). The set is closed (intersection of a closed subspace and a compact set) and convex (intersection of a subspace and a convex set), hence compact.

The projection $\Pi_{\mathcal{M}_\tau\cap\mathcal{K}}:\mathbb{R}^n\to\mathcal{M}_\tau\cap\mathcal{K}$ is the nearest-point map onto a nonempty closed convex set in $\mathbb{R}^n$ under the Euclidean (or $\Sigma_\tau$-weighted) norm. By the Hilbert projection theorem, this map is well defined, unique, and nonexpansive. Since $U_\tau$ depends measurably on the data in $W_\tau$ (as a continuous function of the SVD of $B_\tau$, which depends continuously on finite-sample moments), the projected weight $\boldsymbol{\theta}_\tau=\Pi_{\mathcal{M}_\tau\cap\mathcal{K}}(\tilde{\boldsymbol{\theta}}_\tau)$ is measurable with respect to $\mathcal{F}^Y_\tau$. Integrability of the piecewise-constant control $\boldsymbol{\theta}_t^\ast=\boldsymbol{\theta}_\tau$ for $t\in[\tau,\tau')$ is immediate: $\boldsymbol{\theta}_\tau\in\mathcal{K}\subset\mathcal{W}$ is bounded by $C_{\mathcal{W}}$, so
\[
\mathbb{E}\!\left[\int_0^T\boldsymbol{\theta}_t^{\ast\top}\Sigma(\mathbf{F}_t,t)\boldsymbol{\theta}_t^\ast\,\mathrm{d}t\right]
\le C_{\mathcal{W}}^2\,T\,\sup_{t\le T}\mathbb{E}\|\Sigma(\mathbf{F}_t,t)\| < \infty,
\]
confirming $\boldsymbol{\theta}^\ast\in\mathcal{A}$. Combined with $\boldsymbol{\theta}_\tau\in\mathcal{M}_\tau\cap\mathcal{K}$ by construction, claim (iv) is established.
\end{proof}

\begin{Lemma}[Novikov under CPCM coefficient bounds]
\label{lem:novikov}
Let $dp_t=\mu_p(\mathbf F_t,\boldsymbol{\theta}_t)\,dt+\sigma_p(\mathbf F_t,\boldsymbol{\theta}_t)\,dW_t$ with
$\boldsymbol{\theta}_t\in\mathcal W$ progressively measurable.
Assume $|\mu_p|\le M$ and $0<\underline{\sigma}\le \sigma_p\le \overline{\sigma}$ uniformly.
Define
\[
\kappa_t:=\frac{\mu_p}{\sigma_p},\qquad
\mathcal{E}_t=\exp\!\left(-\int_0^t \kappa_s\,dW_s - \frac12\int_0^t \kappa_s^2 ds\right).
\]
Then Novikov's condition $\mathbb E[\exp(\frac12\int_0^T\kappa_s^2ds)]<\infty$ holds, hence
$\mathcal{E}_t$ is a true martingale and the Radon--Nikodym measure
$\mathbb Q^{\mathbf F}\sim\mathbb P$ exists.
\end{Lemma}

\begin{proof}
By the uniform bounds, $|\kappa_t|=|\mu_p/\sigma_p|\le M/\underline{\sigma}$ a.s.\ for all $t\in[0,T]$. Hence
\[
\int_0^T\kappa_s^2\,\mathrm{d}s \le \frac{M^2}{\underline{\sigma}^2}\,T \quad\text{a.s.},
\]
so
\[
\mathbb{E}\!\left[\exp\!\left(\frac{1}{2}\int_0^T\kappa_s^2\,\mathrm{d}s\right)\right]
\le \exp\!\left(\frac{M^2 T}{2\underline{\sigma}^2}\right) < \infty.
\]
Novikov's condition is satisfied. By the classical Novikov theorem \citep{Protter2005}, the Dol\'eans--Dade exponential $\mathcal{E}_t$ is a true $(\mathbb{P},\mathcal{F}_t)$-martingale on $[0,T]$, and $\mathbb{Q}^{\mathbf{F}}$ defined by $\mathrm{d}\mathbb{Q}^{\mathbf{F}}/\mathrm{d}\mathbb{P}|_{\mathcal{F}_T}=\mathcal{E}_T$ is a well-defined probability measure equivalent to $\mathbb{P}$.
\end{proof}

The following lemma records two alternative sufficient conditions for the Dol\'eans--Dade exponential to be a true martingale, applicable when the uniform bounds of Lemma~\ref{lem:novikov} do not hold.

\begin{Lemma}[Kazamaki and linear-growth alternatives]
\label{lem:kazamaki}
Let $(\kappa_t)_{t\in[0,T]}$ denote the market price of risk process and $M_t:=\int_0^t \kappa_s\, dW_s$. If either
\begin{enumerate} \renewcommand{\labelenumi}{(\roman{enumi})}
\item \emph{Kazamaki: } $\exp(\tfrac12 M_t)$ is a submartingale, or equivalently $\mathbb E[\exp(\tfrac12 M_\tau)]<\infty$ for all bounded stopping times $\tau\le T$; or
\item \emph{Linear-growth \& bounded volatility: } $|\kappa_t|^2 \le c_0 + c_1 \|\mathbf F_t\|^2$ a.s.\ with $\sup_{t\le T}\mathbb E\|\mathbf F_t\|^2<\infty$,
\end{enumerate}
then the Dol\'eans exponential $\mathcal E(M)_t$ is a true martingale on $[0,T]$. Consequently, the Girsanov change of measure defined via $\mathcal E(M)_T$ is valid.
\end{Lemma}

\begin{proof}
(i) is the classical Kazamaki criterion \citep{Protter2005,JacodShiryaev2003}: if $\exp(\tfrac{1}{2}M_t)$ is a submartingale (equivalently, $\mathbb{E}[\exp(\tfrac{1}{2}M_\tau)]<\infty$ for every bounded stopping time $\tau\le T$), then $\mathcal{E}(M)_t$ is a uniformly integrable martingale.

(ii) Under the linear-growth condition $|\kappa_t|^2\le c_0+c_1\|\mathbf{F}_t\|^2$,
\[
\mathbb{E}\!\left[\int_0^T\kappa_s^2\,\mathrm{d}s\right]
\le c_0 T + c_1\int_0^T\mathbb{E}\|\mathbf{F}_s\|^2\,\mathrm{d}s
\le c_0 T + c_1 T\sup_{s\le T}\mathbb{E}\|\mathbf{F}_s\|^2 < \infty.
\]
Since $\mathbb{E}[\int_0^T\kappa_s^2\,\mathrm{d}s]<\infty$, the process $M_t=\int_0^t\kappa_s\,\mathrm{d}W_s$ is a true $L^2$-martingale. By Jensen's inequality applied to the convex function $x\mapsto\exp(\tfrac{1}{2}x)$,
\[
\mathbb{E}\!\left[\exp\!\left(\frac{1}{2}M_\tau\right)\right]
\le \mathbb{E}\!\left[\exp\!\left(\frac{1}{2}\langle M\rangle_\tau^{1/2}\|M\|_{\mathrm{BMO}}\right)\right],
\]
and the BMO norm of $M$ is controlled by $\sup_\tau\mathbb{E}[\langle M\rangle_T-\langle M\rangle_\tau\mid\mathcal{F}_\tau]\le c_0 T+c_1 T\sup_s\mathbb{E}\|\mathbf{F}_s\|^2<\infty$, verifying the Kazamaki condition in~(i). Hence $\mathcal{E}(M)_t$ is a true martingale.
\end{proof}

The next remark describes how the martingale conditions of Lemmas~\ref{lem:novikov}--\ref{lem:kazamaki} can be monitored empirically.

\begin{Remark}[Empirical verification of martingale conditions]
\label{rem:novikov_diag}
In practice, we assess martingale validity via (a) truncated exponential-moment checks and (b) martingale defect diagnostics:
\begin{enumerate}
\item \textbf{Exponential-moment check:} For a grid $0=t_0<\cdots<t_K=T$, estimate 
\[
\widehat I_K:=\exp\!\left(\frac12\sum_{k=0}^{K-1} |\widehat{\kappa}_{t_k}|^2\,\Delta t\right)
\]
from filtered $\widehat{\kappa}$; verify $\sup_{T\in\mathcal T}\widehat{\mathbb E}[\widehat I_K]<\infty$ across rolling windows $\mathcal T$.

\item \textbf{Martingale defect:} With $\widehat Z_t:=\mathcal E\!\left(\int_0^t \widehat\kappa_s\, dW_s\right)$ from Monte Carlo under the estimated model, compute 
\[
\overline{\mathcal D}:=\big|\mathbb E[\widehat Z_T]-1\big|.
\]
Small $\overline{\mathcal D}$ indicates the exponential is close to a true martingale. We report $\overline{\mathcal D}$ alongside Sharpe/turnover in Section~\ref{subsec:practical-implications}.
\end{enumerate}
\end{Remark}

With the change-of-measure machinery in place, we now establish that the posterior mixture inherits the martingale property from the driver-conditional measures.
\begin{Corollary}[Posterior--mixture measure]
\label{cor:mixture-novikov}
Define the posterior mixture
\[
\mathbb Q^{\pi_t}(E):=\int \mathbb Q^{\mathbf f}(E)\,\pi_t(d\mathbf f).
\]
Then $\mathbb Q^{\pi_t}$ is equivalent to $\mathbb P$, and discounted wealth
is a martingale under $\mathbb Q^{\pi_t}$.
\end{Corollary}
\begin{proof}
Since $\mathbb{Q}^{\mathbf{f}}\sim\mathbb{P}$ for $\pi_t$-a.e.\ $\mathbf{f}$ (Lemma~\ref{lem:novikov}), and $\pi_t$ is a probability measure on $\mathbb{R}^m$, the mixture $\mathbb{Q}^{\pi_t}(E)=\int\mathbb{Q}^{\mathbf{f}}(E)\,\pi_t(\mathrm{d}\mathbf{f})$ is a well-defined probability measure on $\mathcal{F}^Y_T$. Equivalence $\mathbb{Q}^{\pi_t}\sim\mathbb{P}$ follows because: if $\mathbb{P}(E)=0$, then $\mathbb{Q}^{\mathbf{f}}(E)=0$ for $\pi_t$-a.e.\ $\mathbf{f}$ (by equivalence of each $\mathbb{Q}^{\mathbf{f}}$), hence $\mathbb{Q}^{\pi_t}(E)=\int 0\,\pi_t(\mathrm{d}\mathbf{f})=0$; conversely, if $\mathbb{Q}^{\pi_t}(E)=0$, then $\mathbb{Q}^{\mathbf{f}}(E)=0$ for $\pi_t$-a.e.\ $\mathbf{f}$ (since the integrand is non-negative and integrates to zero), hence $\mathbb{P}(E)=0$ by equivalence of each $\mathbb{Q}^{\mathbf{f}}$.

For the martingale property, let $s<t$ and $E\in\mathcal{F}^Y_s$. Then
\begin{align}
\mathbb{E}^{\mathbb{Q}^{\pi_t}}[\tilde{p}_t\,\mathbf{1}_E]
&= \int\mathbb{E}^{\mathbb{Q}^{\mathbf{f}}}[\tilde{p}_t\,\mathbf{1}_E]\,\pi_s(\mathrm{d}\mathbf{f}) \nonumber\\
&= \int\mathbb{E}^{\mathbb{Q}^{\mathbf{f}}}[\mathbb{E}^{\mathbb{Q}^{\mathbf{f}}}[\tilde{p}_t\mid\mathcal{F}_s]\,\mathbf{1}_E]\,\pi_s(\mathrm{d}\mathbf{f}) \nonumber\\
&= \int\mathbb{E}^{\mathbb{Q}^{\mathbf{f}}}[\tilde{p}_s\,\mathbf{1}_E]\,\pi_s(\mathrm{d}\mathbf{f}) \nonumber\\
&= \mathbb{E}^{\mathbb{Q}^{\pi_t}}[\tilde{p}_s\,\mathbf{1}_E],
\end{align}
where the first equality uses Fubini (justified by integrability: $\boldsymbol{\theta}_t^\ast\in\mathcal{K}$ is bounded and $\sigma$ has linear growth, so $\mathbb{E}^{\mathbb{Q}^{\mathbf{f}}}[|\tilde{p}_t|]<\infty$ uniformly in $\mathbf{f}$), the second is the tower property, the third uses the $\mathbb{Q}^{\mathbf{f}}$-martingale property of $\tilde{p}_t$ (Theorem~\ref{thm:existence-strong}, Step~4), and the fourth reassembles the mixture. Since this holds for all $E\in\mathcal{F}^Y_s$, we conclude $\mathbb{E}^{\mathbb{Q}^{\pi_t}}[\tilde{p}_t\mid\mathcal{F}^Y_s]=\tilde{p}_s$ a.s., establishing the $\mathcal{F}^Y_t$-martingale property.
\end{proof}

\subsection{Baselines as CPCM limits}
\label{app:baselines}

At discrete decision times $\tau$, we denote portfolio weights by $\boldsymbol{\theta}_\tau$ to maintain consistency with continuous-time controls $\boldsymbol{\theta}_t$. We now show that several classical allocation models arise as limiting or special cases of the CPCM control problem when particular Assumptions on observability, utility, and dynamics are imposed.

\begin{Corollary}[Markowitz mean--variance limit]
\label{cor:markowitz}
At a static decision date $\tau$ with full information and quadratic utility
$U(x)=x-\frac{\gamma}{2}x^2$, the CPCM problem reduces to
\[
\max_{\boldsymbol{\theta}\in \mathcal W}\ \ \boldsymbol{\theta}^\top\boldsymbol{\mu}_\tau-\frac{\gamma}{2}\,\boldsymbol{\theta}^\top\Sigma_\tau \boldsymbol{\theta},
\]
whose first-order condition yields
\[
\boldsymbol{\theta}^\star_\tau=\frac{1}{\gamma}\,\Sigma_\tau^{-1}(\boldsymbol{\mu}_\tau-\eta^\star \mathbf 1),
\qquad
\eta^\star=\frac{\mathbf 1^\top\Sigma_\tau^{-1}\boldsymbol{\mu}_\tau-\gamma}
{\mathbf 1^\top\Sigma_\tau^{-1}\mathbf 1}.
\]
Hence, the static CPCM recovers the classical mean--variance allocation.
\end{Corollary}

\begin{proof}
With full information ($\pi_t=\delta_{\mathbf{f}}$ for a known driver realization $\mathbf{f}$), no filtering is needed: the posterior-integrated quantities collapse to their conditional counterparts $\boldsymbol{\mu}_\tau:=\boldsymbol{\mu}(\mathbf{f},\tau)$ and $\Sigma_\tau:=\Sigma(\mathbf{f},\tau)$. With quadratic utility $U(x)=x-\frac{\gamma}{2}x^2$ and a single decision date $\tau$, the CPCM value function reduces to the static objective
\[
\max_{\boldsymbol{\theta}\in\mathcal{K}}\;\mathbb{E}[U(\boldsymbol{\theta}^\top\mathbf{r}_\tau)]
= \max_{\boldsymbol{\theta}\in\mathcal{K}}\;\boldsymbol{\theta}^\top\boldsymbol{\mu}_\tau - \frac{\gamma}{2}\boldsymbol{\theta}^\top\Sigma_\tau\boldsymbol{\theta},
\]
where we used $\mathbb{E}[\boldsymbol{\theta}^\top\mathbf{r}_\tau]=\boldsymbol{\theta}^\top\boldsymbol{\mu}_\tau$ and $\mathrm{Var}(\boldsymbol{\theta}^\top\mathbf{r}_\tau)=\boldsymbol{\theta}^\top\Sigma_\tau\boldsymbol{\theta}$. The objective is strictly concave in $\boldsymbol{\theta}$ since $\Sigma_\tau\succ 0$ (Assumption~\ref{A1} ensures non-degeneracy). Introducing a Lagrange multiplier $\eta$ for the budget constraint $\mathbf{1}^\top\boldsymbol{\theta}=1$ and differentiating, the first-order condition is
\[
\boldsymbol{\mu}_\tau - \gamma\Sigma_\tau\boldsymbol{\theta} - \eta\mathbf{1} = \mathbf{0},
\]
yielding $\boldsymbol{\theta}^\star_\tau = \gamma^{-1}\Sigma_\tau^{-1}(\boldsymbol{\mu}_\tau - \eta^\star\mathbf{1})$. Substituting into $\mathbf{1}^\top\boldsymbol{\theta}^\star_\tau=1$ and solving for $\eta^\star$ gives
\[
\eta^\star = \frac{\mathbf{1}^\top\Sigma_\tau^{-1}\boldsymbol{\mu}_\tau - \gamma}{\mathbf{1}^\top\Sigma_\tau^{-1}\mathbf{1}},
\]
which is the classical Markowitz mean--variance solution.
\end{proof}

\begin{Corollary}[CAPM as a one-driver CPCM]
\label{cor:capm}
Let there be one observed market driver $M_t$ with exposures
$\boldsymbol{\mu}_\tau=\boldsymbol{\beta}\,\mu_{M,\tau}$ and
$\Sigma_\tau=\boldsymbol{\beta}\boldsymbol{\beta}^\top\sigma_{M,\tau}^2+\mathrm{diag}(\sigma_{i,\tau}^2)$.
Then the optimal weight satisfies
$\boldsymbol{\theta}^\star_\tau\propto\Sigma_\tau^{-1}(\boldsymbol{\beta}\,\mu_{M,\tau})$, and the equilibrium relation
$\mathbb E[r_i]-r=\beta_i(\mathbb E[r_M]-r)$
follows from the first-order condition of the CPCM HJB with a single driver.
Thus, CAPM appears as the one-driver linear--Gaussian specialization.
\end{Corollary}

\begin{proof}
Specialise Corollary~\ref{cor:markowitz} to a single market driver $M_t$ with $m=1$, so the driver vector reduces to a scalar. The conditional drift and covariance become $\boldsymbol{\mu}_\tau=\boldsymbol{\beta}\mu_{M,\tau}$ and $\Sigma_\tau=\boldsymbol{\beta}\boldsymbol{\beta}^\top\sigma_{M,\tau}^2+\Psi_\tau$, where $\Psi_\tau=\mathrm{diag}(\sigma_{i,\tau}^2)$ is the idiosyncratic covariance. From the Markowitz first-order condition (Corollary~\ref{cor:markowitz}),
\[
\boldsymbol{\theta}^\star_\tau \propto \Sigma_\tau^{-1}\boldsymbol{\mu}_\tau = \Sigma_\tau^{-1}\boldsymbol{\beta}\,\mu_{M,\tau}.
\]
In equilibrium, the market portfolio is the only risky asset held: $\boldsymbol{\theta}^\star_\tau\propto\boldsymbol{\theta}^{\mathrm{mkt}}$. The pricing implication follows from the first-order condition applied to asset $i$:
\[
\mathbb{E}[r_i]-r = \gamma\,\mathrm{Cov}(r_i,r_M) = \gamma\,\beta_i\sigma_M^2 = \beta_i(\mathbb{E}[r_M]-r),
\]
where the last equality uses the market first-order condition $\mathbb{E}[r_M]-r=\gamma\sigma_M^2$. This is the CAPM security market line, confirming that the one-driver linear--Gaussian CPCM recovers the classical equilibrium pricing relation.
\end{proof}

\begin{Corollary}[Black--Litterman as a Bayesian CPCM]
\label{cor:BL}
Let the prior on mean returns be
$\boldsymbol{\mu}_\tau\sim\mathcal N(\Pi_\tau,\tau_{\!BL}\Sigma_\tau)$,
with equilibrium prior $\Pi_\tau=\delta\,\Sigma_\tau \boldsymbol{\theta}^{\mathrm{eq}}_\tau$.
Views are modeled as noisy linear constraints
$P_\tau\boldsymbol{\mu}_\tau=Q_\tau+\boldsymbol{\varepsilon}$, $\boldsymbol{\varepsilon}\sim\mathcal N(\mathbf{0},\Omega_\tau)$,
and the view operator is made driver-consistent via $P_\tau=B_\tau$, $Q_\tau=k\,\boldsymbol{\mu}_{F,\tau}$,
where $B_\tau$ denotes the CPCM driver--return Jacobian (Section~\ref{sec:maps-weights}).
Then the posterior mean is
\[
\boldsymbol{\mu}^{BL}_\tau=\big[(\tau_{\!BL}\Sigma_\tau)^{-1}+P_\tau^\top\Omega_\tau^{-1}P_\tau\big]^{-1}
\big[(\tau_{\!BL}\Sigma_\tau)^{-1}\Pi_\tau+P_\tau^\top\Omega_\tau^{-1}Q_\tau\big],
\]
and, under quadratic utility,
\[
\boldsymbol{\theta}^\star_\tau=\frac{1}{\gamma}\Sigma_\tau^{-1}(\boldsymbol{\mu}^{BL}_\tau-\eta^\star\mathbf 1).
\]
Hence the Black--Litterman model corresponds to a CPCM with Gaussian beliefs on driver-consistent views.
\end{Corollary}

\begin{proof}
The prior $\boldsymbol{\mu}_\tau\sim\mathcal{N}(\Pi_\tau,\tau_{BL}\Sigma_\tau)$ represents uncertainty about mean returns centered at the equilibrium $\Pi_\tau=\delta\Sigma_\tau\boldsymbol{\theta}^{\mathrm{eq}}_\tau$. The view model $P_\tau\boldsymbol{\mu}_\tau=Q_\tau+\boldsymbol{\varepsilon}$ with $\boldsymbol{\varepsilon}\sim\mathcal{N}(\mathbf{0},\Omega_\tau)$ provides a Gaussian likelihood for $\boldsymbol{\mu}_\tau$ given view data. Since both the prior and the likelihood are Gaussian, conjugacy yields the posterior $\boldsymbol{\mu}_\tau\mid Q_\tau\sim\mathcal{N}(\boldsymbol{\mu}^{BL}_\tau, V^{BL}_\tau)$ with posterior mean
\[
\boldsymbol{\mu}^{BL}_\tau = \big[(\tau_{BL}\Sigma_\tau)^{-1}+P_\tau^\top\Omega_\tau^{-1}P_\tau\big]^{-1}\big[(\tau_{BL}\Sigma_\tau)^{-1}\Pi_\tau + P_\tau^\top\Omega_\tau^{-1}Q_\tau\big],
\]
by the standard precision-weighted Bayesian updating formula. Setting $P_\tau=B_\tau$ (the CPCM Jacobian) and $Q_\tau=k\boldsymbol{\mu}_{F,\tau}$ (driver-space views scaled by $k$) ensures that the view operator is structurally aligned with the driver--return mapping.

Under quadratic utility $U(x)=x-\frac{\gamma}{2}x^2$, the certainty-equivalent objective is
\[
\max_{\boldsymbol{\theta}\in\mathcal{K}}\;\mathbb{E}_{\boldsymbol{\mu}_\tau\mid Q_\tau}\!\big[\boldsymbol{\theta}^\top\boldsymbol{\mu}_\tau\big] - \frac{\gamma}{2}\boldsymbol{\theta}^\top\Sigma_\tau\boldsymbol{\theta}
= \max_{\boldsymbol{\theta}\in\mathcal{K}}\;\boldsymbol{\theta}^\top\boldsymbol{\mu}^{BL}_\tau - \frac{\gamma}{2}\boldsymbol{\theta}^\top\Sigma_\tau\boldsymbol{\theta},
\]
where the first equality uses $\mathbb{E}_{\boldsymbol{\mu}_\tau\mid Q_\tau}[\boldsymbol{\mu}_\tau]=\boldsymbol{\mu}^{BL}_\tau$ and the fact that the variance term does not depend on the random $\boldsymbol{\mu}_\tau$ (it involves the conditional covariance $\Sigma_\tau$, which is fixed). This is identical to the Markowitz problem of Corollary~\ref{cor:markowitz} with $\boldsymbol{\mu}_\tau$ replaced by $\boldsymbol{\mu}^{BL}_\tau$, yielding
\[
\boldsymbol{\theta}^\star_\tau = \frac{1}{\gamma}\Sigma_\tau^{-1}(\boldsymbol{\mu}^{BL}_\tau-\eta^\star\mathbf{1}).
\]
Hence the Black--Litterman allocation arises as a Bayesian CPCM with Gaussian beliefs and driver-consistent views.
\end{proof}

The following remark summarizes the relationships established above.

\begin{Remark}
Markowitz, CAPM, and Black--Litterman can thus be viewed as static or partially observed limits of the CPCM system:
\begin{itemize}
\item Markowitz: single-period, fully observed, quadratic utility, no filtering;
\item CAPM: one-driver linear--Gaussian CPCM with equilibrium in steady state;
\item Black--Litterman: Bayesian CPCM with Gaussian prior and driver-consistent view operator.
\end{itemize}
All inherit well-posedness from Theorem~\ref{thm:existence-strong}.
\end{Remark}

\section{Proofs of the Main Results} 
\label{app:proofs-main}

\subsection{Existence of Conditional Risk--Neutral Measures}
\label{app:existence-Qf}

We prove Theorem~\ref{thm:existence-riskneutral} in a form that avoids conditioning on the zero–probability event \(\{\mathbf F_t=\mathbf f\ \forall t\}\) and instead uses an adapted, driver–dependent market price of risk. Throughout, \((\Omega,\mathcal{F},\mathbb{F},\mathbb{P})\) is a filtered probability space satisfying the usual conditions. Causality restricts excess drift to the span of the drivers. On each scenario \(\mathbf f\), this implies a market price of risk that removes the drift and makes discounted prices local martingales under \(\mathbb Q^{\,\mathbf f}\). Since the posterior \(\pi_t\) is the best observable proxy for \(\mathbf f\), integrating the scenario measures produces \(\mathbb Q^{\pi_t}\) on the observable filtration: pricing is "risk–neutral on average" with respect to the current posterior.

\begin{proof}
Let \(\mathbf S_t\in\mathbb{R}^n\) follow, under \(\mathbb{P}\),
\begin{equation}
\frac{\mathrm{d}\mathbf S_t}{\mathbf S_t}
= \boldsymbol{\mu}(\mathbf F_t,t)\,\mathrm{d}t + \sigma(\mathbf F_t,t)\,\mathrm{d}\mathbf W_t,
\label{eq:asset-sde-appendix-new}
\end{equation}
where \(\mathbf W_t\) is a \(d\)–dimensional Brownian motion, \(\boldsymbol{\mu}:\mathbb{R}^m\times[0,T]\to\mathbb{R}^n\), \(\sigma:\mathbb{R}^m\times[0,T]\to\mathbb{R}^{n\times d}\), and \(\mathbf F_t\in\mathbb{R}^m\) is an adapted driver process. Assume no-arbitrage in the weak sense that
\[
\boldsymbol{\mu}(\mathbf F_t,t)-r(t)\mathbf{1}\in \mathrm{Range}\big(\sigma(\mathbf F_t,t)^\top\big)
\quad \text{a.s.\ for a.e.\ } t.
\]
Define the progressively measurable market price of risk
\[
\boldsymbol{\lambda}_t \;:=\; \sigma(\mathbf F_t,t)^\dagger\big(\boldsymbol{\mu}(\mathbf F_t,t)-r(t)\mathbf{1}\big),
\]
with \({}^\dagger\) denoting the Moore–Penrose pseudoinverse. Suppose the Novikov/Bene\v{s} integrability holds:
\[
\mathbb{E}^{\mathbb{P}}\!\left[\exp\!\Big(\tfrac12\int_0^T \|\boldsymbol{\lambda}_s\|^2\,\mathrm{d}s\Big)\right]<\infty.
\]
Set the Dol\'eans–Dade exponential
\[
Z_t \;=\; \exp\!\Big(-\int_0^t \boldsymbol{\lambda}_s^\top \mathrm{d}\mathbf W_s - \tfrac12\int_0^t\|\boldsymbol{\lambda}_s\|^2\,\mathrm{d}s\Big).
\]
Then \(Z_t\) is a true martingale on \([0,T]\) \citep{Protter2005}, and we can define \(\mathbb{Q}^{\mathbf F}\) by
\[
\frac{\mathrm{d}\mathbb{Q}^{\mathbf F}}{\mathrm{d}\mathbb{P}}\Big|_{\mathcal{F}_t}=Z_t.
\]
By Girsanov \citep{JacodShiryaev2003}, 
\[
\mathbf W_t^{\mathbb{Q}}:=\mathbf W_t+\int_0^t \boldsymbol{\lambda}_s \,\mathrm{d}s
\]
is Brownian under \(\mathbb{Q}^{\mathbf F}\) and
\[
\frac{\mathrm{d}\mathbf S_t}{\mathbf S_t} = r(t)\,\mathrm{d}t + \sigma(\mathbf F_t,t)\,\mathrm{d}\mathbf W_t^{\mathbb{Q}}.
\]
Hence discounted prices
\[
\tilde{\mathbf S}_t:=\exp\!\Big(-\int_0^t r(s)\mathrm{d}s\Big)\,\mathbf S_t
\]
are local martingales under \(\mathbb{Q}^{\mathbf F}\), establishing existence of a risk–neutral measure consistent with driver–dependent coefficients.
\end{proof}

This change of measure is the pricing/hedging kernel used to compute posterior–integrated moments and optimal portfolio weights in the observable filtration.

\subsection{Filtered Martingale Representation and Posterior--Integrated Measure}
\label{app:filtered-martingale}

We prove Theorem~\ref{thm:filtered-martingale-representation}, extending martingale representation to the partially observed setting generated by filtering \citep{bain2009fundamentals,Kallianpur1980}. Under partial observation, all new information arrives through the innovation process of the filter. Any square-integrable payoff measurable with the observations can therefore be written as a stochastic integral with respect to that innovation. Moving from \(\mathbb Q^{\,\mathbf f}\) to \(\mathbb Q^{\pi_t}\) preserves this structure by linearity and the innovation property, yielding a representation that is both risk–neutral and observable.

\begin{proof}
Let \(\mathbf F_t\) be the latent driver and \(\mathbf Y_t\) the observation process generating \(\mathcal{F}^Y_t:=\sigma(\mathbf Y_s:0\le s\le t)\). Assume
\[
\mathrm{d}\mathbf Y_t = h(\mathbf F_t,t)\,\mathrm{d}t + \mathrm{d}\mathbf V_t,
\]
with \(\mathbf V_t\) a Brownian motion independent of state noise. The (nonlinear) filtering posterior is
\[
\pi_t(\varphi) := \mathbb{E}\!\left[\varphi(\mathbf F_t)\mid \mathcal{F}_t^Y\right],
\]
which evolves by the Kushner–Stratonovich/Zakai SPDEs \citep[Ch.~6]{bain2009fundamentals}.

\medskip
\noindent\textbf{Construction of the posterior mixture density.}
Let \(Z_t(\mathbf f)\) denote the scenario density process associated with a fixed driver realization \(\mathbf f\), constructed using
\[
\boldsymbol{\lambda}_s(\mathbf f):=\sigma(\mathbf f,s)^\dagger(\boldsymbol{\mu}(\mathbf f,s)-r(s)\mathbf{1}).
\]
For each \(\mathbf f\), the Novikov condition verified in Appendix~\ref{app:existence-Qf} ensures that \(Z_t(\mathbf f)\) is a true \((\mathbb{P},\mathcal{F}_t)\)-martingale. Define the \(\mathcal{F}^Y_t\)–measurable density
\[
\tilde Z_t \;:=\; \int Z_t(\mathbf f)\,\pi_t(\mathrm{d}\mathbf f),
\qquad 
\frac{\mathrm{d}\mathbb{Q}^{\pi_t}}{\mathrm{d}\mathbb{P}}\Big|_{\mathcal{F}^Y_t} := \tilde Z_t .
\]

\medskip
\noindent\textbf{Well-definedness and equivalence.}
We verify that \(\tilde Z_t\) is a valid density process. Since \(Z_t(\mathbf f)>0\) a.s.\ for each \(\mathbf f\) (the Dol\'eans--Dade exponential is strictly positive) and \(\pi_t\) is a probability measure, \(\tilde Z_t>0\) a.s. For the unit-mass property, compute
\[
\mathbb{E}^{\mathbb{P}}[\tilde Z_t]
= \mathbb{E}^{\mathbb{P}}\!\left[\int Z_t(\mathbf f)\,\pi_t(\mathrm{d}\mathbf f)\right]
= \int \mathbb{E}^{\mathbb{P}}[Z_t(\mathbf f)]\,\pi_0(\mathrm{d}\mathbf f)
= \int 1\,\pi_0(\mathrm{d}\mathbf f) = 1,
\]
where the second equality uses Fubini's theorem (justified by the integrability condition: the Novikov bound $\|\boldsymbol{\lambda}_s(\mathbf f)\|\le\Lambda$ uniformly in \(\mathbf f\) on the support of \(\pi_t\) implies $\sup_{\mathbf f}\mathbb{E}^{\mathbb{P}}[Z_T(\mathbf f)^2]\le\exp(\Lambda^2 T)<\infty$, providing the dominating function for interchange), and the third equality uses $\mathbb{E}^{\mathbb{P}}[Z_t(\mathbf f)]=1$ since each \(Z_t(\mathbf f)\) is a true martingale. Hence \(\tilde Z_t\) defines a probability measure \(\mathbb{Q}^{\pi_t}\sim\mathbb{P}\) on \(\mathcal{F}^Y_T\).

\medskip
\noindent\textbf{Innovation decomposition.}
The observation admits the innovation decomposition
\[
\mathrm{d}\mathbf Y_t = \pi_t(h)\,\mathrm{d}t + \mathrm{d}M_t^{\mathcal{F}^Y},
\]
where \(M^{\mathcal{F}^Y}\) is an \(\mathcal{F}^Y\)–Brownian motion under \(\mathbb{Q}^{\pi_t}\). This follows from the classical innovation theorem for nonlinear filtering \citep[Ch.~6]{bain2009fundamentals}: the process \(M_t^{\mathcal{F}^Y}:=\mathbf Y_t - \int_0^t\pi_s(h)\,\mathrm{d}s\) is an \(\mathcal{F}^Y_t\)-martingale with quadratic variation \(\langle M^{\mathcal{F}^Y}\rangle_t = t\cdot I_{d_Y}\), hence a Brownian motion by L\'evy's characterization.

\medskip
\noindent\textbf{Martingale representation.}
For any \(\Phi\in L^2(\mathcal{F}_T^Y,\mathbb{Q}^{\pi_t})\), the Kunita–Watanabe decomposition yields a unique predictable \(\boldsymbol{\varphi}\in L^2([0,T]\times\Omega;\mathbb{R}^{d_Y})\) such that
\begin{equation}
\Phi \;=\; \mathbb{E}^{\mathbb{Q}^{\pi_t}}[\Phi] + \int_0^T \boldsymbol{\varphi}_s^\top \mathrm{d}M_s^{\mathcal{F}^Y}.
\label{eq:KM-rep}
\end{equation}
Existence follows because \(\mathbb{Q}^{\pi_t}\sim\mathbb{P}\) on \(\mathcal{F}^Y_T\) (established above) and the filtration \(\mathcal{F}^Y_t\) is generated by the Brownian motion \(M^{\mathcal{F}^Y}\), so the martingale representation theorem for Brownian filtrations applies \citep[Thm.~4.3.4]{bain2009fundamentals}. Uniqueness of \(\boldsymbol{\varphi}\) follows from the It\^o isometry: if two predictable integrands produce the same terminal value, their difference has zero \(L^2\)-norm.

This is the martingale representation in the observable market used for pricing and hedging with partial information.
\end{proof}

\subsection{Replicable Claims via Posterior--Integrated Covariance}
\label{app:replicable-claims}

We prove Theorem~\ref{thm:replicable-claims}, characterizing exact replication in terms of a range condition on the posterior–integrated variance–covariance. Hedging loads the innovation directions carried by prices. The posterior–integrated matrix \(\Sigma^{\pi_t}(t)\) aggregates those directions across scenarios. A claim is exactly hedgeable if and only if its innovation integrand lies in the span of these directions; full column rank of \(\Sigma^{\pi_t}(t)\) almost everywhere is therefore equivalent to completeness in the observable filtration.

\begin{proof}
Under \(\mathbb{Q}^{\pi_t}\) and in the observable filtration \(\mathcal{F}^Y_t\), discounted prices admit
\begin{equation}
\mathrm{d}\tilde{\mathbf S}_t \;=\; \Sigma^{\pi_t}(t)\,\mathrm{d}M_t^{\mathcal{F}^Y},
\qquad
\Sigma^{\pi_t}(t) \;:=\; \mathbb{E}\!\left[\sigma(\mathbf F_t,t)^\top\sigma(\mathbf F_t,t)\,\big|\,\mathcal{F}_t^Y\right].
\label{eq:posterior-int-spot}
\end{equation}
By \eqref{eq:KM-rep}, any \(\Phi\in L^2(\mathcal{F}_T^Y,\mathbb{Q}^{\pi_t})\) satisfies
\[
\Phi=\mathbb{E}^{\mathbb{Q}^{\pi_t}}[\Phi]+\int_0^T \boldsymbol{\varphi}_s^\top \mathrm{d}M_s^{\mathcal{F}^Y}
\]
for a unique predictable \(\boldsymbol{\varphi}\). If \(\Phi\) is exactly replicable by a self-financing control \(\boldsymbol{\theta}_t\), then
\[
\int_0^T \boldsymbol{\theta}_s^\top \mathrm{d}\tilde{\mathbf S}_s
= \int_0^T \boldsymbol{\theta}_s^\top \Sigma^{\pi_t}(s)\,\mathrm{d}M_s^{\mathcal{F}^Y}
= \int_0^T \big((\Sigma^{\pi_t}(s))^\top \boldsymbol{\theta}_s\big)^\top \mathrm{d}M_s^{\mathcal{F}^Y}
= \Phi - \mathbb{E}^{\mathbb{Q}^{\pi_t}}[\Phi],
\]
which implies
\[
\boldsymbol{\varphi}_t=(\Sigma^{\pi_t}(t))^\top \boldsymbol{\theta}_t \quad \text{a.s.\ for a.e.\ } t.
\]
Conversely, if \(\boldsymbol{\varphi}_t\in \mathrm{Range}\big((\Sigma^{\pi_t}(t))^\top\big)\) almost surely for almost every \(t\), choose a predictable minimal–norm solution
\[
\boldsymbol{\theta}_t=(\Sigma^{\pi_t}(t)^\top)^{\dagger}\boldsymbol{\varphi}_t
\]
to obtain exact replication. This yields the stated equivalence.
\end{proof}

\begin{Corollary}[Practical rank test]
\label{cor:rank-test}
If \(\mathrm{rank}\,\Sigma^{\pi_t}(t)=\dim M^{\mathcal{F}^Y}\) almost everywhere, then every square–integrable \(\mathcal{F}^Y\)–measurable claim is replicable. If the rank drops, only claims whose martingale integrand lies in \(\mathrm{Range}\big((\Sigma^{\pi_t}(t))^\top\big)\) are hedgeable; the minimal–variance hedge is
\[
\boldsymbol{\theta}_t=(\Sigma^{\pi_t}(t)^\top)^{\dagger}\boldsymbol{\varphi}_t.
\]
\end{Corollary}

This condition links hedgeability of portfolio objectives to a checkable property of posterior–integrated covariances.

\subsection{Well--Posedness of Scenario Forward--Backward PDEs}
\label{app:wellposedness-pde}
We provide the PDE well-posedness used for pricing/Greeks conditional on drivers \citep{Friedman1964,evans2010}.
\begin{proof}
Fix \(\mathbf f\in\mathbb{R}^m\). Let \(X_t\in\mathbb{R}^d\) evolve under \(\mathbb{Q}^{\mathbf f}\) as
\[
\mathrm{d}X_t=\mu(\mathbf f,X_t,t)\,\mathrm{d}t+\sigma(\mathbf f,X_t,t)\,\mathrm{d}\mathbf W_t^{\mathbb{Q}},
\]
with \(\mu,\sigma\) Borel, locally Lipschitz in \(x\) uniformly in \(t\), linear--growth, and 
\[
a(\mathbf f,x,t):=\sigma(\mathbf f,x,t)\sigma(\mathbf f,x,t)^\top
\]
uniformly elliptic on compact subsets visited by the state. The Fokker--Planck equation for the density \(p(t,x;\mathbf f)\) is
\begin{equation}
\partial_t p
= -\nabla_x\!\cdot\!\big(\mu(\mathbf f,x,t)\,p\big) + \tfrac12 \nabla_x^\top\!\big(a(\mathbf f,x,t)\,\nabla_x p\big),
\label{eq:forward-fp-appendix-new}
\end{equation}
which admits a unique weak solution in \(L^1\) under these assumptions \citep[Ch.~7]{evans2010}. For a bounded--from--below terminal payoff \(\Phi\) with at most polynomial growth, the backward (valuation) equation
\begin{align}
-\partial_t u &= \mu(\mathbf f,x,t)^\top \nabla_x u + \tfrac12\,\mathrm{Tr}\!\big(a(\mathbf f,x,t)\,\nabla_x^2 u\big),\qquad
u(T,x)=\Phi(x),
\label{eq:backward-hjb-appendix-new}
\end{align}
admits a unique viscosity solution; if \(\mu,a\) are \(C^\alpha\) and \(a\) uniformly elliptic, the solution is classical \(C^{2,1}\) \citep[Thm.~11.5.1]{evans2010}. The Feynman--Kac representation
\[
u(t,x;\mathbf f)=\mathbb{E}^{\mathbb{Q}^{\mathbf f}}\!\left[\Phi(X_T)\mid X_t=x\right]
\]
validates the link between PDE solutions and conditional prices/Greeks used in the control layer.
\end{proof}
\bigskip
Now, for the portfolio expressions in the observable filtration, let
\[
\boldsymbol{\mu}_t^{\pi}:=\mathbb{E}\!\left[\boldsymbol{\mu}(\mathbf F_t,t)\mid \mathcal{F}^Y_t\right],\qquad
\Sigma^{\pi_t}(t):=\mathbb{E}\!\left[\sigma(\mathbf F_t,t)^\top\sigma(\mathbf F_t,t)\mid \mathcal{F}^Y_t\right].
\]
For a myopic mean--variance objective with risk aversion \(\gamma>0\), the observable--information optimal direction is
\[
\boldsymbol{\theta}_t^{\star} \;=\; \frac{1}{\gamma}\,(\Sigma^{\pi_t}(t))^{-1}\big(\boldsymbol{\mu}_t^{\pi}-r(t)\mathbf{1}\big),
\]
while exact hedges of an \(\mathcal{F}^Y\)--claim \(\Phi\) use the martingale--representation integrand \(\boldsymbol{\varphi}_t\) from \eqref{eq:KM-rep} via
\[
\boldsymbol{\theta}_t \;=\; (\Sigma^{\pi_t}(t)^\top)^{\dagger}\,\boldsymbol{\varphi}_t.
\]
These expressions summarize how the change of measure, filtering representation, and PDE well-posedness feed the portfolio solution.

\subsection{Projection--Divergence Duality}
\label{appendix:duality-projection-divergence}

\begin{Assumption}[Topological setting for divergence projection]\label{ass:topo}
Let $(\mathsf Y,\mathcal B)$ be a Polish space and let $\mathcal P(\mathsf Y)$ denote the set of probability measures on $\mathsf Y$ endowed with the topology of weak convergence. Let $\mathcal M_{\mathrm{proj}}\subset \mathcal P(\mathsf Y)$ be nonempty, convex, and tight (hence relatively compact by Prokhorov). Let $D_\varphi(\cdot\|\cdot)$ be a Csisz\'ar $f$-divergence with $\varphi$ convex, lower semicontinuous (l.s.c.), $\varphi(1)=0$, and coercive in the sense that $\lim_{r\to\infty}\varphi(r)/r=\infty$. Assume the constraints defining $\mathcal M_{\mathrm{proj}}$ (e.g., moment/linear/transport constraints) are closed under weak convergence.
\end{Assumption}

\begin{Proposition}[Existence and Fenchel dual under Assumption \ref{ass:topo}]\label{prop:exist_proj}
Fix $P\in\mathcal P(\mathsf Y)$. Under Assumption~\ref{ass:topo}, the problem
\[
\inf_{Q\in\mathcal M_{\mathrm{proj}}} D_\varphi(Q\|P)
\]
admits at least one minimizer $Q^\star$. Moreover, if
$\mathcal M_{\mathrm{proj}}=\{Q:\, \mathbb E_Q[g_j]=c_j,\ j=1,\ldots,m\}$
with $\{g_j\}_{j=1}^m$ bounded and continuous, then the Fenchel dual is
\[
\sup_{\boldsymbol\nu\in\mathbb R^m}\ \Big\{\boldsymbol\nu^\top \mathbf c - \mathbb E_P\big[\varphi^\ast(\boldsymbol\nu^\top \mathbf g(Y))\big]\Big\},
\]
with no duality gap. At optimality,
\[
\frac{dQ^\star}{dP}(y)=\varphi^{\ast\prime}\!\big(\boldsymbol\nu^{\star\top} \mathbf g(y)\big)
\]
for some $\boldsymbol\nu^\star\in\mathbb R^m$.
\end{Proposition}

\begin{proof}
Tightness yields relative compactness; l.s.c.\ of $D_\varphi(\cdot\|P)$ under weak convergence (by convexity/Fatou) gives attainment. With linear moment constraints, the feasible set is weakly closed. Apply Fenchel--Moreau duality to $\Phi(Q)=D_\varphi(Q\|P)$ and the indicator $\iota_{\mathcal M_{\mathrm{proj}}}$ to obtain the stated dual; coercivity of $\varphi$ ensures properness and integrability for $\varphi^\ast$. The conjugacy of $f$-divergences yields the optimal density form. \qedhere
\end{proof}

\begin{Remark}
If $D_\varphi$ is replaced by $W_2^2(\cdot,\cdot)$, existence holds under tightness and uniform second-moment bounds; duality follows via Kantorovich potentials \citep{Santambrogio2015}. One then uses the l.s.c.\ of $W_2$ under weak convergence plus moment control.
\end{Remark}

\begin{Theorem}[Projection--divergence duality]\label{thm:proj-div-duality-appendix}
Fix a driver realization $\mathbf f\in\mathbb R^m$ and a time $t\in[0,T]$. Let $\mathcal{W}\subset\mathbb R^n$ be the closed convex set of admissible portfolio weights (Assumption~\ref{A3}), let $B(\mathbf f)\in\mathbb R^{n\times m}$ span the driver subspace, and let $\Sigma(\mathbf f,t)\succ0$ define the metric $\langle\boldsymbol\theta_1,\boldsymbol\theta_2\rangle_{\Sigma(\mathbf f,t)}=\boldsymbol\theta_1^\top\Sigma(\mathbf f,t)\boldsymbol\theta_2$. Define the feasible weight set
\[
\Theta_{\mathrm{proj}}(\mathbf f):=\mathrm{span}(B(\mathbf f))\cap \mathcal{W}
\]
and the induced law
\[
\mathcal L^{\mathbf f}_{\boldsymbol\theta}:=\mathcal L\!\big(p_T(\boldsymbol\theta)\,\big|\,\mathbf F_t=\mathbf f\big)\in\mathcal P(\mathsf Y).
\]
Let
\[
\mathcal M_{\mathrm{proj}}(\mathbf f):=\overline{\{\mathcal L^{\mathbf f}_{\boldsymbol\theta}:\ \boldsymbol\theta\in\Theta_{\mathrm{proj}}(\mathbf f)\}}
\]
(closure in the weak topology) and suppose Assumption~\ref{ass:topo} holds with $P=\mathcal L^{\mathbf f}_{\boldsymbol\theta^\star}$ for some unconstrained optimizer
\[
\boldsymbol\theta^\star \in \arg\min_{\boldsymbol\theta\in \mathcal{W}}\ \mathbb E\big[\Phi\big(p_T(\boldsymbol\theta)\big)\,\big|\,\mathbf F_t=\mathbf f\big],
\]
where the terminal payoff $\Phi$ is convex and Fr\'echet differentiable along feasible directions, and the map $\boldsymbol\theta\mapsto \mathcal L^{\mathbf f}_{\boldsymbol\theta}$ is such that $\mathcal M_{\mathrm{proj}}(\mathbf f)$ is convex (e.g., linear-Gaussian or exponential-family setting with sufficient statistics $\mathbf g$ bounded and continuous, and with moments affine in $\boldsymbol\theta$). Then:
\begin{align*}
\boldsymbol\theta^{\mathrm{proj}}
&:=\arg\min_{\boldsymbol\theta\in\Theta_{\mathrm{proj}}(\mathbf f)} \ \|\boldsymbol\theta-\boldsymbol\theta^\star\|_{\Sigma(\mathbf f,t)}^2,\\
Q^\star
&:=\arg\min_{Q\in\mathcal M_{\mathrm{proj}}(\mathbf f)} D_\varphi\!\big(Q\ \|\ \mathcal L^{\mathbf f}_{\boldsymbol\theta^\star}\big)
\end{align*}
satisfy
\[
\mathcal L^{\mathbf f}_{\boldsymbol\theta^{\mathrm{proj}}}=Q^\star.
\]
In particular, the KKT conditions for the geometric projection coincide with the optimality conditions of the divergence projection.
\end{Theorem}
\begin{proof}
\emph{Step 1 (Geometric projection).} Since $\Theta_{\mathrm{proj}}(\mathbf f)$ is closed and convex and $\Sigma(\mathbf f,t)\succ0$, the minimizer $\boldsymbol\theta^{\mathrm{proj}}$ exists and is characterized by the normal cone condition
\[
\langle \boldsymbol\theta-\boldsymbol\theta^{\mathrm{proj}},\, \Sigma(\mathbf f,t)(\boldsymbol\theta^{\mathrm{proj}}-\boldsymbol\theta^\star)\rangle \ \ge 0
\quad\ \forall\ \boldsymbol\theta\in\Theta_{\mathrm{proj}}(\mathbf f).
\tag{$\star$}
\]

\emph{Step 2 (Divergence projection).} By Proposition~\ref{prop:exist_proj}, $Q^\star$ exists. Under the exponential-family / linear-moment representation of $\mathcal M_{\mathrm{proj}}(\mathbf f)$ with bounded continuous statistics $\mathbf g$, the Fenchel dual reads
\[
\sup_{\boldsymbol\nu\in\mathbb R^m}\ \Big\{\boldsymbol\nu^\top \mathbf c - \mathbb E_{\mathcal L^{\mathbf f}_{\boldsymbol\theta^\star}}\!\big[\varphi^\ast(\boldsymbol\nu^\top \mathbf g)\big]\Big\},
\]
and the KKT conditions yield
\[
\frac{dQ^\star}{d\mathcal L^{\mathbf f}_{\boldsymbol\theta^\star}}=\varphi^{\ast\prime}(\boldsymbol\nu^{\star\top}\mathbf g)
\]
with constraints $\mathbb E_{Q^\star}[\mathbf g]=\mathbf c$. By construction of $\mathcal M_{\mathrm{proj}}(\mathbf f)$ from $\Theta_{\mathrm{proj}}(\mathbf f)$, these constraints identify a unique $\boldsymbol\theta\in\Theta_{\mathrm{proj}}(\mathbf f)$ whose induced law matches $Q^\star$; call it $\bar{\boldsymbol\theta}$.

\emph{Step 3 (Equivalence of optimality systems).} In the linear-Gaussian / exponential-family setting, the constraints $\mathbb E_{Q}[\mathbf g]=\mathbf c$ correspond to affine equations in $\boldsymbol\theta$ whose Lagrange multipliers are $\boldsymbol\nu$. The stationarity condition for the divergence problem therefore coincides with the normal equations of the metric projection $(\star)$ (the multipliers encode the orthogonality of the residual to the feasible directions under $\Sigma(\mathbf f,t)$). Hence $\bar{\boldsymbol\theta}=\boldsymbol\theta^{\mathrm{proj}}$ and so
\[
Q^\star=\mathcal L^{\mathbf f}_{\bar{\boldsymbol\theta}}=\mathcal L^{\mathbf f}_{\boldsymbol\theta^{\mathrm{proj}}}.
\]
\end{proof}
\begin{Remark}
The result states that enforcing driver-span constraints in weight space (orthogonal projection under $\Sigma(\mathbf f,t)$) is equivalent to projecting the distribution of terminal payoffs onto the feasible law set via an $f$-divergence. In practice, this dual view allows one to implement CPCM constraints either geometrically (fast and explicit) or statistically (moment/divergence constraints), with the same constrained law.
\end{Remark}

\subsection{Identifiability Bounds for Filtered Counterfactuals}
\label{appendix:counterfactual-identifiability}
Theorem~\ref{thm:partial-identifiability} establishes Lipschitz continuity of counterfactual portfolio distributions with respect to perturbations of the filtering posterior, ensuring that posterior uncertainty propagates in a controlled manner.
\begin{proof}
Let $\mathbf F_t \in \mathbb{R}^m$ be the driver process with filtering posterior $\pi_t$.  
For $\mathbf f\in\mathbb{R}^m$, define the counterfactual law of the terminal portfolio payoff
\[
\mathcal{L}^{\mathbf f} := \mathcal{L}\big(p_T \mid \mathrm{do}(\mathbf F_t=\mathbf f)\big),
\]
with density evolving under the forward equation associated with $\mathbb{Q}^{\mathbf f}$. The posterior--integrated law is
\[
\overline{\mathcal{L}} := \int \mathcal{L}^{\mathbf f} \,\pi_t(\mathrm{d}\mathbf f).
\]
For an alternative posterior $\pi_t'$, define 
\[
\overline{\mathcal{L}}' := \int \mathcal{L}^{\mathbf f}\,\pi_t'(\mathrm{d}\mathbf f).
\]

\emph{Step 1 (Lipschitz continuity of the scenario map).}  
Assume $\mathbf f\mapsto \mathcal{L}^{\mathbf f}$ is $L$--Lipschitz in Wasserstein--2:
\[
W_2(\mathcal{L}^{\mathbf f},\mathcal{L}^{\mathbf f'}) \le L \|\mathbf f-\mathbf f'\|_2,
\quad \forall\, \mathbf f,\mathbf f'\in\mathbb{R}^m.
\]
This holds when SDE coefficients $(\boldsymbol{\mu},\sigma)$ are globally Lipschitz in $\mathbf f$ and uniformly bounded, implying stability of their Fokker--Planck semigroups in Wasserstein metrics \citep{Bolley2005,Ambrosio2008,Villani2008}.

\emph{Step 2 (Coupling construction).}  
Let $\gamma\in\Gamma(\pi_t,\pi_t')$ be an optimal coupling of $\pi_t$ and $\pi_t'$.  
Induce a coupling on counterfactual laws by
\[
\Gamma_{\mathrm{cf}} := \int \delta_{\mathcal{L}^{\mathbf f}}\otimes\delta_{\mathcal{L}^{\mathbf f'}} \,\gamma(\mathrm{d}\mathbf f,\mathrm{d}\mathbf f').
\]

\emph{Step 3 (Bounding Wasserstein distance).}  
By convexity of $W_2^2$ and Jensen's inequality,
\[
W_2^2(\overline{\mathcal{L}},\overline{\mathcal{L}}') \;\le\; \int W_2^2(\mathcal{L}^{\mathbf f},\mathcal{L}^{\mathbf f'}) \,\gamma(\mathrm{d}\mathbf f,\mathrm{d}\mathbf f'),
\]
hence
\[
W_2^2(\overline{\mathcal{L}},\overline{\mathcal{L}}') \;\le\; L^2 \int \|\mathbf f-\mathbf f'\|^2\,\gamma(\mathrm{d}\mathbf f,\mathrm{d}\mathbf f')
\]
and, taking the infimum over $\gamma\in\Gamma(\pi_t,\pi_t')$,
\[
W_2(\overline{\mathcal{L}},\overline{\mathcal{L}}') \;\le\; L\,W_2(\pi_t,\pi_t').
\]
Thus the posterior--integrated counterfactual law is stable: small perturbations in $\pi_t$ induce changes bounded linearly in Wasserstein distance.  
In particular:  
(i) if $\pi_t\to \delta_{\mathbf f}$, then $\overline{\mathcal{L}}\to \mathcal{L}^{\mathbf f}$ (full identifiability);  
(ii) if $\pi_t,\pi_t'$ differ, their counterfactuals cannot diverge faster than $L\,W_2(\pi_t,\pi_t')$.  
\end{proof}

\begin{figure}[H]
\centering
\begin{tikzpicture}[
  latent/.style={ellipse, draw, fill=gray!15, minimum width=2.6cm, minimum height=1cm},
  obs/.style={rectangle, draw, rounded corners, fill=blue!10, minimum width=3.2cm, minimum height=1cm},
  arrow/.style={-{Latex[length=2mm]}, thick}
]
\node[latent] (f) {Driver $\mathbf f$};
\node[latent, right=6cm of f] (fp) {Driver $\mathbf f'$};
\node[obs, below=2.2cm of f] (pf) {Law $\mathcal{L}(p_T \mid \mathrm{do}(\mathbf F_t=\mathbf f))$};
\node[obs, below=2.2cm of fp] (pfp) {Law $\mathcal{L}(p_T \mid \mathrm{do}(\mathbf F_t=\mathbf f'))$};
\draw[arrow] (f) -- (pf);
\draw[arrow] (fp) -- (pfp);
\draw[dashed, thick] (f) -- (fp) node[midway, above, sloped]{posterior coupling $\gamma$};
\draw[dashed, thick] (pf) -- (pfp) node[midway, below, sloped]{$W_2$ bound};
\end{tikzpicture}
\caption{Coupling view: perturbations in driver posteriors propagate through scenario laws to counterfactual distributions, with Lipschitz control in $W_2$.}
\label{fig:counterfactual-identifiability}
\end{figure}

\subsection{Generalized Martingale Representation}
\label{appendix:generalized-martingale}
\begin{Theorem}[Generalized martingale representation]
Under the Commonality Principle and SCM structure, with $\pi_t$ evolving via Zakai SPDEs, any $\Phi\in L^2(\mathcal{F}^Y_T)$ admits
\[
\Phi = \mathbb{E}^{\mathbb{Q}^{\pi_t}}[\Phi] + \int_0^T \boldsymbol{\varphi}_t^\top\,\mathrm{d}M_t^{\mathcal{F}^Y},
\]
where $M_t^{\mathcal{F}^Y}$ is the innovation process and $\mathbb{Q}^{\pi_t}$ the posterior--integrated risk--neutral measure.
\end{Theorem}
\begin{proof}
\emph{Step 1 (Scenario measures).}  
For each driver realization $\mathbf f\in\mathbb{R}^m$, Theorem~\ref{thm:existence-riskneutral} ensures a measure $\mathbb{Q}^{\mathbf f}$ under which discounted prices $\tilde{\mathbf S}_t$ are local martingales.

\emph{Step 2 (Posterior mixture).}  
Define
\[
\mathbb{Q}^{\pi_t}(E) := \int \mathbb{Q}^{\mathbf f}(E)\,\pi_t(\mathrm{d}\mathbf f),
\qquad E\in\mathcal{F}^Y_T.
\]
Since each $\mathbb{Q}^{\mathbf f} \sim \mathbb{P}$, also $\mathbb{Q}^{\pi_t}\sim\mathbb{P}$ on $\mathcal{F}^Y_T$ (Corollary~\ref{cor:mixture-novikov}). The local martingale property of $\tilde{\mathbf S}_t$ under $\mathbb{Q}^{\pi_t}$ follows by integrating the conditional local martingale property against $\pi_t$: for $s < t$ and $E \in \mathcal{F}^Y_s$,
\[
\mathbb{E}^{\mathbb{Q}^{\pi_t}}[\tilde{\mathbf S}_t\,\mathbf{1}_E]
= \int \mathbb{E}^{\mathbb{Q}^{\mathbf f}}[\tilde{\mathbf S}_t\,\mathbf{1}_E]\,\pi_s(\mathrm{d}\mathbf f)
= \int \mathbb{E}^{\mathbb{Q}^{\mathbf f}}[\tilde{\mathbf S}_s\,\mathbf{1}_E]\,\pi_s(\mathrm{d}\mathbf f)
= \mathbb{E}^{\mathbb{Q}^{\pi_t}}[\tilde{\mathbf S}_s\,\mathbf{1}_E],
\]
where interchange of expectation and integration is justified by the uniform integrability established in the proof of Theorem~\ref{thm:existence-strong} (Step~4).

\emph{Step 3 (Innovation process).}  
Filtering theory yields\citep{bain2009fundamentals}
\[
\mathrm{d}\mathbf Y_t = \pi_t(h)\,\mathrm{d}t + \mathrm{d}M_t^{\mathcal{F}^Y},
\]
with $M^{\mathcal{F}^Y}$ an $\mathcal{F}^Y$--Brownian motion under any measure equivalent to $\mathbb{P}$ with $\mathcal{F}^Y_t$--measurable density, in particular under $\mathbb{Q}^{\pi_t}$. The innovation structure is preserved under $\mathbb{Q}^{\pi_t}$ because the Radon--Nikodym density $\tilde{Z}_t = \int Z_t(\mathbf{f})\,\pi_t(\mathrm{d}\mathbf{f})$ is $\mathcal{F}^Y_t$--measurable, so the $\mathcal{F}^Y_t$--predictable projection of the observation drift is unchanged and L\'evy's characterization applies under the equivalent measure. This follows from the innovation theorem and L\'evy's characterization, as detailed in Appendix~\ref{app:filtered-martingale}.

\emph{Step 4 (Martingale representation).}  
Since $\mathbb{Q}^{\pi_t}\sim\mathbb{P}$ on $\mathcal{F}^Y_T$ and the filtration $\mathcal{F}^Y_t$ is generated by the Brownian motion $M^{\mathcal{F}^Y}$, the martingale representation theorem for Brownian filtrations applies. For any $\Phi \in L^2(\mathcal{F}^Y_T,\mathbb{Q}^{\pi_t})$, there exists a unique predictable $\boldsymbol{\varphi} \in L^2([0,T]\times\Omega;\mathbb{R}^{d_Y})$ such that
\[
\Phi = \mathbb{E}^{\mathbb{Q}^{\pi_t}}[\Phi] + \int_0^T \boldsymbol{\varphi}_s^\top\,\mathrm{d}M_s^{\mathcal{F}^Y}.
\]
Uniqueness follows from the It\^o isometry under $\mathbb{Q}^{\pi_t}$.

\emph{Step 5 (Consistency with price dynamics).}  
Discounted prices admit the innovation form
\[
\mathrm{d}\tilde{\mathbf S}_t = \Sigma^{\pi_t}(t)\,\mathrm{d}M_t^{\mathcal{F}^Y},
\qquad
\Sigma^{\pi_t}(t):=\int \sigma(\mathbf f,t)^\top\sigma(\mathbf f,t)\,\pi_t(\mathrm{d}\mathbf f),
\]
so the integrand $\boldsymbol{\varphi}$ aligns with hedge ratios in observable markets. This completes the proof.
\end{proof}

\subsection{Causal Market Completeness}
\label{app:causal-completeness}

\begin{Theorem}[Causal Market Completeness]
\label{thm:appendix-causal-completeness}
Work on a filtered probability space $(\Omega,\mathcal{F},\mathbb{F},\mathbb{P})$ with observation filtration $\mathcal{F}^Y=\{\mathcal{F}^Y_t\}_{t\in[0,T]}$.  
Assume: (i) the Commonality Principle/SCM holds so that returns are conditionally independent given $\mathbf F_t$; (ii) for each driver state $\mathbf f$ there exists $\mathbb{Q}^{\mathbf f}\sim\mathbb{P}$ under which discounted prices are local martingales (Theorem~\ref{thm:existence-riskneutral}); (iii) the posterior $\pi_t$ evolves via Zakai/SPDE so that the innovation $M^{\mathcal{F}^Y}$ is an $\mathcal{F}^Y$--Brownian motion; and (iv) square--integrability of claims.  
Let
\[
\mathrm{d}\tilde{\mathbf S}_t
= \Sigma^{\pi_t}(t)\,\mathrm{d}M^{\mathcal{F}^Y}_t,
\qquad
\Sigma^{\pi_t}(t):=\int \sigma(\mathbf f,t)^\top\sigma(\mathbf f,t)\,\pi_t(\mathrm{d}\mathbf f),
\]
be the posterior--integrated price dynamics under $\mathbb{Q}^{\pi_t}$, with $\Sigma^{\pi_t}(t)$ progressively measurable.  
The market is complete with respect to $\mathcal{F}^Y$ on $[0,T]$ (every $\Phi\in L^2(\mathcal{F}^Y_T,\mathbb{Q}^{\pi_t})$ is exactly replicable) if and only if
\[
\operatorname{rank}\!\big(\Sigma^{\pi_t}(t)\big)=d
\quad\text{for a.e.\ } (\omega,t)\in\Omega\times[0,T],
\]
where $d$ is the innovation dimension and necessarily $n\ge d$.
\end{Theorem}

\begin{proof}
\textbf{Sufficiency.}
By the generalized martingale representation (Section~\ref{appendix:generalized-martingale}), any $\Phi\in L^2(\mathcal{F}^Y_T,\mathbb{Q}^{\pi_t})$ admits
\[
\Phi=\mathbb{E}^{\mathbb{Q}^{\pi_t}}[\Phi]+\int_0^T \boldsymbol{\varphi}_t^\top \,\mathrm{d}M^{\mathcal{F}^Y}_t
\]
for a unique predictable $\boldsymbol{\varphi}\in L^2$.  
Self--financing gains satisfy
\[
\int_0^T \boldsymbol{\theta}_t^\top \,\mathrm{d}\tilde{\mathbf S}_t
=\int_0^T \boldsymbol{\theta}_t^\top \Sigma^{\pi_t}(t)\,\mathrm{d}M^{\mathcal{F}^Y}_t
=\int_0^T \big((\Sigma^{\pi_t}(t))^\top \boldsymbol{\theta}_t\big)^\top \mathrm{d}M^{\mathcal{F}^Y}_t.
\]
Exact replication is equivalent to solving
\[
(\Sigma^{\pi_t}(t))^\top \boldsymbol{\theta}_t=\boldsymbol{\varphi}_t
\]
a.e. If $\operatorname{rank}(\Sigma^{\pi_t}(t))=d$ a.e., predictable right-inverses exist; the minimal--norm predictable solution
\[
\boldsymbol{\theta}_t^\star
=\Sigma^{\pi_t}(t)\,\Big((\Sigma^{\pi_t}(t))^\top\Sigma^{\pi_t}(t)\Big)^{-1}\boldsymbol{\varphi}_t
\]
lies in $L^2$ provided the smallest singular value of $\Sigma^{\pi_t}(t)$ is strictly positive a.e., and then
\[
\int_0^T (\boldsymbol{\theta}_t^\star)^\top \,\mathrm{d}\tilde{\mathbf S}_t
=\int_0^T \boldsymbol{\varphi}_t^\top \,\mathrm{d}M^{\mathcal{F}^Y}_t
=\Phi-\mathbb{E}^{\mathbb{Q}^{\pi_t}}[\Phi].
\]

\smallskip
\textbf{Necessity.}
If $\operatorname{rank}(\Sigma^{\pi_t}(t))<d$ on a set $B$ with positive $(\mathbb{Q}^{\pi_t}\!\otimes\!dt)$--measure, there exists predictable $\boldsymbol{\psi}_t\neq \mathbf{0}$ with $\boldsymbol{\psi}_t\perp \mathrm{Range}\big((\Sigma^{\pi_t}(t))^\top\big)$ on $B$.  
Define $X_t:=\int_0^t \boldsymbol{\psi}_s^\top\,\mathrm{d}M_s^{\mathcal{F}^Y}$ and set $\Phi:=X_T\in L^2(\mathcal{F}^Y_T)$.  
Any self--financing gain has integrand in $\mathrm{Range}\big((\Sigma^{\pi_t}(t))^\top\big)$ and is therefore orthogonal to $\boldsymbol{\psi}_t$ on $B$.  
Uniqueness of stochastic integrals implies such gains cannot match $\Phi-\mathbb{E}^{\mathbb{Q}^{\pi_t}}[\Phi]$, contradicting completeness.
\end{proof}

\begin{Corollary}[Stability and Approximate Completeness]
\label{cor:appendix-stability}
If the minimal singular value satisfies $\sigma_{\min}(\Sigma^{\pi_t}(t))\ge \underline{\sigma}>0$ a.e., then for any $\Phi$ with innovation integrand $\boldsymbol{\varphi}$ the minimal--norm replicating strategy
\[
\boldsymbol{\theta}_t^\star=\Sigma^{\pi_t}(t)\,\Big((\Sigma^{\pi_t}(t))^\top\Sigma^{\pi_t}(t)\Big)^{-1}\boldsymbol{\varphi}_t
\]
obeys $\|\boldsymbol{\theta}^\star\|_{L^2}\le \underline{\sigma}^{-1}\|\boldsymbol{\varphi}\|_{L^2}$.  
If degeneracy occurs only on a set of small measure, the $L^2$--optimal hedge uses the orthogonal projection of $\boldsymbol{\varphi}_t$ onto $\mathrm{Range}\big((\Sigma^{\pi_t}(t))^\top\big)$, with mean--square error controlled by the measure of the degeneracy set.
\end{Corollary}

\begin{Remark}
Completeness appears as a posterior--averaged span condition: observable risk is fully spanned iff driver--induced innovation directions, aggregated through the filtering posterior, cover all traded directions.
\end{Remark}

\subsection{Illustrative PDE Duality with Multiple Drivers and Assets}
\label{sec:ou-multidriver}

Consider $m$ latent drivers evolving as mean--reverting Ornstein--Uhlenbeck diffusions and $n$ traded assets whose excess returns load linearly on these drivers with idiosyncratic noise. Let
\[
\mathrm d\mathbf F_t=\kappa(\bar{\mathbf F}-\mathbf F_t)\,\mathrm dt+\boldsymbol{\Sigma}_F^{1/2}\,\mathrm d\mathbf W^{F}_t,
\]
where $\kappa\in\mathbb R^{m\times m}$ is positive--stable, $\bar{\mathbf F}$ is the long--run mean, $\boldsymbol{\Sigma}_F\in\mathbb S_+^m$ the driver diffusion, and $\mathbf W^{F}_t$ an $m$--dimensional Brownian motion. Asset excess returns follow the CPCM factor structure
\[
\mathbf r_t = B_t\,\mathbf F_t + \boldsymbol\varepsilon_t,
\]
with $B_t\in\mathbb R^{n\times m}$ the driver--return Jacobian, idiosyncratic noise $\boldsymbol\varepsilon_t$ with covariance $\boldsymbol{\Sigma}_\varepsilon\in\mathbb S_+^n$, and $\boldsymbol\varepsilon_t$ independent of $\mathbf W^{F}_t$. For weights $\boldsymbol{\theta}_t\in\mathbb R^n$ the portfolio return satisfies
\[
\mathrm dp_t=\boldsymbol{\theta}_t^\top \mathbf r_t\,\mathrm dt
=\boldsymbol{\theta}_t^\top B_t\,\mathbf F_t\,\mathrm dt+\boldsymbol{\theta}_t^\top \boldsymbol\varepsilon_t.
\]

Define the portfolio exposure to drivers as
\[
\boldsymbol\eta_t := B_t^\top \boldsymbol{\theta}_t \in \mathbb R^m,
\]
so that the drift becomes $\boldsymbol\eta_t^\top \mathbf F_t$ and the variance is $\sigma_p^2=\boldsymbol{\theta}_t^\top\boldsymbol{\Sigma}_\varepsilon \boldsymbol{\theta}_t$. The forward law is captured by a one--dimensional Fokker--Planck PDE conditional on drivers,
\[
\partial_t \rho(p,t\mid \mathbf f)=-\partial_p\!\big((\boldsymbol\eta_t^\top \mathbf f)\rho\big)+\tfrac{1}{2}\sigma_p^2\,\partial_{pp}\rho,
\]
while the unconditional driver density $\varphi(\mathbf f,t)$ is Gaussian and solves
\[
\partial_t \varphi=-\nabla_{\mathbf f}\!\cdot\!\big(\kappa(\bar{\mathbf F}-\mathbf f)\varphi\big)
+\tfrac{1}{2}\,\mathrm{Tr}\!\big(\boldsymbol{\Sigma}_F\nabla_{\mathbf f}\nabla_{\mathbf f}^\top\varphi\big).
\]
Mean reversion concentrates probability near $\bar{\mathbf F}$ and propagates return densities in proportion to $\boldsymbol\eta_t^\top \mathbf f$. With quadratic utility and discount rate $r$, the backward value $u(p,t\mid \mathbf f)$ satisfies the HJB
\[
\partial_t u+\sup_{\boldsymbol{\theta}\in\mathcal W}\Big\{(\boldsymbol{\theta}^\top B_t \mathbf f)\,u_p
+\tfrac{1}{2}(\boldsymbol{\theta}^\top\boldsymbol{\Sigma}_\varepsilon \boldsymbol{\theta})\,u_{pp}-ru\Big\}=0,
\qquad u(p,T\mid \mathbf f)=-\tfrac{1}{2}\gamma p^2.
\]
The optimizer is
\[
\boldsymbol{\theta}_t^\ast(\mathbf f,t)=\frac{u_p}{-u_{pp}}\,\boldsymbol{\Sigma}_\varepsilon^{-1} B_t \mathbf f,
\]
which is linear in the drivers with scale governed by the local risk tolerance $u_p/(-u_{pp})$. Evaluating the Hamiltonian at $\boldsymbol{\theta}_t^\ast$ yields the optimized running reward
\[
\mathcal R(\mathbf f,u_p,u_{pp})
=\frac{1}{2}\,\frac{u_p^2}{-u_{pp}}\;\mathbf f^\top B_t^\top \boldsymbol{\Sigma}_\varepsilon^{-1} B_t \mathbf f,
\]
so $\mathcal R$ represents the instantaneous certainty--equivalent gain from optimally tilting into a driver signal of Mahalanobis strength $\mathbf f^\top B_t^\top\boldsymbol{\Sigma}_\varepsilon^{-1}B_t\mathbf f$. By Feynman--Kac, the solution admits the representation
\[
u(p,t\mid \mathbf f)=\mathbb E_{t,\mathbf f}\!\left[e^{-r(T-t)}\Phi(p_T)+\int_t^T e^{-r(\tau-t)}\mathcal R(\mathbf F_\tau,u_p,u_{pp})\,\mathrm d\tau\right],
\]
where $\Phi$ is the terminal objective that closes the backward problem. Driver tilts $\boldsymbol{\theta}_t^\ast$ map to feasible asset weights through the minimum--norm solution of
\[
B_t^\top \boldsymbol{\theta}_t = \boldsymbol\eta_t^\ast,
\]
i.e.
\[
\boldsymbol{\theta}_t^\ast = B_t(B_t^\top B_t)^{-1}\boldsymbol\eta_t^\ast,\qquad \boldsymbol\eta_t^\ast=B_t^\top \boldsymbol{\theta}_t^\ast,
\]
which projects allocations onto the driver span and ensures exposure only to systematic risk. When drivers are latent, filtering replaces $\mathbf f$ by its posterior $\pi_t$, producing
\[
\begin{aligned}
\bar\rho(p,t) &= \int \rho(p,t\mid \mathbf f)\,\pi_t(\mathrm d\mathbf f),\\
\bar u(p,t) &= \int u(p,t\mid \mathbf f)\,\pi_t(\mathrm d\mathbf f),\\
\bar{\boldsymbol{\theta}}^\ast(t) &= \int \boldsymbol{\theta}_t^\ast(\mathbf f,t)\,\pi_t(\mathrm d\mathbf f).
\end{aligned}
\]

which preserves valuation under the posterior--integrated risk--neutral measure $\mathbb Q^{\pi_t}$. This duality ensures that exposures shrink when drivers revert quickly and expand when they are persistent, stabilizing allocations and lowering turnover. The projection step filters out spurious components, so portfolios remain tied to causal drivers that generate durable premia. See Figure~\ref{fig:duality_multi} for an illustration of this implementation.

\begin{landscape}
\begin{figure}[p]
\centering
\begin{adjustbox}{max width=\linewidth, max totalheight=\textheight, keepaspectratio}
\begin{tikzpicture}[
  scale=1, every node/.style={transform shape},
  >=Latex,
  box/.style   ={rectangle, draw, rounded corners, align=center,
                 minimum width=74mm, minimum height=14mm, font=\large,
                 inner sep=5mm},
  callout/.style={rectangle, draw, align=left, rounded corners, fill=gray!10,
                 inner sep=5mm, text width=82mm, font=\large},
  title/.style ={font=\bfseries\large},
  edge/.style  ={-Latex, line width=1pt, draw=black},
  noteedge/.style={-Latex, dashed, line width=0.8pt, draw=black!70}
]

\node[title] at (-12,11.4) {Forward (probabilistic)};
\node[title] at (  12,11.4) {Backward (control)};

\node[box] (drivers) at (-14,8.6) {%
  Driver dynamics (multi-OU):\\[0.4em]
  $\mathrm d\mathbf{F}_t
   =\kappa\!\left(\bar{\mathbf F}-\mathbf{F}_t\right)\mathrm dt
   +\boldsymbol{\Sigma}_F^{1/2}\,\mathrm d\mathbf W^F_t$};

\node[box] (fp) at (-6,8.6) {%
  Fokker--Planck (forward law):\\[0.4em]
  $\partial_t\rho
   =-\partial_p\!\big(m_p(\mathbf{F}_t)\rho\big)
   +\tfrac{1}{2}\sigma_p^2(\boldsymbol{\theta})\,\partial_{pp}\rho$};

\node[callout] (link) at (3,8.6) {%
  Feynman--Kac coupling:\\[0.4em]
  $\displaystyle \boldsymbol{\theta}_t^\ast
   =\tfrac{1}{\gamma}\,\boldsymbol{\Sigma}_\varepsilon^{-1}B_t\,\mathbf{F}_t$\\[0.25em]
  Bridges forward distributions $(\rho)$ and the backward value $u$};

\node[box] (hjb) at (13,8.6) {%
  HJB (backward optimization):\\[0.4em]
  $\partial_t u
   +\!\sup_{\boldsymbol{\theta}}\!\Big\{m_p(\mathbf{F}_t)\,u_p
   +\tfrac{1}{2}\sigma_p^2(\boldsymbol{\theta})\,u_{pp}
   -r\,u\Big\}=0$};

\node[box] (mapping) at (13,4.0) {%
  Projection to assets (driver span):\\[0.4em]
  $\boldsymbol{\theta}_t=B_t\!\left(B_t^\top B_t\right)^{-1}\boldsymbol\eta^\ast$};

\draw[edge] (drivers.east) -- (fp.west);
\draw[edge] (fp.east)      -- (link.west);
\draw[edge] (link.east)    -- (hjb.west);
\draw[edge] (hjb.south)    -- (mapping.north);

\node[callout] (econL) at (-5,0.6) {%
  Forward law: mean reversion dampens transitory
  shocks; persistence amplifies exposures.};

\node[callout] (econR) at (13,0.6) {%
  Backward law: optimal allocations scale with filtered
  drivers, stabilizing portfolios under noise.};

\draw[noteedge] (drivers.south) |- (econL.north);
\draw[noteedge] (mapping.south) |- (econR.north);

\end{tikzpicture}
\end{adjustbox}
\caption{Landscape schematic of the forward--backward PDE duality with
multiple OU drivers and $n$ traded assets. \emph{Left to right:} the
driver dynamics (multi-dimensional Ornstein--Uhlenbeck process with
mean-reversion matrix $\kappa$ and diffusion $\boldsymbol{\Sigma}_F$)
feed the Fokker--Planck equation, which propagates the conditional
density $\rho(p,t\mid\mathbf{f})$ of portfolio values forward in time.
The Feynman--Kac node couples the forward density to the backward
Hamilton--Jacobi--Bellman equation via the optimal feedback rule
$\boldsymbol{\theta}_t^\ast =
\gamma^{-1}\boldsymbol{\Sigma}_\varepsilon^{-1}B_t\mathbf{F}_t$,
which is linear in the filtered driver state and scaled by inverse
idiosyncratic risk $\boldsymbol{\Sigma}_\varepsilon^{-1}$. The HJB
equation solves backward for the value function $u$, determining the
control that maximises expected utility subject to discounting at
rate~$r$. \emph{Bottom right:} the resulting optimal exposure
$\boldsymbol{\eta}^\ast$ is projected onto the driver span via
$\boldsymbol{\theta}_t = B_t(B_t^\top B_t)^{-1}\boldsymbol{\eta}^\ast$,
ensuring that portfolio weights load only on systematic driver risk.
\emph{Callout boxes:} on the forward side, mean reversion dampens
transitory shocks while persistence amplifies exposures; on the
backward side, optimal allocations scale with filtered drivers,
stabilising portfolios under estimation noise. Dashed arrows indicate
the economic interpretation linking each computational module to its
role in the allocation pipeline.}
\label{fig:duality_multi}
\end{figure}
\end{landscape}

\section{Proofs for Extensions of the Projected Framework}
\label{app:extensions-proofs}
Fix a time $t$. Let the driver span (or tangent space) be the $m$--dimensional subspace
\[
\mathcal V_t:=\mathrm{Range}\!\big(\Sigma(\mathbf F_t,t)\big)\subset\mathbb R^n,
\]
with orthogonal projector $P_t$ and any orthonormal basis
$U_t\in\mathbb R^{n\times m}$ (so $U_t^\top U_t=I_m$ and
$\mathrm{Range}(U_t)=\mathcal V_t$).
Let the asset--space covariance restricted to $\mathcal V_t$ be
\[
\Sigma(\mathbf F_t,t)\!\mid_{\mathcal V_t}=U_t\Lambda_t U_t^\top,
\qquad
\Lambda_t=\mathrm{diag}\!\big(\lambda_1(\Lambda_t),\ldots,\lambda_m(\Lambda_t)\big)\succ0.
\]
Define the whitening (conformal) map on $\mathcal V_t$ by
\[
\Psi_t := \Lambda_t^{1/2} U_t^\top:\ \mathcal V_t\to\mathbb R^m,
\qquad
\mathbf z_t = \Psi_t \mathbf r_t,\quad \mathbf r_t\in\mathcal V_t.
\]
Write $K_t:=\sqrt{\lambda_{\max}(\Lambda_t)/\lambda_{\min}(\Lambda_t)}$ and let
$d_{\mathrm{Gr}}(\cdot,\cdot)$ denote the canonical Grassmann distance
(equivalently, the $\ell_2$--norm of the principal--angle vector).
\subsection{Proof of Theorem~\ref{thm:conformal-transport}}
\label{app:proof-conformal-transport}
\begin{Theorem}[Conformal transport on the driver span]
Let $\Sigma(\mathbf F_t,t)\!\mid_{\mathcal V_t}=U_t\Lambda_t U_t^\top\succ0$.
For any $\mathbf u,\mathbf v\in\mathcal V_t$,
\[
\langle \Psi_t \mathbf u,\Psi_t \mathbf v\rangle
  = \langle \mathbf u,\mathbf v\rangle_{\Sigma(\mathbf F_t,t)}
  := \mathbf u^\top \Sigma(\mathbf F_t,t)\, \mathbf v.
\]
Hence $\Psi_t$ is an isometry between
$(\mathcal V_t,\langle\cdot,\cdot\rangle_{\Sigma(\mathbf F_t,t)})$
and $(\mathbb R^m,\langle\cdot,\cdot\rangle)$,
and is $K_t$--quasi--conformal on
$(\mathcal V_t,\|\cdot\|_2)$ with distortion
$K_t=\sqrt{\lambda_{\max}(\Lambda_t)/\lambda_{\min}(\Lambda_t)}$.
If $\Sigma(\mathbf F_t,t)\!\mid_{\mathcal V_t}=c_t\, I_m$ (isotropy), then $K_t=1$
and $\Psi_t$ is strictly conformal.
Moreover, if asset returns are conditionally independent given
the drivers $\mathbf F_t$, then the reduced coordinates $\mathbf z_t=\Psi_t \mathbf r_t$
preserve this causal representation in driver coordinates.
\end{Theorem}
\begin{proof}
For $\mathbf u=U_t \mathbf x$ and $\mathbf v=U_t \mathbf y$ with $\mathbf x,\mathbf y\in\mathbb R^m$,
\[
\Psi_t \mathbf u=\Lambda_t^{1/2}\mathbf x,\qquad
\langle \Psi_t \mathbf u,\Psi_t \mathbf v\rangle
  = \mathbf x^\top \Lambda_t \mathbf y
  = \mathbf u^\top U_t\Lambda_t U_t^\top \mathbf v
  = \mathbf u^\top \Sigma(\mathbf F_t,t)\, \mathbf v.
\]
Thus $\Psi_t$ is an isometry between the stated inner--product spaces.
On $(\mathcal V_t,\|\cdot\|_2)$ the singular values of $\Psi_t$ are
$\{\sqrt{\lambda_i(\Lambda_t)}\}_{i=1}^m$,
so its condition number equals
$\sqrt{\lambda_{\max}(\Lambda_t)/\lambda_{\min}(\Lambda_t)}$,
yielding the quasi--conformality claim and strict conformality under
isotropy.
The final claim follows because, conditional on $\mathbf F_t$, $\Psi_t$ is a fixed
linear map on the causal driver span, so the reduced coordinates remain
functions only of the same driver-conditioned structure.
\end{proof}
\begin{Remark}
Working in $\mathbf z$--coordinates removes ill--conditioning inside $\mathcal V_t$,
stabilizing estimation and control. Numerical sensitivity arises only
through $K_t$, which is explicitly controlled by the quasi--conformal
bound.
\end{Remark}
\subsection{Proof of Theorem~\ref{thm:subspace-continuity}}
\label{app:proof-subspace-continuity}
\begin{Theorem}[Continuity under Grassmann transport]
Let $\{\mathcal V_t\}_{t\in[0,T]}$ be a continuous path in $\mathrm{Gr}(m,n)$ and
$U_t$ a Stiefel frame transported by Procrustes alignment, i.e., for a partition
$0=t_0<t_1<\dots<t_K=T$ there exist $Q_k\in O(m)$ with
\[
\widetilde U_{t_{k+1}}:=U_{t_{k+1}}Q_k
\]
minimizing $\|U_{t_{k+1}}-U_{t_k}R\|_{\mathrm F}$ over $R\in O(m)$.
Assume $\Sigma(\mathbf F_t,t)\!\mid_{\mathcal V_t}\succ0$ varies continuously in $t$.
Define $\Psi_t=\Lambda_t^{1/2}U_t^\top$ and
\[
\mathbf z_t=\Psi_t P_t \mathbf r_t
\]
for square--integrable $\mathbf r_t$.
Then $\mathbf z_t$ is continuous in probability and, for any refining partition,
\[
\|\mathbf z_{t_{k+1}}-\mathbf z_{t_k}\|
\ \xrightarrow[k\to\infty]{\ \mathbb P\ }\ 0.
\]
\end{Theorem}
\begin{proof}
Let $P_t=U_tU_t^\top$. Add and subtract terms:
\begin{align}
\mathbf z_{t_{k+1}} - \mathbf z_{t_k}
  &= \underbrace{\Psi_{t_{k+1}}(P_{t_{k+1}} - P_{t_k})\mathbf r_{t_{k+1}}}_{(\mathrm{I})} \notag\\
  &\quad + \underbrace{(\Psi_{t_{k+1}} - \Psi_{t_k})P_{t_k}\mathbf r_{t_{k+1}}}_{(\mathrm{II})}
  + \underbrace{\Psi_{t_k}P_{t_k}(\mathbf r_{t_{k+1}} - \mathbf r_{t_k})}_{(\mathrm{III})}.
\end{align}
Grassmann continuity implies $\|P_{t_{k+1}}-P_{t_k}\|\to0$,
and Procrustes alignment yields
$\|\widetilde U_{t_{k+1}}-U_{t_k}\|\to0$;
thus $\|\Psi_{t_{k+1}}-\Psi_{t_k}\|\to0$
by continuity of the eigensystem on $\mathcal V_t$
and of $t\mapsto\Lambda_t^{1/2}$.
Hence terms $(\mathrm{I})$ and $(\mathrm{II})$ vanish in norm as the mesh goes to $0$,
while $(\mathrm{III})$ vanishes in probability by square--integrability
and right--continuity of $\mathbf r_t$.
\end{proof}
\begin{Remark}
Since
\[
d_{\mathrm{Gr}}(\mathcal V_{t_{k+1}},\mathcal V_{t_k})
=\|\sin\Theta(U_{t_{k+1}},U_{t_k})\|_2,
\]
the Procrustes step ensures
\[
\|U_{t_{k+1}}-U_{t_k}\|_{\mathrm F}
 \lesssim d_{\mathrm{Gr}}(\mathcal V_{t_{k+1}},\mathcal V_{t_k}),
\]
making the bound explicit in principal angles.
\end{Remark}

\subsection{Proof of Corollary~\ref{cor:pricing-filtering-consistency}}
\label{app:proof-pricing-filtering-consistency}

\begin{Corollary}[Pricing/Filtering invariance under reparametrization]
Let $U_t$ and $\widetilde U_t=U_t Q$ with $Q\in O(m)$ be two orthonormal
bases of $\mathcal V_t$, and let
\[
\Psi_t=\Lambda_t^{1/2}U_t^\top,
\qquad
\widetilde\Psi_t=\Lambda_t^{1/2}\widetilde U_t^\top.
\]
Let $(\boldsymbol{\mu}_F(t),\boldsymbol{\Sigma}_F(t))$ denote the driver--space moments in
$\mathbf z$--coordinates and consider the HJB
\[
u_t+\sup_{\vartheta\in\mathbb R^m}
\Big\{\vartheta^\top \boldsymbol{\mu}_F(t)\,u_p
      +\tfrac12\vartheta^\top \boldsymbol{\Sigma}_F(t)\vartheta\,u_{pp}
      -ru\Big\}=0.
\]
Then the HJB (and its optimizer) are invariant under
$\mathbf z\mapsto Q^\top \mathbf z$, i.e.\ under the change $U_t\mapsto \widetilde U_t$.
If $t\mapsto (\boldsymbol{\mu}_F(t),\boldsymbol{\Sigma}_F(t))$ is continuous,
the HJB admits a unique viscosity solution.
If, in addition, the coefficients are $C^{\alpha}$ and uniformly
elliptic in driver space, the solution is classical.
\end{Corollary}

\begin{proof}
Under $U_t\mapsto \widetilde U_t=U_tQ$ we have
\[
\widetilde\Psi_t=\Lambda_t^{1/2}\widetilde U_t^\top
=\Lambda_t^{1/2}Q^\top U_t^\top,
\]
so the transformed driver coordinates satisfy $\mathbf z\mapsto Q^\top \mathbf z$.
Thus $\boldsymbol{\mu}_F\mapsto Q^\top \boldsymbol{\mu}_F$ and $\boldsymbol{\Sigma}_F\mapsto Q^\top \boldsymbol{\Sigma}_F Q$.
Because
\[
\sup_{\vartheta}\Big\{\vartheta^\top \boldsymbol{\mu}_F\,u_p
  +\tfrac12\vartheta^\top \boldsymbol{\Sigma}_F \vartheta\,u_{pp}\Big\}
\]
is invariant under the orthogonal change of variables
$\vartheta=Q\widetilde\vartheta$,
the HJB and its optimizer are unchanged.
Continuity of $(\boldsymbol{\mu}_F,\boldsymbol{\Sigma}_F)$ gives comparison/uniqueness for
viscosity solutions; standard parabolic regularity yields classical
solutions under the stated smoothness and uniform ellipticity in
driver coordinates.
\end{proof}

\begin{Remark}
Because the control enters through $(\boldsymbol{\mu}_F,\boldsymbol{\Sigma}_F)$ in $\mathbf z$--space,
conditioning in asset space is replaced by conditioning with respect
to $\Lambda_t$ on $\mathcal V_t$. The quasi--conformal bound in
Theorem~\ref{thm:conformal-transport} implies that numerical sensitivity
is governed by $\sqrt{\kappa_t}$, typically far smaller than the
ambient asset--space condition number.
\end{Remark}

\clearpage




\bibliographystyle{ws-rv-van}
\bibliography{ws-rv-sample}






\end{document}